\documentclass[referee,sn-basic]{sn-jnl}

\usepackage{xurl} 

\usepackage{amsmath}  
\usepackage{soul} 
\usepackage{color} 
\usepackage{booktabs} 
\usepackage{multirow}
\usepackage[caption=false]{subfig} 

\jyear{2021}%

\newcommand{\hashtag}[1]{\texttt{\##1}}
\newcommand{\aurl}{{\small $\langle$URL$\rangle$}}
\newcommand{\redacted}[1]{{\small $\langle$#1$\rangle$}}

\begin{document}

\title[Online Polarisation During Australia's 2019-2020 Bushfires]{Promoting and countering misinformation during Australia's 2019-2020 bushfires: A case study of polarisation}

\author*[1,2]  {\fnm{Derek}   \sur{Weber}}   \email{derek.weber@\{adelaide.edu.au,dst.defence.gov.au\}}
\author[3,4]   {\fnm{Lucia}   \sur{Falzon}}  \email{lucia.falzon@unimelb.edu.au}
\author[4,5]   {\fnm{Lewis}   \sur{Mitchell}}\email{lewis.mitchell@adelaide.edu.au}
\author[4,5,6] {\fnm{Mehwish} \sur{Nasim}}   \email{mehwish.nasim@flinders.edu.au}

\affil*[1]{
    \orgdiv{School of Computer Science}, \orgname{University of Adelaide}, 
    \orgaddress{\street{North Terrace}, \city{Adelaide}, \postcode{5005}, \state{South Australia}, \country{Australia}}%
}
\affil[2]{
    \orgname{Defence Science and Technology Group}, 
    \orgaddress{\street{West Terrace}, \city{Edinburgh}, \postcode{5111}, \state{South Australia}, \country{Australia}}%
}
\affil[3]{
    \orgdiv{School of Psychological Sciences}, \orgname{University of Melbourne}, 
    \orgaddress{\street{Parkville Campus}, \city{Melbourne}, \postcode{3010}, \state{Victoria}, \country{Australia}}%
}
\affil[4]{
    \orgdiv{School of Mathematical Sciences}, \orgname{University of Adelaide}, 
    \orgaddress{\street{North Terrace}, \city{Adelaide}, \postcode{5005}, \state{South Australia}, \country{Australia}}%
}
\affil[5]{
    \orgname{ARC Centre of Excellence for Mathematical and Statistical Frontiers}, 
    \orgaddress{\street{North Terrace}, \city{Adelaide}, \postcode{5005}, \state{South Australia}, \country{Australia}}%
}
\affil[6]{
    \orgdiv{College of Science and Engineering}, \orgname{Flinders University}, 
    \orgaddress{\street{South Road}, \city{Tonsley}, \postcode{5042}, \state{South Australia}, \country{Australia}}%
}

\abstract{
During Australia's unprecedented bushfires in 2019-2020, misinformation blaming arson resurfaced on Twitter using \hashtag{ArsonEmergency}. 
The extent to which bots were responsible for disseminating and amplifying this misinformation has received scrutiny in the media and academic research.
Here we study Twitter communities spreading this misinformation during the population-level event, and investigate the role of online communities and bots. 
Our in-depth investigation of the dynamics of the discussion uses a phased approach -- before and after reporting of bots promoting the hashtag was broadcast by the mainstream media.

Though we did not find many bots, the most bot-like accounts were \emph{social bots}, which present as genuine humans. 

Further, we distilled meaningful quantitative differences between two polarised communities in the Twitter discussion, resulting in the following insights. 

First, \emph{Supporters} of the arson narrative promoted misinformation by engaging others directly with replies and mentions using hashtags and links to external sources. In response, \emph{Opposers} retweeted fact-based articles and official information.

Second, Supporters were embedded throughout their interaction networks, but Opposers obtained high centrality more efficiently despite their peripheral positions. 
By the last phase, Opposers and unaffiliated accounts appeared to coordinate, potentially reaching a broader audience.

Finally, unaffiliated accounts shared the same URLs as Opposers over Supporters by a ratio of 9:1 in the last phase, having shared mostly Supporter URLs in the first phase. 
This foiled Supporters' efforts, highlighting the value of exposing misinformation campaigns. 
We speculate that the communication strategies observed here could be discoverable in other misinformation-related discussions and could inform counter-strategies.

}


\maketitle

\section{Introduction}

People share an abundance of useful information on social media during crises \citep{BrunsLiang2012,bruns2012qldfloods}. 
This information, if analysed correctly, can rapidly reveal population-level events such as imminent civil unrest, natural disasters, or accidents \citep{tuke2020pachinko}. 
Not all content is helpful, however: different entities may try to popularise false 
narratives using sophisticated social bots and/or engaging humans. 
The spread of 
such mis- and disinformation not only makes it difficult for analysts to use Twitter data for public benefit \citep{nasim2018real} but may also encourage large numbers of people to adopt the false narratives causing social disruption and polarisation, which may then influence public policy and action, and thus can be particularly dangerous during crises \citep{SingerB2019likewar,kuvsen2020you,soufan2021qanon,Scott2021capitolriots}. 

This paper expands our previous work \citep{WeberNFM2020bushfiresspringer} presenting deeper analysis of a case study of the dynamics of misinformation propagation, and the communities which promote or counter it, during one such crisis. 
We demonstrate that polarised groups can communicate/use social media in very different ways even when they are discussing the same issue, and in effect these can be considered communication strategies, as they are promoting their narrative and trying to convince others to accept their position. 

\subsection{The ``Black Summer'' bushfires and misinformation on Twitter}
The 2020 Australian `Black Summer' bushfires (a.k.a., wildfires) burnt over $16$ million hectares, destroyed over $3{,}500$ homes, and caused at least $33$ human and a billion animal  fatalities,\footnote{\url{https://www.abc.net.au/news/2020-02-19/australia-bushfires-how-heat-and-drought-created-a-tinderbox/11976134}} and attracted global media attention.
During the bushfires, as in other crises, social media provided a mechanism for people in the fire zones to provide on-the-ground reports of what was happening around them, a way for those outside to get insight into the events as they occurred (including authorities and media), but also a way for the broader community to connect and process the imagery and experiences through discussion. 
The lack of the traditional information mediator or gatekeeper role played by the mainstream media on social media permits factual errors, mis-interpretation and outright bias to proliferate without check in a way it could not in decades past.
Our analysis of online discussion at this time shows:

\begin{itemize}
    \item Significant Twitter discussion activity accompanied the Australian bushfires, influencing media coverage. 
    \item Clearly discernible communities in the discussion had very different interpretations of the ongoing events.
    \item In the midst of the discussion, false narratives and misinformation circulated on social media, much of it seen during previous crises, including specific statements that: \begin{itemize}
        \item the bushfires were mostly caused by arson; 
        \item preventative backburning efforts had been reduced due to green activism (previously presented in 2009\footnote{\url{https://www.smh.com.au/national/green-ideas-must-take-blame-for-deaths-20090211-84mk.html}});
        \item Australia commonly experiences such bushfires (previously put forward in 2013\footnote{\url{https://www.theguardian.com/world/2013/oct/24/greg-hunt-wikipedia-climate-change-bushfires}}); and
        \item climate change is not related to bushfires.
    \end{itemize}
\end{itemize}

\sloppy All of these statements and their associated narratives were refuted officially, including via a state government inquiry which found that of $11{,}744$ fires, only 
``$11$ were lit with intention to cause a bush fire'' \cite[][p.29]{nsw_bushfire_inquiry2020}. 
In particular, the arson figures being disseminated online were incorrect,\footnote{\url{https://www.abc.net.au/radionational/programs/breakfast/victorian-police-reject-claims-bushfires-started-by-arsonists/11857634}} preventative backburning has increasingly limited effectiveness,
\footnote{\url{https://www.theguardian.com/australia-news/2020/jan/08/hazard-reduction-is-not-a-panacea-for-bushfire-risk-rfs-boss-says}} its use has not been curbed to appease environmentalists,\footnote{\url{https://theconversation.com/theres-no-evidence-greenies-block-bushfire-hazard-reduction-but-heres-a-controlled-burn-idea-worth-trying-129350}}  
the fires are ``unprecedented'',\footnote{The Australian Academy of Science's statement: \url{https://www.science.org.au/news-and-events/news-and-media-releases/statement-regarding-australian-bushfires}} and climate change is, in fact, increasing the frequency and severity of the fires.
\footnote{Science Brief, on 14 January 2020, reports on a survey of 57 papers on the matter conducted by researchers from the University of East Anglia, Imperial College, London, Australia's CSIRO, the Univerity of Exeter and the Met Office Hadley Centre, Exeter: \url{https://sciencebrief.org/briefs/wildfires}}
The Twitter discussion surrounding the bushfires made use of many hashtags, but according to research by \cite{GrahamKeller2020conv} 
reported on ZDNet \citep{Stilgherrian2020zdnet}, the arson narrative was over-represented on 
\hashtag{ArsonEmergency}, likely created as a counter to the pre-existing 
\hashtag{ClimateEmergency} \citep{Barry2020mw}.
Furthermore, their research indicated that \hashtag{ArsonEmergency} was being boosted by bots and trolls.
This 
attracted widespread media attention, with most coverage 
debunking the arson conspiracy theory.\footnote{The BBC's Ros Atkins's video on the matter was one of the most highly shared URLs: \url{https://www.youtube.com/watch?v=aDvmAMsYwNY}} 
This case thus presents an interesting natural experiment: the nature of the 
online narrative, and the communities that formed in the related discussions, before the publication of the ZDnet article 
and then 
after these conspiracy theories were 
debunked.

We present an exploratory mixed-method analysis of the Twitter activity using the term `ArsonEmergency' 
approximately a week before and after 
the publication of the ZDNet article \citep{Stilgherrian2020zdnet}, making use of social network analysis (SNA), behavioural and content analyses. Comparisons are made with activity related to 
another prominent contemporaneous bushfire-related hashtag, \hashtag{AustraliaFire}, and a prominent but unrelated hashtag, \hashtag{brexit}. 
A timeline analysis revealed two points in time that define 
three phases of activity. 
SNA of retweeting behaviour identifies two 
polarised groups of Twitter users: those promoting the arson narrative, and those exposing and arguing against it. 
These polarised groups, along with the 
unaffiliated accounts, provide a further lens through which to examine the behaviour observed. 
Analysis of the networks of different interactions in the data reveal how central these groups became and to what degree they connected to each other and the broader discussion.
Content and co-activity analyses highlight how the different groups used hashtags, external articles and 
other sources to promote their narratives. 
Finally, an analysis of bot-like behaviour then seeks to replicate \citeauthor{GrahamKeller2020conv}'s findings (\citeyear{GrahamKeller2020conv}) and explores the most bot-like contributors in detail, including their contribution to the overall discussion.

\subsection{Expansion from conference version}
This paper expands upon our original work, which was presented at the 2\textsuperscript{nd} Multidisciplinary International Symposium on Disinformation in Open Online Media (MISDOOM) in $2020$ \citep{WeberNFM2020bushfiresspringer} 
by providing:

\begin{itemize}
    \item An examination of polarised and unaffiliated accounts' behaviour and content over time at the group level, which shows how how the Opposers were mostly active only in Phase~2 and the majority of Supporter and Unaffiliated activity appeared in response to that in Phase~3;
    \item A specific research question addressing the polarised accounts' behaviours, other than retweeting, via SNA measures and visualisations, to explore how central to the discussion they were, and to what degree they interacted with each other and the broader discussion community;
    \item A specific research question addressing apparent coordinated retweeting, hashtag use and link sharing behaviour at the group level, via the analysis of new visualisations;
    \item A specific research question addressing the country of origin of polarised accounts and other active participants and to what degree the groups received `external' support, which is addressed with manual examination and categorisation of accounts' self-reported location descriptions, finding that a significant minority of non-Australians were present in the discussion; 
    \item An exploration of inauthentic behaviour via hashtag use and tweet text patterns, finding Supporters engaged in aggressive trolling behaviour more than Opposers;
    \item An examination of the contribution of most bot-like accounts to the discussion, with close examination of five in particular, raising questions regarding the distinction between bot behaviour and highly repetitive human behaviour;
    \item Comparison with a further contemporaneous contentious discussion, namely the \hashtag{brexit} discussion at a time when the United Kingdom was in the final stages of separating from the European Union; and
    \item An expanded literature review and updated sources, including independent reviews of the bushfires that have occurred since the publication of the original conference paper.

\end{itemize}

\subsection{Contribution}

The contribution of this work includes: \begin{enumerate}
    \item Insights into the evolution of a misinformation campaign deliberately exaggerating the role of arson and downplaying the role of climate change in a catastrophic weather event;  
    \item Characterisation of two polarised communities active in the discussion with distinct agendas and communication strategies, also considered within the context of the broader discussion; and 
    \item A further dataset contemporaneous with the original period, augmenting those published in \cite{WeberNFM2020bushfiresspringer};
    \item Methods and approaches for examining the behaviour and interaction of polarised communities in the context of the broader discussion, including co-activity analysis and statistical measures of community homophily.
\end{enumerate}


\subsection{Related Work}

The study of 
Twitter during crises and times of political significance is well established \citep{BrunsLiang2012,bruns2012qldfloods,FlewBBCS2014,Marozzo2017,graham2020virus}, and has provided recommendations to governments and social media platforms alike 
regarding its exploitation for timely community outreach. The social media response of the Australian Queensland State Government was praised for its use of social media to manage communication during devastating floods \citep{bruns2012qldfloods}, and analyses of coordinated behaviour have revealed significant organised anti-lockdown behaviour during the COVID pandemic \citep{graham2020virus,magelinski2020,loucaides2021} and in the lead up to the January 6 Capitol Riots in America \citep{Scott2021capitolriots,Ng2021}.
The continual presence of trolling and bot behaviour diverts attention and can confuse the public at times of political significance, whether it is to generate artificial support for policies and their proponents \citep{KellerICWSM2017,rizoiu2018debatenight,woolley2018us}, harass opponents \citep{KellerICWSM2017,crest2017} or just pollute existing communication channels \citep{woolley2016autopower,nasim2018real,kuvsen2020you}. Malign actors can also foster online community-based conflict \citep{kumar2018conflict,DattaA19conflictnetwork,MaricontiSBCKLS2019cscw} and polarisation \citep{conover2011,garimella2018,morstatter2018alt,Villa2021}.

Misinformation on social media has also been studied \citep{Kumar2018FalseIO,Starbird2019,StarbirdWilson2020,SingerB2019likewar,graham2020virus}, with growing attention to its overall effect on society \citep{Starbird2019,Carley2020}, but many relevant current events are yet to be explored in the peer-reviewed literature. 
Because social media has become such a mainstay of modern communication, misinformation on social media is often amplified on the mainstream media (MSM), or by prominent individuals, often when it aligns with their ideological outlook, which then feeds back into social media as people discuss it further.\footnote{\url{https://www.abc.net.au/triplej/programs/hack/spread-of-arson-disinformation-us-wildfires-similar-to-australia/12666336}} Such cycles have been known to be deliberately fostered \citep{Benkler2018,StarbirdWilson2020,vanbadham2021nyt}.
Patterns of fire-related misinformation similar to those observed on \hashtag{ArsonEmergency} were repeated in the US during Californian wildfires in mid-2020, even causing armed vigilante gangs to form to counter non-existent Antifa activists who were blamed for the fires on social media.\footnote{\url{https://www.theguardian.com/us-news/2020/sep/16/oregon-fires-armed-civilian-roadblocks-police}} 
Arson has again been blamed for the 2021 fires around the Mediterranean, throughout southern Europe and in northern Africa,\footnote{\url{https://edition.cnn.com/2021/08/11/world/wildfires-climate-change-arson-explainer-intl/index.html}} even as the United Nations' Intergovernmental Panel on Climate Change released its sixth Assessment Report stating that humans' effect on climate is now ``unequivocal'' \citep{ipcc2021}.
Furthermore, when the misinformation involved relates to conspiracy theories involving public health measures during a global pandemic, the risk is that adherents will turn away from other evidence-based policies \citep{Ball2020,brazil2020flatearth,soufan2021qanon}.

A particular mixed-method investigation of the disinformation campaign against the White Helmets rescue group in Syria is useful to consider here \citep{StarbirdWilson2020}. 
Starbird \& Wilson identified two clear corresponding clusters of pro- and anti-White Helmet Twitter accounts 
and used them to frame an investigation of how external references to YouTube videos and channels compared with videos embedded in Twitter. 
They found the anti-White Helmet narrative was consistently sustained through ``sincere activists'' and concerted efforts from Russian 
and alternative news sites. These particularly exploited YouTube to spread critical videos, while the pro-White Helmet activity 
relied 
on the White Helmets' own online activities and sporadic media attention. 
Other researchers have found similar patterns \citep{Benkler2018,Jamieson2020,Pacheco2020www}. 
This interaction between supporter and detractor groups 
and the media may offer insight into activity surrounding similar crises.


\subsection{Research Questions}\label{sec:research_questions}

We propose the following research questions to guide our exploration of Twitter activity over an $18$ day period during the 2019-20 Australian ``Black Summer'' bushfires:
\begin{description}
    \item [\textbf{RQ1}] To what extent can online misinformation campaigns be discerned? Are there discernible groups of accounts driving the misinformation, and if so how are they doing it?
    \item [\textbf{RQ2}] How did the spread of arson narrative-related misinformation differ between 
    phases, and did the spread of 
    the hashtag \hashtag{ArsonEmergency} differ from 
    other emergent discussions (e.g., \hashtag{AustraliaFire} and \hashtag{brexit})?
    \item [\textbf{RQ3}] How did the online behaviour of those who 
    prefer the arson narrative 
    differ from those who refute or question it? 
    How was it affected by media coverage exposing how the \hashtag{ArsonEmergency} hashtag was being used?
    \item [\textbf{RQ4}] How central were the communities to the discussion and how insular were they from each other and the broader discussion?
    \item [\textbf{RQ5}] How did the communities make use of retweets, hashtags and URLs to promote their narrative? What evidence is there of coordination?
    \item [\textbf{RQ6}] To what degree did the polarised groups receive support from outside Australia?
    \item [\textbf{RQ7}] To what degree was the spread of misinformation facilitated or aided by trolls and/or automated bot behaviour engaging in inauthentic behaviour? 
\end{description}


In the remainder of this paper, we describe our mixed-method analysis and the datasets used. A timeline analysis is followed by the polarisation analysis. The revealed polarised communities are compared from behavioural and content perspectives, as well as through bot analysis. Answers to the research questions are summarised and we conclude with observations and proposals for further study of polarised communities. 


\section{The Dataset and its Timeline}

The primary dataset, 
`ArsonEmergency', consists of $27{,}546$ tweets containing this term 
posted by $12{,}872$ unique accounts 
from 31 December 2019 to 17 January 2020. 
The tweets were obtained using Twitter's Standard search Application Programming Interface (API)\footnote{\url{https://developer.twitter.com/en/docs/tweets/search/api-reference/get-search-tweets}} by combining the results of searches conducted with Twarc\footnote{\url{https://github.com/DocNow/twarc}} on 8, 12, and 17 January 2020.
As a contrast, the `AusFire' dataset comprises tweets containing the term `AustraliaFire' over the same period, made from the results of Twarc searches on 8 and 17 January 2020. `AusFire' contains $111{,}966$ tweets by $96{,}502$ accounts. 
Broader searches using multiple related terms were not conducted due to time constraints and in the interests of comparison with Graham and Keller's findings \citep{GrahamKeller2020conv}. Due to the use of Twint\footnote{\url{https://github.com/twintproject/twint}} in that study, differences in dataset were likely but expected to be minimal. 
Differences in datasets collected simultaneously with different tools have been previously noted \citep{WeberNMF2021reliability}. 
Live filtering was also not employed, as the research started after Graham and Keller's findings were reported. Twitter may have removed inauthentic content in the time between it being posted and us conducting searches as part of data cleaning routines. For these reasons, some of the content observed by Graham and Keller were expected to be missing from our dataset. This lack of consistency between social media datasets for comparative analyses is a growing challenge recently identified in the benchmarking literature \citep{Assenmacher2021}.
A final contrast dataset was obtained via the Real-Time Analytics Platform for Interactive Data Mining (RAPID) \citep{rapid2017} consisting of tweets containing the term `\hashtag{brexit}' during the same period. It contains $187{,}792$ tweets by $78{,}216$ accounts. 

\subsection{The Timeline}
This study focuses on 
about a week of Twitter activity before and after the publication of the ZDNet article \citep{Stilgherrian2020zdnet}. 
Prior to its publication, the narratives that arson was the primary cause of 
the bushfires and that 
fuel load 
caused the extremity of the blazes were well known in the conservative media \citep{Barry2020mw}. 
The ZDnet article was published at 6:03am GMT (5:03pm AEST\footnote{Australian Eastern Standard Time.}) 
on 7 January 2020, and was then reported more widely in the MSM morning news, starting around 13 hours later. 
We use these temporal markers to define three dataset phases:
\begin{itemize}
    \item \textit{Phase~1}: Before 6am GMT, 7 January 2020; 
    \item \textit{Phase~2}: From 6am to 7pm GMT, 7 January 2020; and
    \item \textit{Phase~3}: After 7pm GMT, 7 January 2020.
\end{itemize}

\begin{figure}[t!]
    \centering
    \includegraphics[width=0.99\columnwidth]{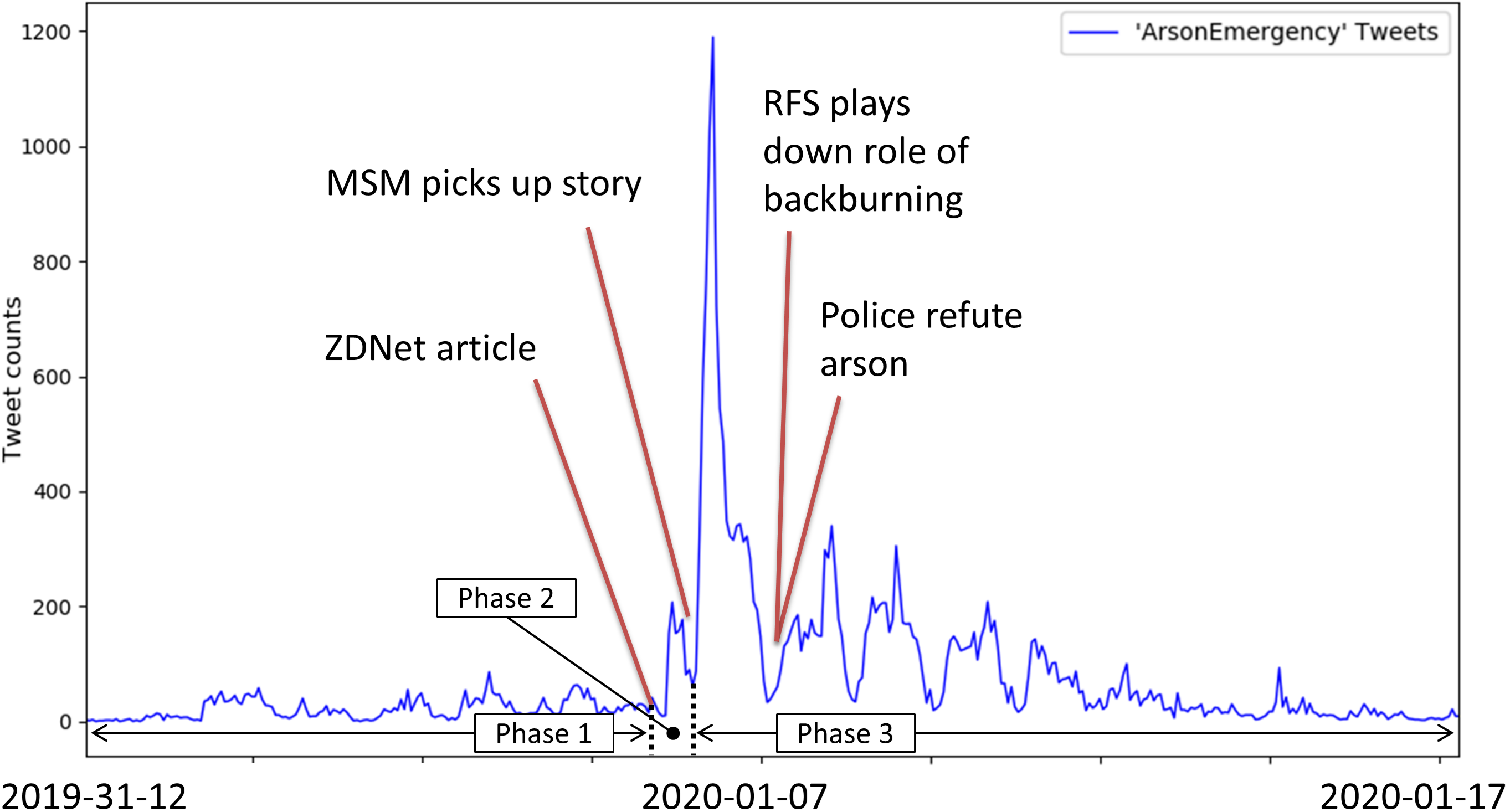}
    \caption{Tweet activity in the `ArsonEmergency' dataset, annotated with notable real-world events and the identified phases.}
    \label{fig:arson-timeline}
\end{figure}

Figure~\ref{fig:arson-timeline} shows the number of tweets posted each hour in the `ArsonEmergency' dataset, and 
highlights the 
phases and notable events including: the publication of the ZDNet article; when the story hit the MSM; the time at which the Rural Fire Service (RFS) and Victorian Police countered the narratives promoted on the \hashtag{ArsonEmergency} hashtag; and the clear subsequent day/night cycle. 
The RFS and Victorian Police announcements countered the false narratives promoted in political discourse in the days prior.

Since late September 2020, Australian and international media had reported on the bushfires around Australia, including 
stories and photos drawn directly from social media, as those caught in the fires shared their experiences. 
No one hashtag had emerged to dominate the online conversation 
and many were in use, including  \hashtag{AustraliaFires}, \hashtag{ClimateEmergency}, \hashtag{bushfires}, and \hashtag{AustraliaIsBurning}. 

The use of \hashtag{ArsonEmergency} was limited in Phase~1, with the busiest hour having around $100$ tweets, but 
there was an influx of new accounts in Phase~2. Of all $927$ accounts active in Phase~2 (responsible for $1{,}207$ tweets), $824$ ($88.9\%$) of them had not posted in Phase~1 (which had $2{,}061$ active accounts). 
$1{,}014$ ($84\%$) of the tweets in Phase~2 were retweets, more than $60\%$ of which were retweets promoting the ZDNet article and the findings it reported. Closer examination of the timeline  
revealed that the majority of the discussion occurred between 9pm and 2am AEST, possibly inflated by a single tweet referring to the ZDNet article (at 10:19 GMT), which was retweeted $357$ times. In Phase~3, more new accounts joined the conversation, but the day/night cycle indicates that the majority of discussion was local to Australia (or at least its major timezones).



\subsection{Growth of the Discussion Community}

\begin{figure}[t!]
    \centering
    \includegraphics[width=0.99\columnwidth]{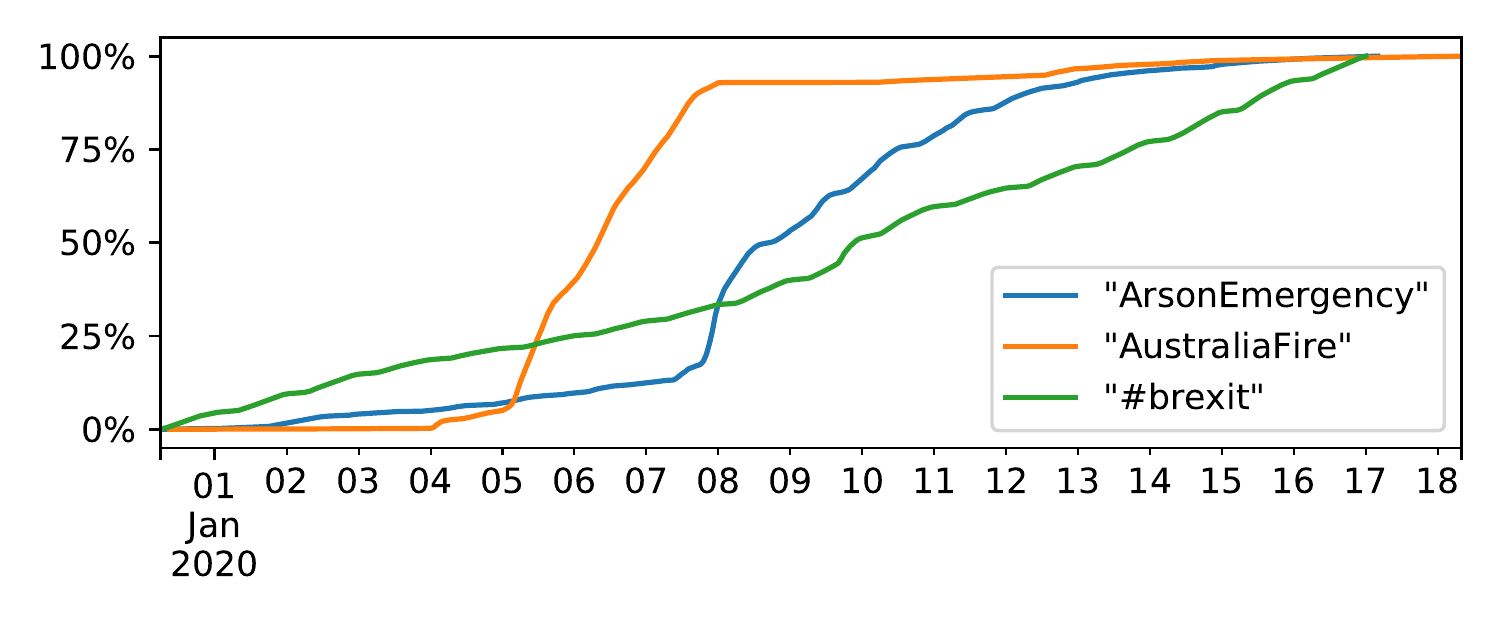}
    \caption{Growth patterns in the number of users using the terms `ArsonEmergency' and `AustraliaFire' (which includes use with the `\#' prefix as well as without), and `\hashtag{brexit}' over the same period.}
    \label{fig:arson-ausfire-brexit-user-growth}
\end{figure}

The figures above raise the question: is this growth in accounts using a term typical? As a contrast, we considered tweets in the same period containing the term `AustraliaFire' and compared the growth in the accounts using the term over time. \hashtag{AustraliaFire} was one of the terms used in Graham's analysis as a contrast to \hashtag{ArsonEmergency}. Figure~\ref{fig:arson-ausfire-brexit-user-growth} shows that the patterns of growth of the users of the two terms differed considerably. One notable difference between the use of these terms is that `AustraliaFire', though employed more than `ArsonEmergency', did not receive the same degree of media exposure around the 7\textsuperscript{th} of January. As a further contrast, the same growth in uses of the term `brexit' in a collection based on just that keyword was also used. Given the use of \hashtag{brexit} is well-established and the period did not include any notable Brexit-related events, we offer it as an example of steady activity. Use of `AustraliaFire' clearly has a significant period of growth in use early in January at a different point to `ArsonEmergency'.

\subsection{Meta-discussion: Avoiding promotion of the hashtag}
\begin{figure}[t!]
    \centering
    \includegraphics[width=0.99\columnwidth]{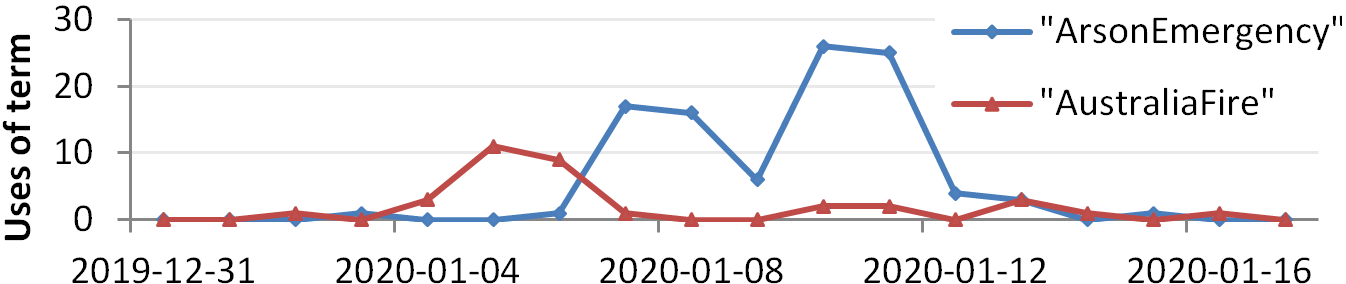}
    \caption{Counts of tweets using the terms `ArsonEmergency' and `AustraliaFire' without a `\#' symbol from the period 2--15 January 2020 in meta-discussion regarding each term's use as a hashtag (counts outside were zero).}
    \label{fig:arson-ausfire-term_only-counts}
\end{figure}

The term `ArsonEmergency' (sans `\#') was used for the Twarc searches, rather than `\#ArsonEmergency', to capture tweets that did not include the hashtag symbol but were relevant to the discussion.
This was done to capture discussions of the term, in which participants deliberately chose to avoid using the term in a way that would contribute to the hashtag discussion (i.e., by including the hashtag symbol). We refer to this as meta-discussion, i.e., discussion \emph{about} the discussion. We sought to understand how much of the discussion relating to \hashtag{ArsonEmergency} was, in fact, meta-discussion. 
Of the $27{,}546$ tweets in the `ArsonEmergency' dataset, only $100$ did not use it with the `\#' symbol ($0.36\%$), and only $34$ of the $111{,}966$ `AustraliaFire' tweets did the same ($0.03\%$), so it is clear that very little of the discussion was meta-discussion. That said, there were several days on which tens of tweets seemed to be involved in meta-discussion, as shown in Figure~\ref{fig:arson-ausfire-term_only-counts}. These coincide with Phase~2, when the story reached the MSM, and then again a few days later, possibly as a secondary reaction to the story (commenting on the initial reaction to the story on the MSM).


The small number of uses in the meta-discussion imply that most use of the term `ArsonEmergency' without the hash or pound symbol is a deliberate, rather than an incidental, part of the discussion. Examination of these particular tweets confirms this; we present examples in Table~\ref{tab:metadiscussion_examples}.

\begin{table}[ht]
    \centering
    \caption{Examples of meta-discussion referring to the \#ArsonEmergency hashtag without including it directly by removing or separating the leading `\#' character.}
    \label{tab:metadiscussion_examples}
    \resizebox{\columnwidth}{!}{%
        \begin{tabular}{p{1.1\linewidth}}  
            \toprule
            Research from QUT shows that `some kind of a disinformation campaign' is pushing the Twitter hashtag \# ArsonEmergency. There is no arson emergency. https://t.co/\aurl \\ 
            \midrule
            @\redacted{ACADEMIC} @\redacted{JOURNALIST} Venn Diagram of ``ArsonEmergency'' with ``Qanon'' and ``Agenda21'' conspiracies could be interesting \redacted{UNIMPRESSED EMOJI} \\ 
            \midrule
            suggest @AFP  @NSWpolice ,@Victoriapolice as this misinformation is likely to cause panic \& distress in Bushfire hit communties.  \newline
            This link is US news but it contains saliant facts about  arrests. https://t.co/\aurl \\ 
            When retweeting, remove  hashtag from `arsonemergency' https://t.co/\aurl \\ 
            \midrule
            @\redacted{JOURNALIST} \#!ArsonEmergency - a notag. \\
            \bottomrule
        \end{tabular}
    } 
\end{table}

\section{Polarised Communities} \label{sec:polarisation}

\begin{figure}[ht!]
    \centering
	\includegraphics[width=0.99\columnwidth]{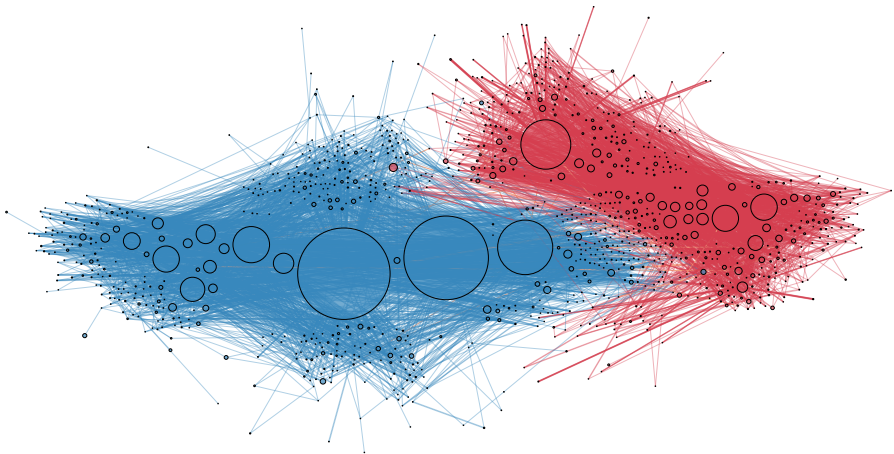}
	\caption{
	Retweet network of the \hashtag{ArsonEmergency} discussion showing clear polarisation. 
	On the left in blue is the \emph{Opposer} community, which countered the arson narrative promoted by the red \emph{Supporter} community on the right.
	Nodes represent users. 
	An edge from node A to B means that account A retweeted one of B's tweets. 
	Node size corresponds to indegree centrality, indicating how often the account was retweeted.}
	\label{fig:retweets-polarised}
\end{figure}

As our aim is to learn about who is promoting the \hashtag{ArsonEmergency} and its related misinformation, we first looked to the retweets. 
Retweets are the primary mechanism for Twitter users to reshare tweets to their own followers. Retweets reproduce a tweet unmodified, except to include an annotation indicating which account retweeted them. 
There is no agreement on whether retweets imply endorsement or alignment. \cite{metaxas2015retweets} studied retweeting behaviour in detail by conducting user surveys and studying over $100$ relevant papers referring to retweets. Their findings conclude that when users retweet, it indicates interest and agreement as well as trust in not only the message content but also in the originator of the tweet. This opinion is not shared by some celebrities and journalists who put a disclaimer on their profile: ``retweets $\neq$ endorsements''. \cite{metaxas2015retweets} also indicated that inclusion of hashtags strengthens the agreement, especially for political topics. Other motivations, such as the desire to signal to others to form bonds and manage appearances \citep{falzon2017representation}, serve to further imply that even if retweets are not endorsements, we can assume they represent agreement or an appeal to likemindedness at the very least. 

Given the highly connected nature of Twitter data and our aim of exploring human social behaviour, using networks to model our data facilitate social network analysis is a logical step \citep{brandes2005}. Using nodes to represent individuals, edges can be used to represent the flow of information and influence and the strength of those connections. 
We conducted an exploratory analysis on the retweet network built from the `ArsonEmergency' dataset, which shown in Figure~\ref{fig:retweets-polarised}. The nodes represent Twitter accounts and are sized by indegree (i.e., frequency of being retweeted). An edge between two accounts shows that one retweeted a tweet of the other. Using conductance cutting \citep{BrandesGW2015clustering}, we discovered two distinct well-connected communities, with a very low number of edges between the two communities. Next, we selected the top ten most retweeted accounts from each community, 
manually checked their profiles, and hand labelled them as 
\emph{Supporters} and \emph{Opposers} of the arson narrative accordingly.\footnote{Labelling was conducted by the first two authors independently and then compared. Account labelling is available on request.} 
The accounts have been coloured accordingly in Figure~\ref{fig:retweets-polarised}: the $497$ red nodes are accounts that promoted the narrative (the Supporters), while the $593$ blue nodes are accounts that opposed them (the Opposers).

The term \hashtag{ArsonEmergency} had different connotations for each community. Supporters used the hashtag to reinforce and promote their existing beliefs about climate change, while Opposers used this hashtag to refute the arson theory. 
The arson theory was a topic on which people held strong opinions resulting in the formation of the two strongly connected communities. Such polarised communities typically do not admit much information flow between them, hence members of such communities are repeatedly exposed to similar narratives, which then further strengthens their existing beliefs. Such closed communities are also known as \emph{echo chambers}, and they limit people's information space. The retweets tend to coalesce within communities, as has also been shown for Facebook comments \citep{nasim2013commenting}. 

These two groups, Supporters and Opposers, and those users \emph{Unaffiliated} with either group, are used to frame the remainder of the analysis in this paper.




\subsection{Community Timelines}\label{sec:community_timelines}

\begin{figure}[th!]
    \centering
	\includegraphics[width=0.99\columnwidth]{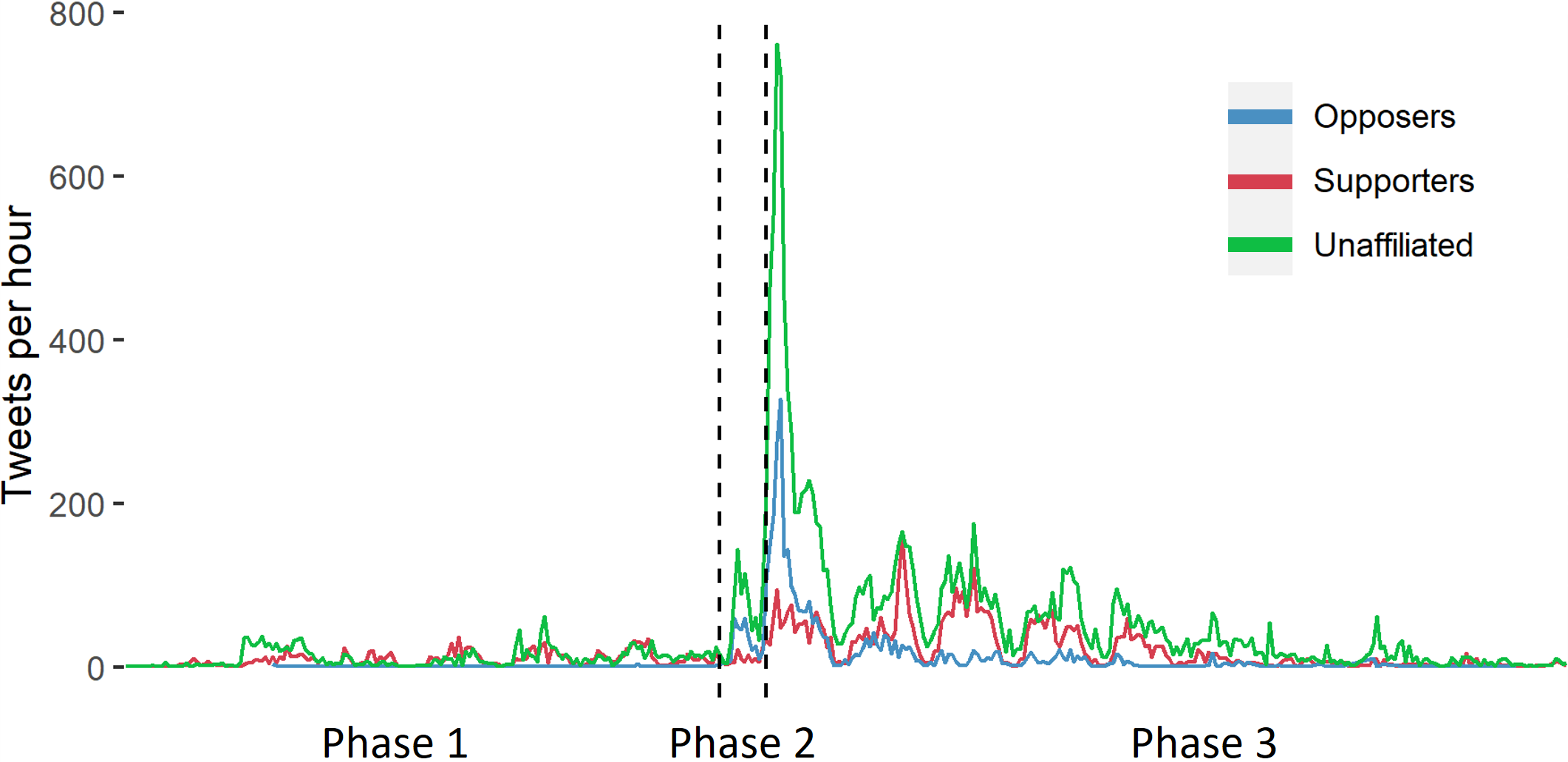}
	\caption{A timeline of each communities' activity over the collection period.}
	\label{fig:community-timelines}
\end{figure}

The relative behaviour of the communities over the collection period, shown in Figure~\ref{fig:community-timelines}, informs several key observations. The first is the impact of the story reaching the MSM: the peaks of both Opposer and Unaffiliated contributions is on the morning of Phase 3, immediately after the story appeared on the morning bulletins. Despite the much greater number of Unaffiliated accounts ($11{,}782$), 
their peak is only a little more than twice that of the $593$ Opposer accounts. Unaffiliated and Supporter accounts are active during the entire collection, but Supporters' activity is prominent each day in Phase~3, and peaks on the second day of Phase~3. That peak might have occurred as a response to the previous peak, as by that time the news would have had a full day to disseminate around the world. By reaching a broader audience via the MSM, more Supporter accounts may have been drawn into the online discussion.

\begin{figure}[th]
    \centering
    \includegraphics[width=0.99\textwidth]{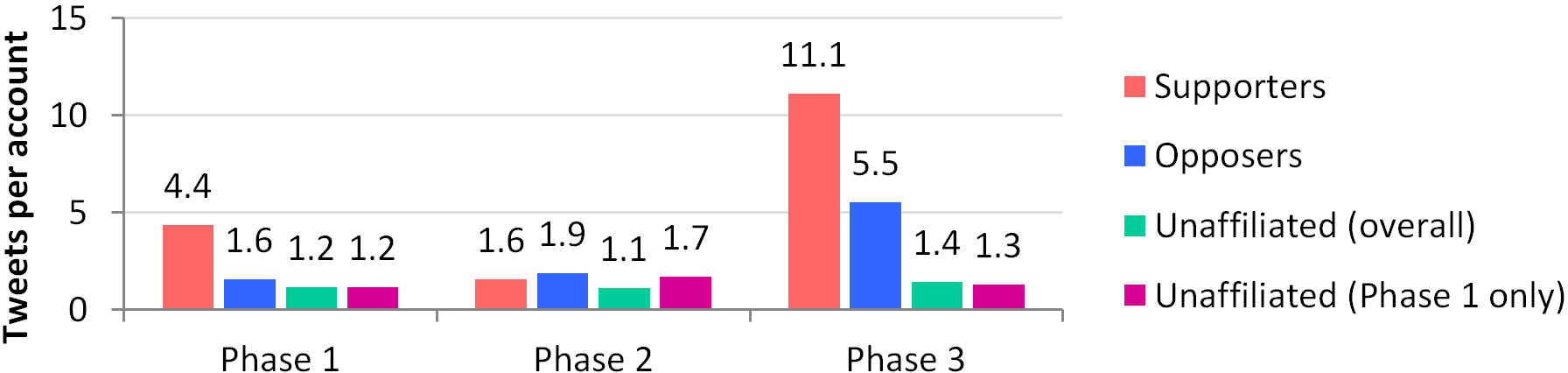}
    \caption{Tweets per account in each phase for the Supporters, Opposers, Unaffiliated accounts overall, and Unaffiliated accounts active in Phase~1.}
    \label{fig:tweets_per_account_by_phase}
\end{figure}

Analysis confirms that the composition of the Unaffiliated participants did change across the phases. Of the $1{,}680$ Unaffiliated accounts active in Phase~1, only $30$ participated in Phase~2 (of $678$ Unaffiliated accounts) and $427$ in Phase~3 (of $10{,}074$ Unaffiliated accounts). Furthermore, the contributions of those Phase~1 Unaffiliated accounts in the later phases were not markedly different from the other Unaffiliated accounts, with the Phase~1 accounts contributing $51$ (of $759$) and $561$ (of $14{,}267$) Unaffiliated tweets in Phases~2 and~3, respectively. This suggests that new accounts joined the discussion with similar enthusiasm, and neither the new nor old accounts dominated the Unaffiliated contribution (Figure~\ref{fig:tweets_per_account_by_phase}).

Finally, a clear diurnal effect can be seen with daily peaks of activity occurring during Australian daytime hours, implying that the majority of the activity is 
domestic and analysis of the `lang' field in the tweets\footnote{The `lang' field is automatically populated by Twitter based on language detection. If insufficient content is available (e.g., the tweet is empty, or only contains URLs or mentions, `und' is used to mean `undefined'.} confirmed that over $99\%$ of tweets used `en' (English, $90.5\%$) or `und' (undefined, $8.7\%$).

\subsection{Behaviour}

User behaviour on Twitter can be examined through the features used to connect with others and through content. Here we consider how active the different groups were across the phases of the collection, and then how that activity manifested itself in the use of mentions, hashtags, URLs, replies, quotes and retweets. 

\begin{table}[th]
    \centering
    \caption{Activity of the polarised retweeting accounts, by interaction type 
    in phases.
    }
    \label{tab:group-activity-by-interaction-type-and-phase}
    \resizebox{0.99\textwidth}{!}{%
        \begin{tabular}{@{}lllrrrrrrrr@{}}
            \toprule
                & Group        &                      & Tweets & Accounts & Hashtags & Mentions & Quotes & Replies & Retweets & URLs  \\ 
            \midrule
            \multirow{6}{*}{\rotatebox[origin=c]{90}{\textbf{Phase 1}}} 
                &              & \textit{Raw count}   & 1,573  & 360      & 2,257    & 1,020    & 185    & 356     & 938      & 405   \\
                & Supporters   & \textit{Per account} & 4.37   & --       & 6.27     & 2.83     & 0.51   & 0.99    & 2.61     & 1.13  \\
                &              & \textit{Per tweet}   & --     & --       & 1.43     & 0.65     & 0.12   & 0.23    & 0.60     & 0.26  \\
                \cmidrule{3-11}
                &              & \textit{Raw count}   & 33     & 21       & 100      & 5        & 8      & 2       & 20       & 9     \\
                & Opposers     & \textit{Per account} & 1.57   & --       & 4.76     & 0.24     & 0.38   & 0.10    & 0.95     & 0.43  \\
                &              & \textit{Per tweet}   & --     & --       & 3.03     & 0.15     & 0.24   & 0.06    & 0.61     & 0.27  \\
            \midrule
            \multirow{6}{*}{\rotatebox[origin=c]{90}{\textbf{Phase 2}}} 
                &              & \textit{Raw count}   & 121    & 77       & 226      & 64       & 11     & 29      & 74       & 24    \\
                & Supporters   & \textit{Per account} & 1.57   & --       & 2.94     & 0.83     & 0.14   & 0.38    & 0.96     & 0.31  \\
                &              & \textit{Per tweet}   & --     & --       & 1.87     & 0.53     & 0.09   & 0.24    & 0.61     & 0.20  \\
                \cmidrule{3-11}
                &              & \textit{Raw count}   & 327    & 172      & 266      & 34       & 7      & 14      & 288      & 31    \\
                & Opposers     & \textit{Per account} & 1.90   & --       & 1.55     & 0.20     & 0.04   & 0.08    & 1.67     & 0.18  \\
                &              & \textit{Per tweet}   & --     & --       & 0.81     & 0.10     & 0.02   & 0.04    & 0.88     & 0.09  \\
            \midrule
            \multirow{6}{*}{\rotatebox[origin=c]{90}{\textbf{Phase 3}}} 
                &              & \textit{Raw count}   & 5,278  & 474      & 7,414    & 2,685    & 593    & 1,159   & 3,212    & 936   \\
                & Supporters   & \textit{Per account} & 11.14  & --       & 15.64    & 5.66     & 1.25   & 2.45    & 6.78     & 1.97  \\
                &              & \textit{Per tweet}   & --     & --       & 1.40     & 0.51     & 0.11   & 0.22    & 0.61     & 0.18  \\
                \cmidrule{3-11}
                &              & \textit{Raw count}   & 3,227  & 585      & 3,997    & 243      & 124    & 95      & 2,876    & 359   \\
                & Opposers     & \textit{Per account} & 5.52   & --       & 6.83     & 0.42     & 0.21   & 0.16    & 4.92     & 0.61  \\
                &              & \textit{Per tweet}   & --     & --       & 1.24     & 0.08     & 0.04   & 0.03    & 0.89     & 0.11  \\
            \midrule
            \multirow{9}{*}{\rotatebox[origin=c]{90}{\textbf{Overall}}} 
                &              & \textit{Raw count}   & 6,972  & 497      & 9,897    & 3,769    & 789    & 1,544   & 4,224    & 1,365 \\
                & Supporters   & \textit{Per account} & 14.03  & --       & 19.91    & 7.58     & 1.59   & 3.11    & 8.50     & 2.75  \\
                &              & \textit{Per tweet}   & --     & --       & 1.42     & 0.54     & 0.11   & 0.22    & 0.61     & 0.20  \\
                \cmidrule{3-11}
                &              & \textit{Raw count}   & 3,587  & 593      & 4,363    & 282      & 139    & 111     & 3,184    & 399   \\
                & Opposers     & \textit{Per account} & 6.05   & --       & 7.36     & 0.48     & 0.23   & 0.19    & 5.37     & 0.67  \\
                &              & \textit{Per tweet}   & --     & --       & 1.22     & 0.08     & 0.04   & 0.03    & 0.89     & 0.11  \\
                \cmidrule{3-11}
                &              & \textit{Raw count}   & 16,987 & 11,782   & 22,192   & 3,474    & 615    & 1,377   & 14,119   & 1,790 \\
                & Unaffiliated & \textit{Per account} & 1.44   & --       & 1.88     & 0.29     & 0.05   & 0.12    & 1.20     & 0.15  \\
                &              & \textit{Per tweet}   & --     & --       & 1.31     & 0.20     & 0.04   & 0.08    & 0.83     & 0.11  \\ 
            \bottomrule

        \end{tabular}
    }
\end{table}

In Phase~1, Supporters used \hashtag{ArsonEmergency} nearly fifty times more often than Opposers ($2{,}086$ to $43$), which accords with Graham and Keller's findings that the false narratives were significantly more prevalent on that hashtag compared with others in use at the time \citep{Stilgherrian2020zdnet,GrahamKeller2020conv}. This use is roughly proportional to the number of tweets posted by the two groups, however  (Table~\ref{tab:group-activity-by-interaction-type-and-phase}). Overall in that Phase, Supporters used $22$ times as many hashtags as Opposers. In Phase~2, during the Australian night, Opposers countered with three times as many tweets as Supporters, including fewer hashtags, more retweets, and half the number of replies, demonstrating different behaviour to Supporters, which actively used hashtags in conversations. 
Manual inspection and content analysis confirmed this to be the case. This is evidence that Supporters wanted to promote the hashtag as a way to promote the narrative. Interestingly, Supporters, having been relatively quiet in Phase~2, responded strongly, producing 64\% more tweets in Phase~3 than Opposers. 
They used proportionately more of all interactions except retweeting, including many more replies, quotes, and tweets spreading the narrative with multiple hashtags, URLs and mentions. In short, Opposers tended to rely more on retweets, while Supporters engaged directly 
and were more active in the longer phases.

Overall, as shown in the bottom section of Table~\ref{tab:group-activity-by-interaction-type-and-phase}, Supporter accounts tweeted much more often than other accounts, and used more hashtags, mentions, quotes, replies and URLs, but retweeted less often than both Opposers and Unaffiliated accounts. This suggests that Supporters were generating their own content (not just retweeting it), and attempting to engage with others through the use of platform features, implying a high degree of motivation on their part.

\subsubsection{Other interaction networks}

\begin{figure}[!th]
    \centering
    \subfloat[Replies.\label{fig:arson-replies-network}]{%
        \includegraphics[width=0.49\textwidth]{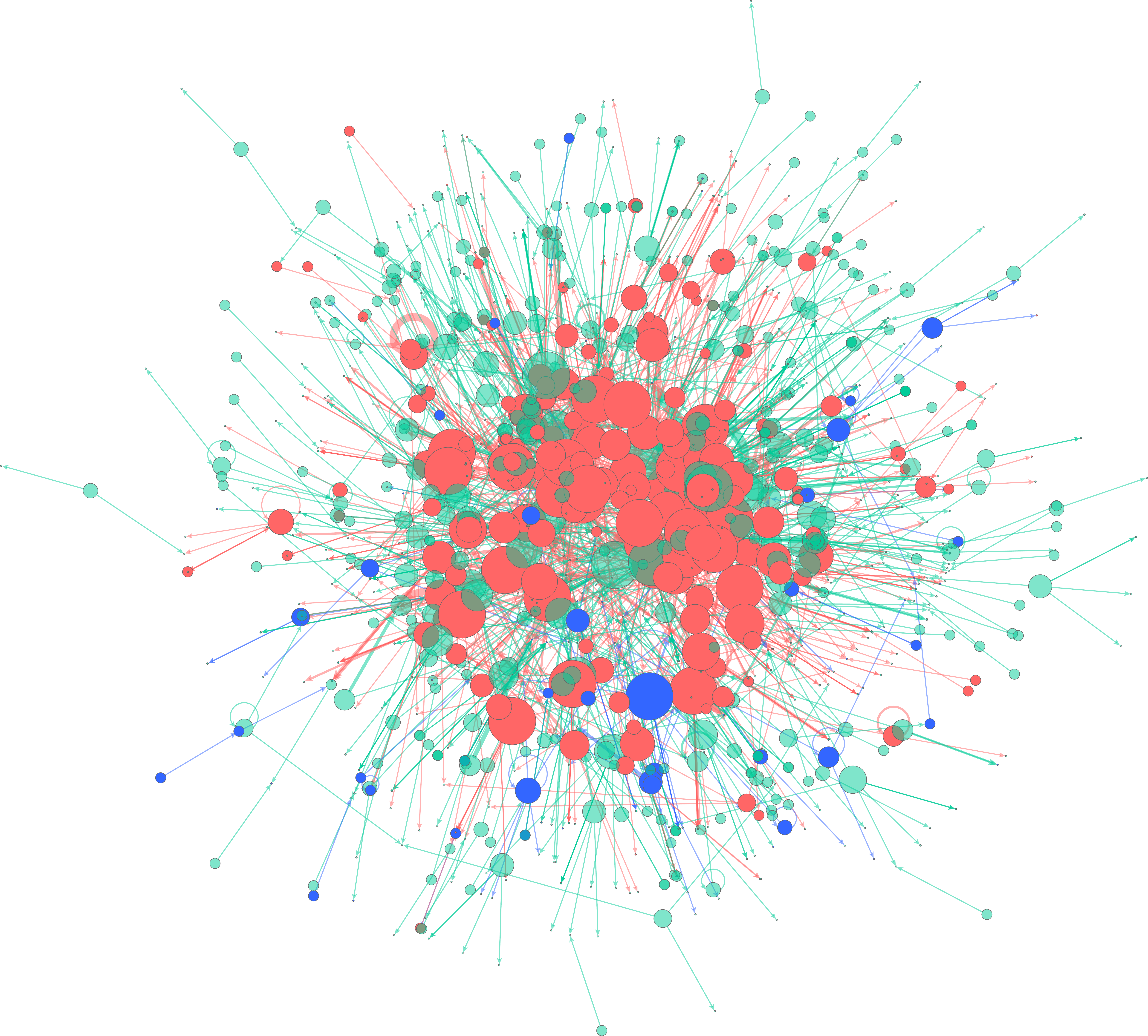}
    }
    \hfill
    \subfloat[Mentions.\label{fig:arson-mentions-network}]{%
        \includegraphics[width=0.49\textwidth]{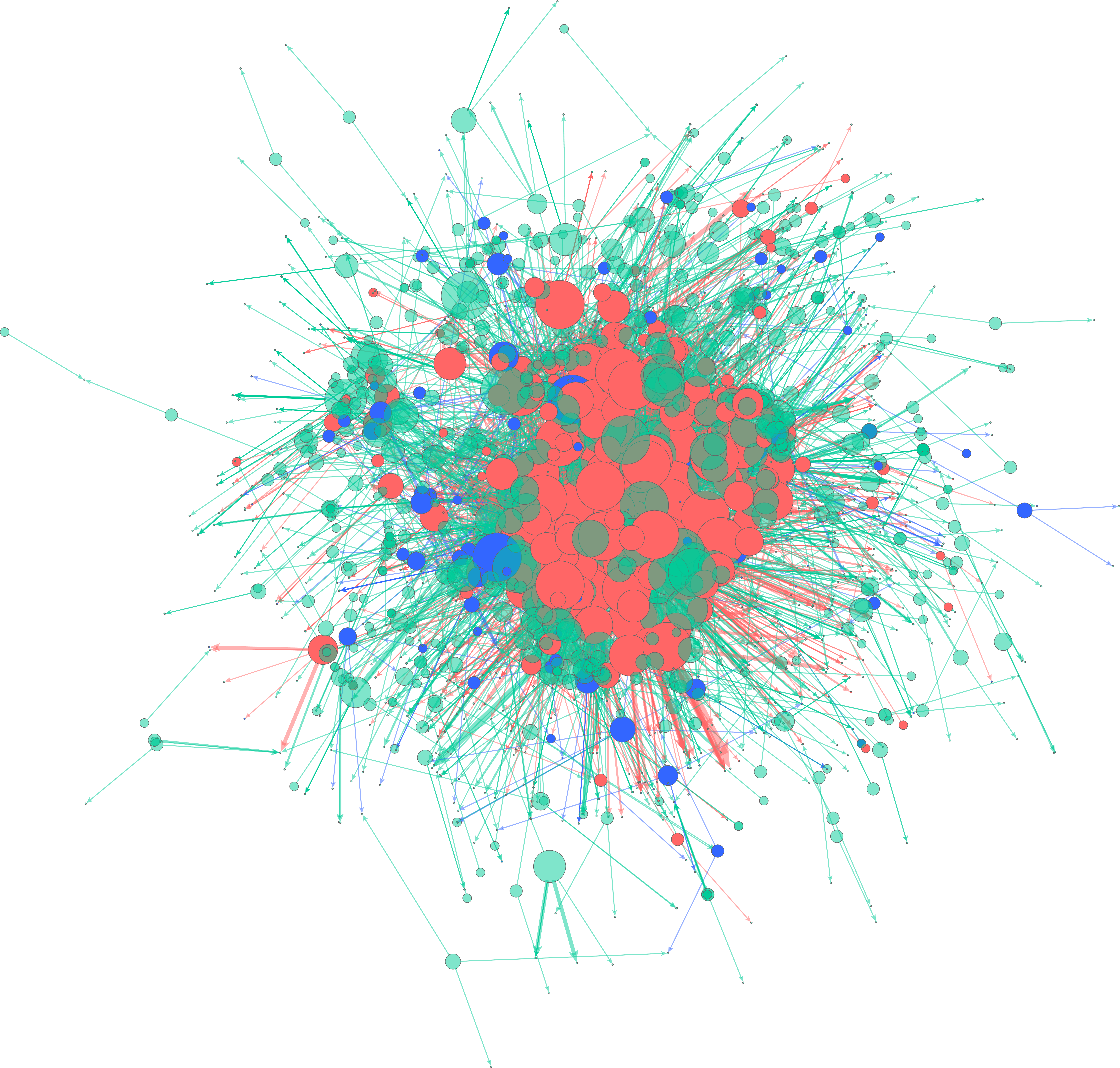}
    }
    \\ 
    \subfloat[Quotes.\label{fig:arson-quotes-network}]{%
        \includegraphics[width=0.49\textwidth]{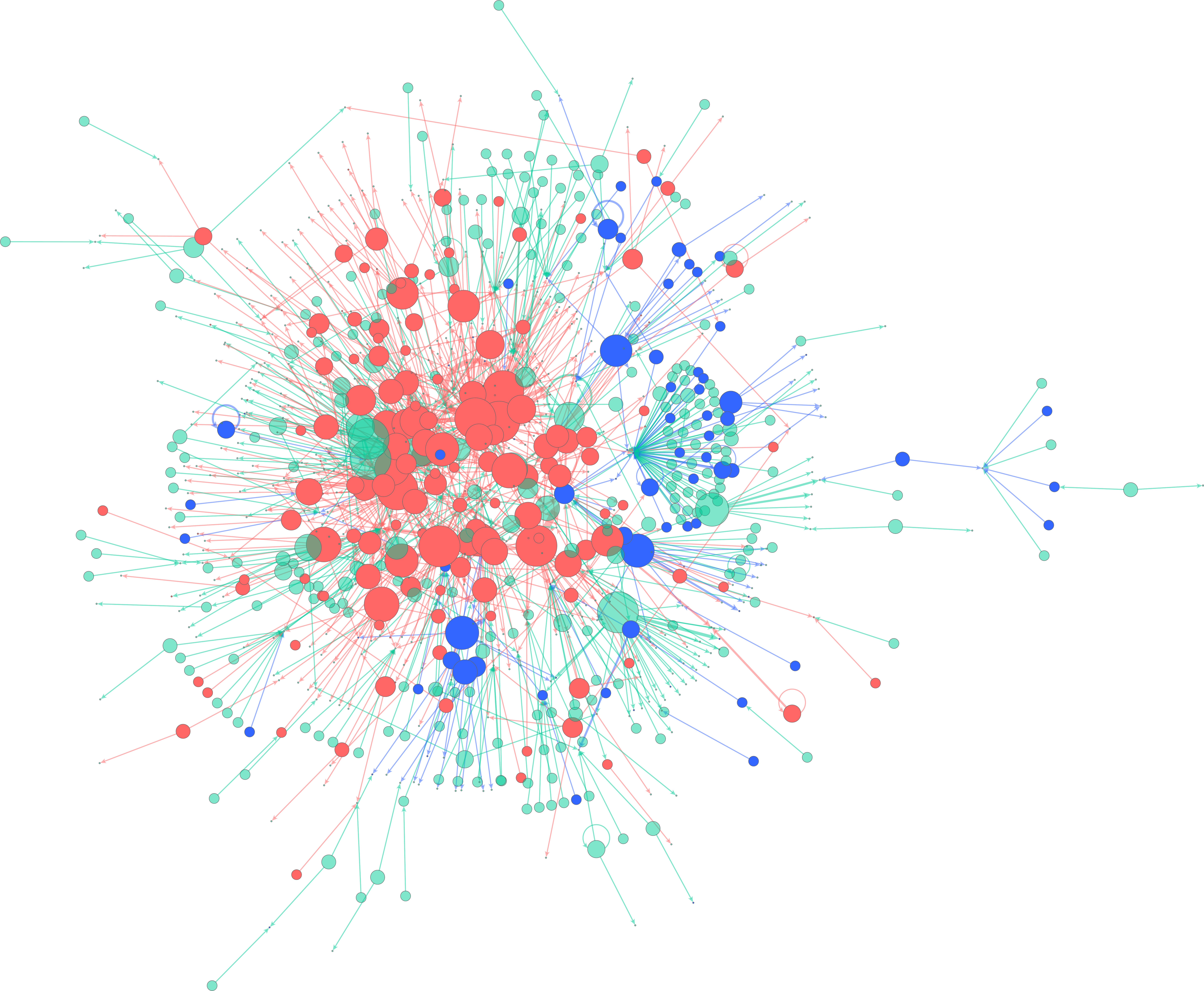}
    }
    \caption{The largest connected components from directed, weighted networks built from the replies, mentions, and quotes, linking from one account to another when it replied, mentioned, or quoted the other, laid out by extracting the quadrilateral Simmelean backbone~\protect\citep{Serrano2009backbone,nocaj2014untangling}. Edges are sized by weight, indicating the frequency of connections, and coloured by source node affiliation. Thicker edges have greater weight. Nodes are sized by outdegree (indicating the replies, mentions and quotes they used) and coloured by affiliation: red nodes are Supporters, blue are Opposers, and green are Unaffiliated. The replies component has 1,580 nodes and 2,308 edges, the mentions component has 2,984 nodes and 5,670 edges, and the quotes component has 915 nodes and 1,230 edges.}
    \label{fig:arson-networks}
\end{figure}

If Supporters employed a variety of interaction mechanisms, while Opposers relied primarily on retweeting, then Supporters should be deeply embedded in networks constructed from those other interaction mechanisms. This is exactly what we find when we examine the largest components of networks constructed from replies (Figure~\ref{fig:arson-replies-network}), mentions (Figure~\ref{fig:arson-mentions-network}), and quotes (Figure~\ref{fig:arson-quotes-network}). These largest components include 77.4\%, 92.0\%, and 72.2\% of the reply, mention and quote networks' nodes, respectively. Supporters had more connections (correspondingly represented by larger nodes) and are clearly more active than Opposers using these interactions, engaging with each other and others in the network. They are particularly tightly and centrally clustered in the mention network, which is a reflection of their attempts to actively engage directly (rather than only indirectly, such as with hashtags). They are more diffusely located in the reply network, and the quote network, sharing similar network positions to Unaffiliated accounts. This is less to do with the amount of activity (i.e., the number of replies or tweets) and more to do with how they connect with others. The Opposer accounts that appear in the networks are not as centrally located nor as tightly clustered. 

To provide a more objective analysis of the structural properties of these networks and the accounts within them, we employ a variety of centrality measures and $k$-core analysis. We also use the assortativity coefficient and a variation of Krackhardt and Stern's E-I Index \citep{Krackhardt1988} as measures of homophily. Centrality measures provide an indication of the importance of a node within a network, while the $k$-core of a node describes how deeply embedded it is within its network based on its connectivity,\footnote{``A $k$-core is a maximal subset of vertices such that each is connected to at least $k$ others in the subset.'' \cite[p.196]{newman2010networks}} and assortativity measures the degree to which accounts in the same groups connect to each other (i.e., their degree of homophily). The reader is referred to \cite{newman2010networks} for an introduction to these concepts. The E-I~Index is a simple ratio of the \emph{internal} edges that connect members of a labeled group to each other, $I$, compared with \emph{external} edges connecting to nodes outside the group, $E$:

\begin{equation}
    E\textnormal{-}I \; Index = \frac{
        \lvert E \rvert - \lvert I \rvert
    }{
        \lvert E \rvert + \lvert I \rvert
    }.
\end{equation}

Both the assortativity coefficient and the E-I Index lie within $[-1,1]$ but the values are reversed: high assortativity coefficients and low E-I indices indicate highly homophilous networks in which nodes connect mostly with others in the same group. Our E-I Index implementation addresses both the availability of edge weights\footnote{Edge weights are ignored in the implementation of the E-I Index in the version of \texttt{NetworkX} \citep{Hagberg2008nx} that we used, version 2.5, which is why we implemented our own.} and imbalances in the size of the polarised groups of interest. It does this by summing the weights of edges (rather than just their number) and then normalising the sums, so what is considered is the proportion of the edge weight sum that connects outside the group compared to inside the group. We refer to this measure as the modified E-I Index in the remainder of this work.

\paragraph{Centrality} Though the location of Supporter and Opposer accounts in the networks in Figure~\ref{fig:arson-networks} gives the impression that Supporters are more central in each network, the statistics presented in Table~\ref{tab:mean_int_cents_by_affiliation} facilitate a more nuanced interpretation. In the reply, mention and quote networks, Supporters and Opposers make up only a small fraction of the overall networks (shown as a percentage in the Nodes column). Supporter betweenness scores are much higher than Opposers' in the reply and mention networks and even twice as high in the quote network (though still very low). Closeness scores are more weighted towards the Opposers, implying that even though they are not centrally positioned, they remain directly linked to more of the network than the Supporters. The mean degree centrality of Supporters is again higher than Opposers' for all networks, reflecting their tendency to directly reach out to a wider audience than Opposers, who relied mostly on retweets to disseminate their message. The eigenvector centrality scores are higher for Opposers in the reply and mention networks, suggesting they are more connected to important nodes in the network and perhaps were more efficient at selecting their interaction targets, while their lower scores for the quotes network is probably reflective of the fact they used them a lot less ($139$ uses to Supporters' $789$). The centrality scores suggest that the Opposers were less centrally located, but well connected, while Supporters were more centrally positioned (reflected in their relatively high betweenness scores).

\begin{table}[th!]
    \centering
    \caption{Mean centrality scores for Supporter and Opposer nodes in the largest components of the reply, mention, and quote networks, omitting Unaffiliated node scores.}
    \label{tab:mean_int_cents_by_affiliation}
    \resizebox{0.99\textwidth}{!}{%
    \begin{tabular}{@{}llrrrrrr@{}}
        \toprule
                                  &             &              &            & \multicolumn{4}{c}{\textbf{Centrality}}          \\
        Network                   & Group       & \multicolumn{2}{c}{Nodes} & Betweenness & Closeness & Degree   & Eigenvector \\
        \cmidrule(r){1-1} \cmidrule(lr){2-2} \cmidrule(lr){3-4} \cmidrule(lr){5-5} \cmidrule(lr){6-6} \cmidrule(lr){7-7} \cmidrule(l){8-8}
        \multirow{2}{*}{Replies}  & Supporters  & 231          & (14.6\%)   & 0.000181    & 0.002871  & 0.004551 & 0.001307    \\
                                  & Opposers    & 82           & (5.2\%)    & 0.000019    & 0.003453  & 0.002757 & 0.001811    \\
        \cmidrule(lr){2-2} \cmidrule(lr){3-4} \cmidrule(lr){5-5} \cmidrule(lr){6-6} \cmidrule(lr){7-7} \cmidrule(l){8-8}
        \multirow{2}{*}{Mentions} & Supporters  & 284          & (9.6\%)    & 0.000304    & 0.005525  & 0.004207 & 0.006575    \\
                                  & Opposers    & 140          & (4.7\%)    & 0.000018    & 0.005067  & 0.001997 & 0.006625    \\
        \cmidrule(lr){2-2} \cmidrule(lr){3-4} \cmidrule(lr){5-5} \cmidrule(lr){6-6} \cmidrule(lr){7-7} \cmidrule(l){8-8}
        \multirow{2}{*}{Quotes}   & Supporters  & 169          & (18.5\%)   & 0.000012    & 0.001876  & 0.006170 & 0.016033    \\
                                  & Opposers    & 80           & (8.7\%)    & 0.000005    & 0.003334  & 0.004171 & 0.007302    \\
        \bottomrule
    \end{tabular}%
    }
\end{table}

\begin{figure}[!ht]
    \centering
    \includegraphics[width=0.99\textwidth]{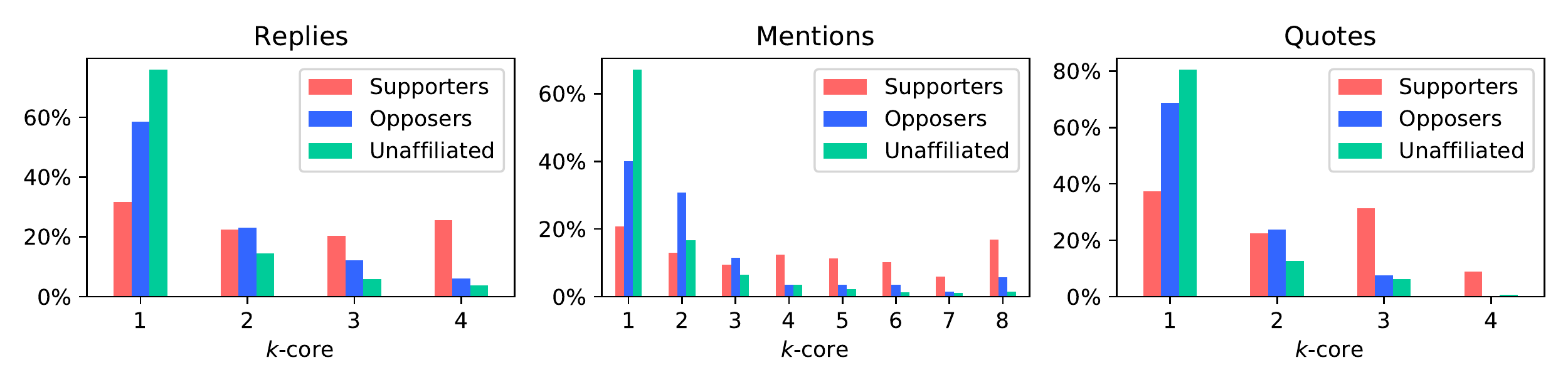}
    \caption{The distributions of $k$-core values for accounts in the reply, mention and quote networks. Nodes with higher $k$-core values are more deeply embedded in their network. The percentage refers to the proportion of each group's accounts with a given $k$-core value.}
    \label{fig:int_networks_k-core_proportions}
\end{figure}

\paragraph{\texorpdfstring{$k$}{k}-core analysis} The question of how tightly clustered the nodes are can be addressed with $k$-core analysis. This analysis progressively breaks a network down to sets of nodes that have at least $k$ neighbours, so nodes on the periphery are discarded first, while highly connected nodes form the `core' of the network. The result is that the higher the $k$-core for a particular node (i.e., the highest $k$-core of which they are a member), the more embedded in the network they are. Figure~\ref{fig:int_networks_k-core_proportions} shows the proportions of each groups' members (of those present in each network) in each core. We can immediately see that across all networks, more Supporters have higher $k$-core values than both Opposers and the Unaffiliated. In fact, while the majority of Opposers and Unaffiliated are on the periphery of the networks, Supporters are relatively evenly spread throughout the networks' cores. This implies more of the Supporters were more active in reaching out to many alters, something that is also reflected in their higher use of mentions, replies and quotes per account, as shown in Table~\ref{tab:group-activity-by-interaction-type-and-phase}.

\paragraph{Homophily measures} The homophily measures introduced provide an indication of how insular the groups were with their interactions, and here we also apply them to the retweet network for comparison (Table~\ref{tab:int_networks_homophily}). As expected, the vast majority of edges involving Supporters and Opposers in the retweet network are homophilic, leading to a very low E-I Index and a very high assortativity coefficient. Even when the broader network is introduced (i.e., re-introducing all Unaffiliated nodes), the E-I Index remains very low, dominated by the polarised groups. Polarisation is maintained between Supporters and Opposers in the other interaction networks, but to a lesser degree, with the most separation observed in the quote network, again drawing our attention to the fact that Supporters used quotes more than Opposers.

\begin{table}[!ht]
    \centering
    \caption{Homophily measures calculated with just Supporters and Opposers and then all nodes within interaction networks.}
    \label{tab:int_networks_homophily}
    \resizebox{0.99\textwidth}{!}{%
        \begin{tabular}{@{}lrrrrrrrrrr@{}}
            \toprule
            \textbf{Network} & \multicolumn{2}{c}{\textbf{Nodes}} & \multicolumn{3}{c}{\textbf{Edges}} & \multicolumn{2}{c}{\textbf{Homophily}} & \multicolumn{3}{c}{\textbf{Broader Network}} \\
                             & Supporters        & Opposers       & Homophilic & Heterophilic & Total  & E-I Index        & Assortativity       & Nodes        & Edges        & E-I Index      \\
            \cmidrule(r){1-1} \cmidrule(lr){2-3} \cmidrule(lr){4-6} \cmidrule(lr){7-8} \cmidrule(l){9-11}
            Retweet          & 493               & 592            & 99.5\%     & 0.5\%        & 5,418  & -0.98935         & 0.98819             & 12,076       & 19,755       & -0.78718       \\
            Reply            & 247               & 105            & 70.8\%     & 29.2\%       & 388    & -0.41667         & 0.19642             & 2,041        & 2,646        & 0.60565        \\
            Mention          & 288               & 149            & 57.0\%     & 43.0\%       & 710    & -0.14032         & 0.10702             & 3,206        & 5,825        & 0.70202        \\
            Quote            & 190               & 104            & 86.0\%     & 14.0\%       & 293    & -0.72093         & 0.61272             & 1,268        & 1,462        & 0.40240        \\

            \bottomrule
        \end{tabular}%
    }
\end{table}

\begin{figure}[!ht]
    \centering
    \subfloat[Raw counts of outgoing connections.\label{fig:arson-int-count-heatmaps-raw}]{%
        \includegraphics[width=0.48\columnwidth]{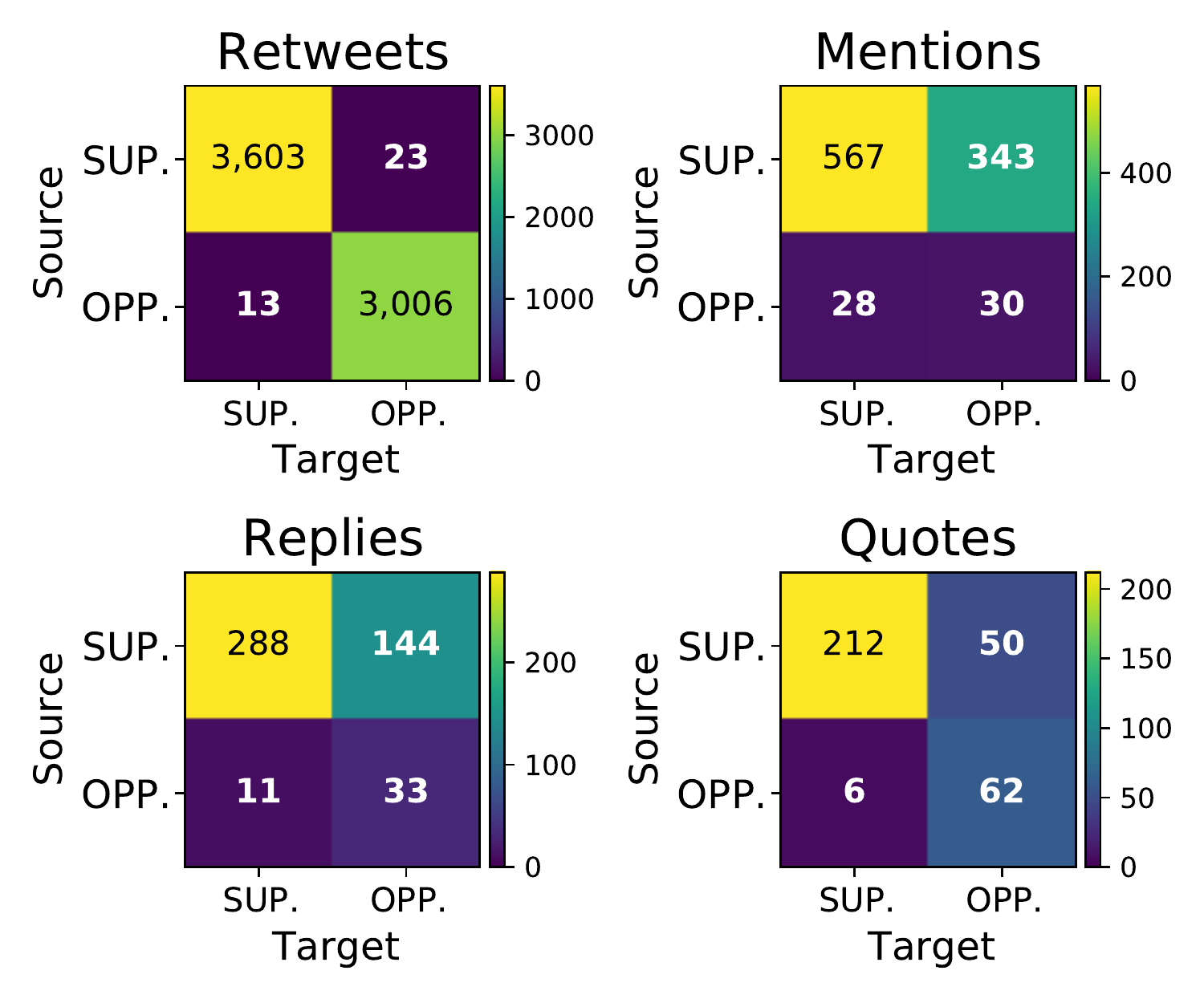}
    }
    \hfill 
    \subfloat[Proportion of outgoing connections.\label{fig:arson-int-count-heatmaps-proportional}]{%
        \includegraphics[width=0.48\columnwidth]{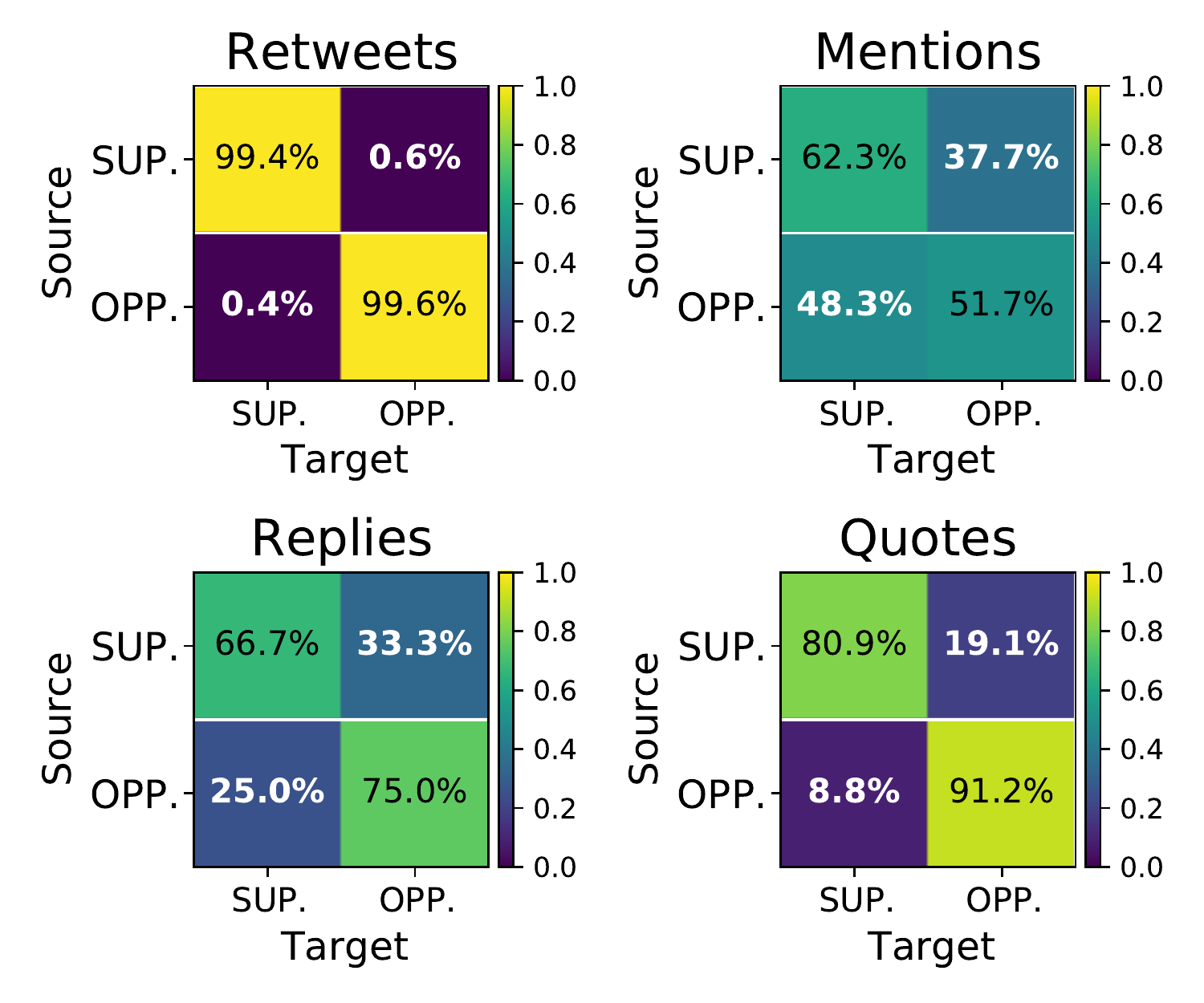}
    }
    \caption{Directed interactions by Supporters (SUP.) and Opposers (OPP.) as \emph{sources} to other Supporters and Opposers as \emph{targets}. Figure~\ref{fig:arson-int-count-heatmaps-raw} shows raw counts of interactions, while Figure~\ref{fig:arson-int-count-heatmaps-proportional} shows the proportions of interactions from each source group to each target group. }
    \label{fig:arson-int-count-heatmaps}
\end{figure}

These strong pointers to polarisation across both groups raise the question of whether there is a difference between the groups: for each group, what proportion of their connections are homophilic or heterophilic? Figure~\ref{fig:arson-int-count-heatmaps} shows two representations of the interactions between Supporters and Opposers. The first (Figure~\ref{fig:arson-int-count-heatmaps-raw}) shows the raw numbers of retweets, mentions, replies and quotes from a member of one group (the \emph{source} of the interaction) to another account (the \emph{target} of the interaction). The second (Figure~\ref{fig:arson-int-count-heatmaps-proportional}) shows the proportion of interactions from each source that is directed to Supporters or Opposers, effectively presenting a normalised view of the source group's interactive behaviour. It is immediately clear that, outside of retweets, Supporters were much more active, and were biased towards connecting to members of their own group. The degree of activity is notable, because there were fewer Supporters ($497$) than Opposers ($593$) though their numbers were similar. Opposers were also heavily biased to connect to other Opposers via replies and quotes, but not so for mentions. The proportional view makes clear the bias in connectivity: while raw numbers of interactions may be low from Opposers, they strongly preferred to connect to themselves, while Supporter bias is less pronounced for mentions, replies and quotes, despite the raw numbers of interactions being much higher.

\subsubsection{The concentration of voices}

The concentration of narrative from certain voices requires attention. 
To consider this, Table~\ref{tab:retweet-concentration-by-phase-and-group} shows the degree to which accounts were retweeted by the different groups by phase and overall. 
Unaffiliated accounts relied on a smaller pool of accounts to retweet than both Supporters and Opposers in each phase and overall, which is reasonable to expect as the majority of Unaffiliated activity occurred in Phase~3, once the story reached the mainstream news, and therefore had access to tweets about the story from the media and prominent commentators. 
Of the top $41$ accounts that were retweeted, each of which was retweeted $100$ times or more in the dataset, $17$ were Supporters and $20$ Opposers. Supporters were retweeted $5{,}487$ times ($322.8$ retweets per account), while Opposers were retweeted $8{,}833$ times ($441.7$ times per account). Together, affiliated accounts contributed $93.3\%$ of the top $41$'s $15{,}350$ retweets, in a dataset with $21{,}526$ retweets overall, and the top $41$ accounts were retweeted far more often than most. 
This pattern was also apparent in the $25$ accounts most retweeted by Unaffiliated accounts in Phase~3 (accounts retweeted at least $100$ times): $8$ were Supporters and $14$ were Opposers.
Thus Supporters and Opposers made up the majority of the most retweeted accounts, and arguably influenced the discussion more than Unaffiliated accounts.

\begin{table}[th]
    \centering
    \caption{Retweeting activity in the dataset, by phase and group.}
    \label{tab:retweet-concentration-by-phase-and-group}
    \resizebox{\textwidth}{!}{%
        \bgroup
        \setlength{\tabcolsep}{3pt}
        \begin{tabular}{@{}crrrrrrrrr@{}}
            \toprule
                    & \multicolumn{3}{c}{\textbf{Supporters}}      & \multicolumn{3}{c}{\textbf{Opposers}}        & \multicolumn{3}{c}{\textbf{Unaffiliated}}    \\
            Phase   & Retweets & Retweeted & Retweets per & Retweets & Retweeted & Retweets per & Retweets & Retweeted & Retweets per \\
                    &          & Accounts  & account      &          & Accounts  & account      &          & Accounts  & account      \\
            \cmidrule(r){1-1} \cmidrule(lr){2-4} \cmidrule(l){5-7} \cmidrule(l){8-10} 
            1       &      938 &        77 &       12.182 &       20 &         8 &        2.500 &    1,659 &       105 &       15.800 \\
            2       &       74 &        21 &        3.524 &      288 &        31 &        9.290 &      652 &        60 &       10.867 \\
            3       &    3,212 &       290 &       11.076 &    2,876 &       228 &       12.614 &   11,807 &       532 &       22.194 \\ 
            \cmidrule(r){1-1} \cmidrule(lr){2-4} \cmidrule(l){5-7} \cmidrule(l){8-10} 
            Overall &    4,224 &       327 &       12.917 &    3,184 &       243 &       13.103 &   14,118 &       613 &       23.030 \\
            \bottomrule        
        \end{tabular}
        \egroup
    }
\end{table}

\subsection{Content Dissemination}

When contrasting the content of the two affiliated groups, we considered the hashtags and external URLs used. 
A hashtag can provide a proxy for a tweet's
topic, and an external URL can refer a tweet's reader to further 
information relevant to the tweet, and therefore tweets that use the same URLs and
hashtags can be considered related.



\subsubsection{Hashtags}\label{sec:hashtag_analysis}

\begin{figure}[!ht]
    \centering
    \subfloat[Supporter hashtags.\label{fig:ccd_hashtag_co-mentions}]{%
        \includegraphics[height=0.41\textheight]{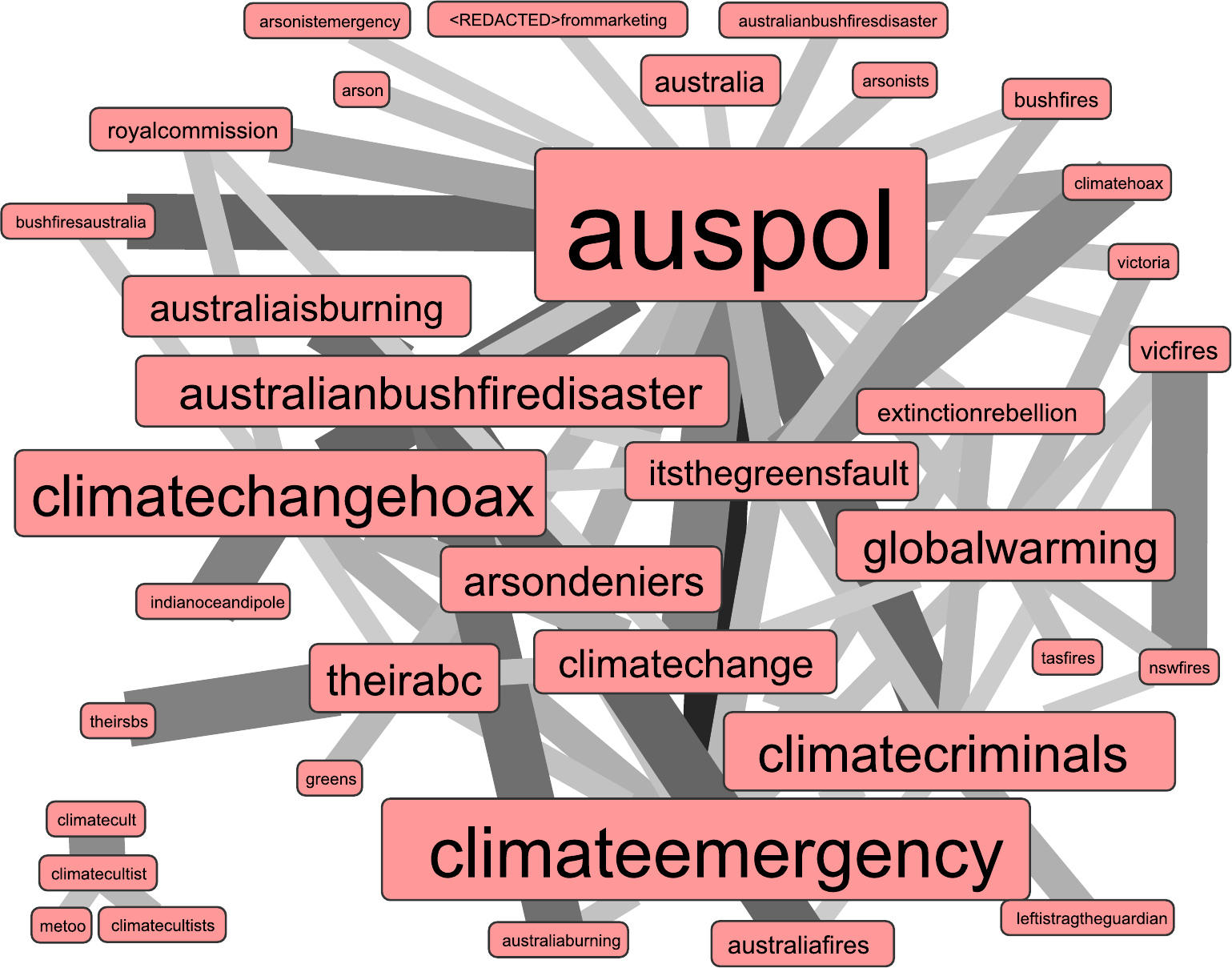}
    }
    \hfill
    \subfloat[Opposer hashtags.\label{fig:cca_hashtag_co-mentions}]{%
        \includegraphics[height=0.41\textheight]{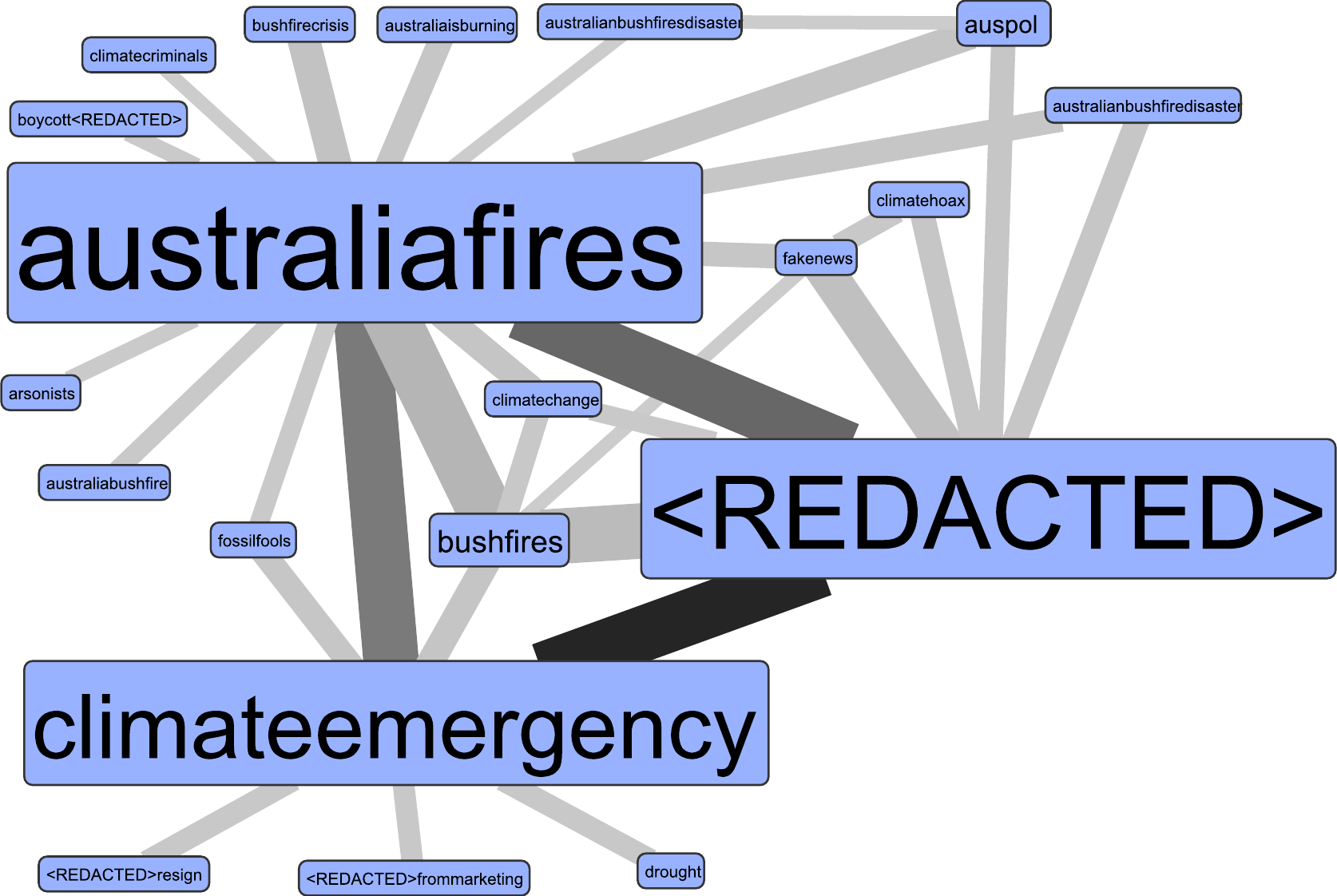}
    }
    \caption{Co-mentioned hashtags of Supporters and Opposers. Hashtag nodes are linked when five or more accounts tweeted both hashtags, and are coloured by the affiliation of the accounts that used them. \textless REDACTED\textgreater~hashtags include identifying information. Heavy edges (with high weight) are thicker and darker. The hashtag \hashtag{ArsonEmergency} has been removed from each network, as it occurred in every tweet in the dataset.}
    \label{fig:hashtag_co-mentions}
\end{figure}

To discover \emph{how} hashtags were used, rather than simply \emph{which} were used, we developed co-mention networks (visualised in Figure~\ref{fig:hashtag_co-mentions}). 
In these networks: each node is a hashtag in its lower case form, sized by degree centrality; 
edges represent 
an account using both hashtags (not necessarily in the same tweet); the edge weight represents the number of such accounts in the dataset. 
Nodes are coloured according to 
the affiliation of the accounts that used them. 
We removed the \hashtag{ArsonEmergency} hashtag (as nearly each tweet in the dataset contained it) as well as edges having weight less than $5$. 
Opposers used a smaller set of hashtags, predominantly linking \hashtag{AustraliaFires}\footnote{Capitals are re-introduced to hashtags used in the discussion for readability.} with \hashtag{ClimateEmergency} and a hashtag referring to a well-known publisher. 
In contrast, Supporters used a variety of hashtags in a variety of combinations, mostly focusing on terms related to `fire', but only a few with `arson' or `hoax', and linking to \hashtag{auspol} and \hashtag{ClimateEmergency}.
Manual inspection of Supporter tweets included many containing only a string of hashtags, unlike the Opposer tweets. Notably, the \hashtag{ClimateChangeHoax} node 
has a similar degree to the \hashtag{ClimateChangeEmergency} node, 
indicating Supporters' skepticism of climate science, but perhaps also that Supporters were attempting to join or merge the discussion communities defined by those hashtags in order to pollute the predominant hashtag of the \hashtag{ClimateChangeEmergency} community with a counter-narrative \citep{woolley2016autopower,nasim2018real}.

\begin{figure}[ht]
    \centering
    \includegraphics[width=0.99\textwidth]{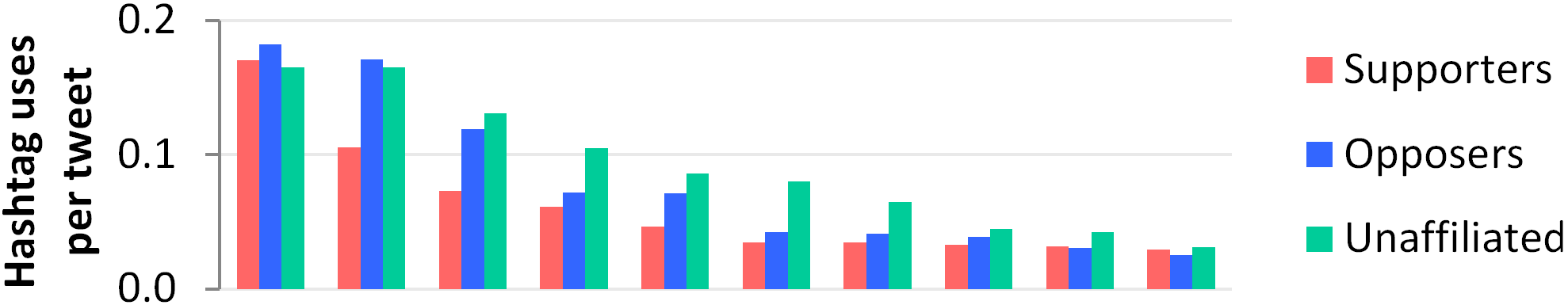}
    \caption{Hashtag uses per tweet for the ten most used hashtags for Supporters, Opposers and the Unaffiliated, ommitting \hashtag{ArsonEmergency}. Opposers used hashtags more frequently than Supporters, but after the second hashtag, Unaffiliated accounts used more than either polarised group.}
    \label{fig:hashtag_uses_per_tweet}
\end{figure}

Even though Supporters used approximately the same number of hashtags per tweet as Opposers ($2.92$ compared with $2.89$), they used $40.9$ hashtags per account, including $1.30$ unique tweets per account. 
In contrast, Opposers only used $17.5$ hashtags per account, including $0.36$ unique ones. 
This indicates the pool of hashtags used by the Opposers was much smaller than that of Supporters. 
The distribution of hashtag uses for the ten most frequently used by each group (which overlap but are not identical), omitting the ever-present \hashtag{ArsonEmergency}, is shown in Figure~\ref{fig:hashtag_uses_per_tweet} It indicates that Opposers focused slightly more strongly on a small set of hashtags, while Supporters spread their use of hashtags over a broader range (and thus their use of even their most frequently used hashtags is less than for Opposers). 
Unaffiliated accounts used their frequently used hashtags more often than both groups by the $4$\textsuperscript{th} hashtag, possibly due to the much greater number of accounts being active but less focused in their hashtag use. 
A second hashtag appeared in fewer than $20\%$ of each groups' tweets.

Manual inspection of Supporter tweets revealed that many replies consisted solely of ``\hashtag{ArsonEmergency}'' (e.g., one Supporter replied to an Opposer $26$ times in under $9$ minutes with a tweet just consisting of the hashtag). 
This kind of behaviour, in addition to inflammatory language in other Supporter replies, suggests a degree of aggression, though aggressive language was also noted among Opposers. 
The tweets that included more than $5$ hashtags made up only $1.7\%$ of Opposer tweets, but $2.8\%$ of Supporter tweets and $2.1\%$ Unaffiliated tweets.
Further analysis of inauthentic behaviour is addressed in Section~\ref{sec:inauthentic_behaviour}.

\paragraph{Polarisation in hashtag use} 

A statistical examination of how Supporters and Opposers used hashtags also revealed significant levels of homophily when considering only Supporters and Opposers, but less so when the hashtags use of Unaffiliated accounts was included. We created an account network by linking accounts that use the same hashtag. For accounts $u$ and $v$, which used a set of hashtags $\{h_1, h_2, ..., h_n\}$ in common, and each account $x$ used a hashtag $h$ with a frequency of $h^x$, the weight of the undirected edge $\{u,v\}$ between $u$ and $v$ is given by

\begin{equation}
    w_{\{u,v\}} = \sum_{i=1}^{n} h^u_i \cdot h^v_i.
\end{equation}
This formulation provides the maximal number of ways that $u$ and $v$'s hashtag uses could be combined. An alternative would be to use the minimum of $h^u$ and $h^v$ for all hashtags, $h$, as per \cite{magelinski2021}'s consideration of hashtags in their search for coordinated behaviour. In their study, however, their aim is to constrain processing requirements, while we do not have that limitation, given the size of our dataset.

Not all hashtags were used by each group, however. In order to determine to what extent their hashtag use overlapped without the influence of widely used hashtags (which connect the majority of accounts in the network), we created a set of hashtags to focus on beginning with the ten most frequently used hashtags unique to each of the Supporters and Opposers. Using this set of twenty hashtags,\footnote{Supporters' ten most frequently used unique hashtags were \hashtag{ItsTheGreensFault} (326), \hashtag{Victoria} (123), \hashtag{GlobalWarming} (107), \hashtag{TheirABC} (78), \hashtag{ClimateCultist} (66), \hashtag{IndianOceanDipole} (66), \hashtag{Greens} (62), \hashtag{ecoterrorism} (54), \hashtag{Melbourne} (53), and \hashtag{NotMyABC} (52), while those of the Opposers were \hashtag{BlackSummer} (74), \hashtag{FossilFools} (34), \hashtag{KoalasNotCoal} (29), \hashtag{ArsonHoax} (28), \hashtag{ArsonMyArse} (23), \hashtag{DontGetDerailed} (23), \hashtag{Smoko} (20), \hashtag{bots} (19), \hashtag{ArseholeEmergency} (17), and \hashtag{FossilFuel} (14).} we extracted the tweets containing them and created the account network from all the hashtag uses they included (i.e., including the co-occurring hashtags). We then removed uses of the ten most frequently occurring hashtags in the overall dataset, however, 
producing a final set of $245$ hashtags.

\begin{figure}[!ht]
    \centering
    \subfloat[Accounts using partisan and co-occurring hashtags.\label{fig:accts_by_partisan_hashtags}]{%
        \includegraphics[height=0.21\textheight]{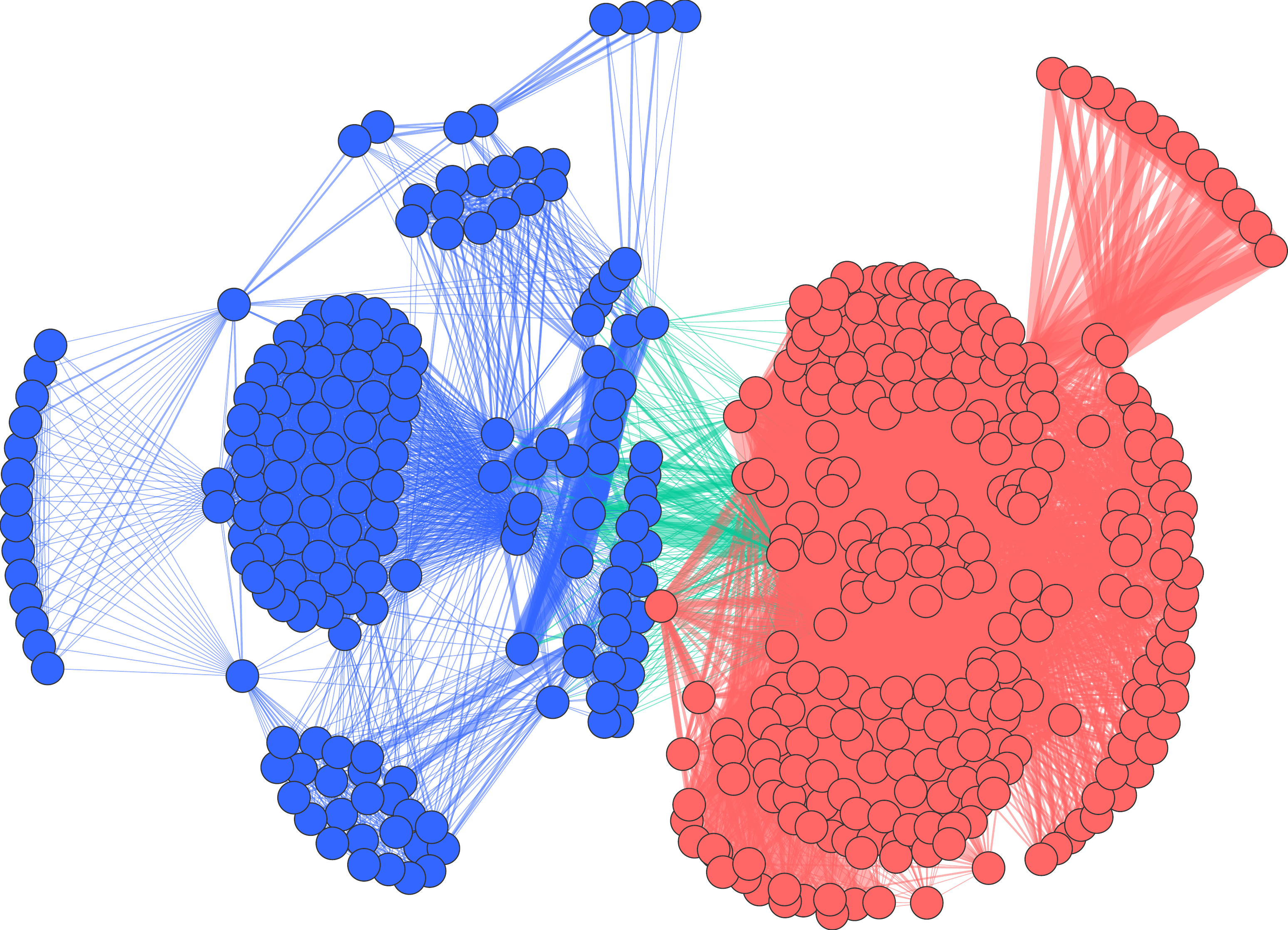}
    }%
    \hfill
    \subfloat[Partisan and co-occurring hashtags.\label{fig:partisan_hashtag_comentions}]{%
        \includegraphics[height=0.21\textheight]{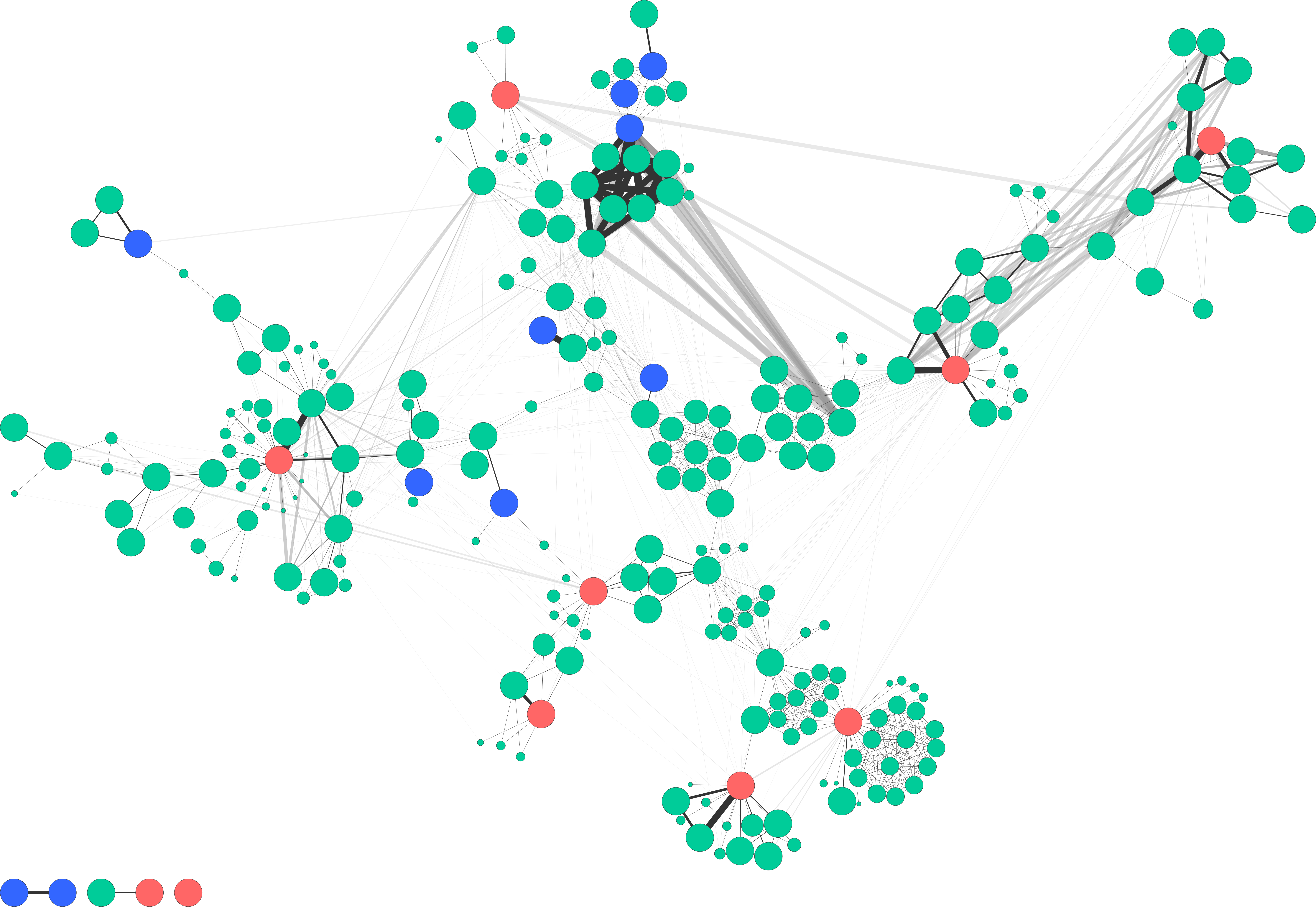}
    }%
    \caption{Two networks built from the tweets containing `partisan' hashtags (omitting uses of the ten most common hashtags). Left: Supporter (red) and Opposer (blue) nodes are linked when they mention the same hashtag, and are laid out with a force-directed algorithm. Red edges connect Supporters, blue connect Opposers, while green edges connect across the groups. Edge width is proportional to edge weight. Isolates have been removed. Though some polarisation should be expected given the partisan hashtags provide a natural axis of polarisation, it is notable quite how little overlap there is in use of the co-occurring hashtags. Right: Supporter partisan hashtags (red), Opposer partisan hashtags (blue) and co-occurring hashtags (green) are linked when mentioned by the same account (potentially in different tweets). Nodes are laid out with the quadrilateral Simmelean backbone algorithm~\protect\citep{Serrano2009backbone,nocaj2014untangling}, and edges are coloured according to their contribution (backbone strength). Edge widths represent edge weights. The separate clusters in the bottom left are included as they had been linked to the most common hashtags prior to their removal. The clusters apparent in the account network (left) are caused by the fact that partisan hashtags are rarely co-mentioned (right). Instead they are clearly co-mentioned with a variety of distinct hashtags, implying that although Supporters and Opposers were polarised in their hashtag use, they also had distinct sub-communities within their discussions (using hashtags as a proxy for discussion topic).}
    \label{fig:partisan_hashtag_networks}
\end{figure}

The resulting account network consists of $12{,}867$ nodes (including the $493$ Supporter and $597$ Opposer accounts) and $90{,}641$ edges. The combined modified E-I Index of the network, which only considers edges internal and external to Supporters and Opposers rather than also including edges between Unaffiliated accounts, 
was $0.250$, 
implying that together the groups expressed a small but solid preference for outside connections (i.e., due to the co-occurring hashtags). When we consider only the $14{,}777$ edges between or within the Supporter and Opposer groups (excluding all edges to adjacent Unaffiliated accounts), the modified E-I Index falls to $-0.964$ 
with a corresponding assortativity coefficient of $0.966$, 
which indicates the great majority of such edges were homophilic (i.e., within groups). Given we started with hashtags unique to each group, a degree of homophily is not surprising, however these very strong results imply that not many of the co-occurring hashtags each group used overlapped either. These results are clearly evident in a visualisation of the network (Figure~\ref{fig:accts_by_partisan_hashtags}).

Quickly returning to the network of hashtags co-mentioned in the partisan tweets, we can see the clusters in the account network (Figure~\ref{fig:accts_by_partisan_hashtags}) are caused by the fact that the accounts rarely used multiple partisan hashtags together (otherwise there would be clusters of partisan hashtags); instead, whenever a tweet included a partisan hashtag, they also included one or a few of a variety of non-partisan hashtags, which are represented by clusters of green nodes in Figure~\ref{fig:partisan_hashtag_comentions}.

\subsubsection{External URLs}

URLs 
in tweets can be categorised as \emph{internal} or \emph{external}. 
Internal URLs refer to other tweets in retweets or quotes, 
while external URLs are often included to highlight something about their content, 
e.g., as a source to support a claim. By analysing the URLs, 
it is possible to gauge the intent of the tweet's author by considering the reputation of the source or the 
argument offered. 

We categorised\footnote{Categorisation was conducted by two authors and confirmed by the others.} 
the ten URLs used most each by the Supporters, Opposers, and Unaffiliated accounts across the three phases, 
and found a 
significant difference between the groups. URLs were assigned to one of these four categories:
\begin{description}
    \item[\textbf{NARRATIVE}] 
    Articles used to emphasise the conspiracy narratives by prominently reporting arson figures and fuel load discussions.
    \item[\textbf{CONSPIRACY}] Articles and web sites that 
    take extreme positions on climate change (typically arguing against predominant scientific opinion).
    \item[\textbf{DEBUNKING}] News articles providing authoritative information about the bushfires and related misinformation on social media.
    \item[\textbf{OTHER}] Other web pages.
\end{description}

\begin{figure}[!ht]
    \centering
    \subfloat[Opposer URLs.\label{fig:cca_urls_by_phase}]{%
        \includegraphics[height=0.13\textheight]{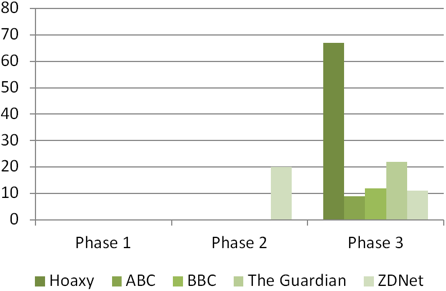}
    }
    \hfill
    \subfloat[Supporter URLs.\label{fig:ccd_urls_by_phase}]{%
        \includegraphics[height=0.13\textheight]{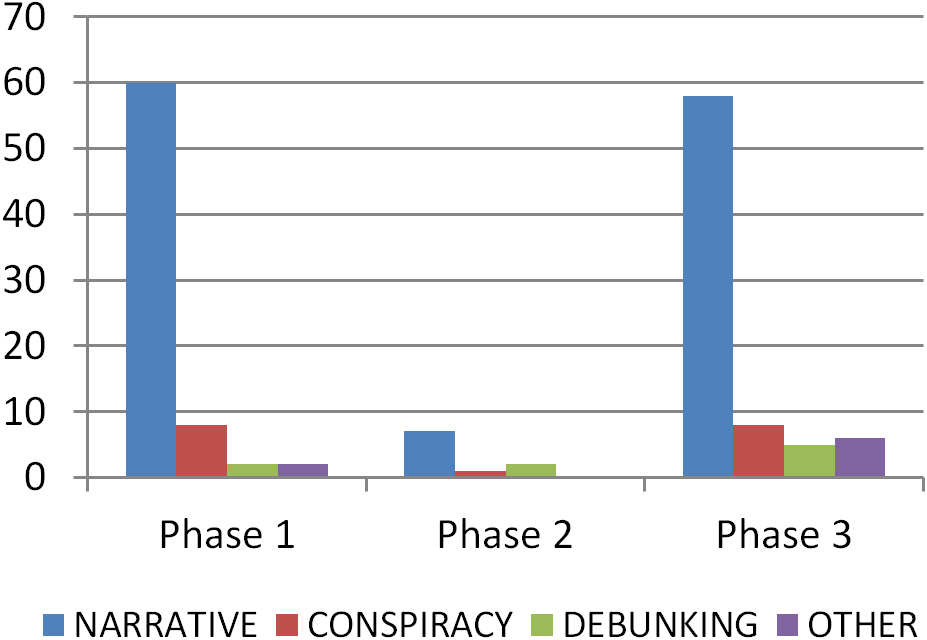}
    }
    \hfill
    \subfloat[Unaffiliated URLs.\label{fig:unaff_urls_by_phase}]{%
        \includegraphics[height=0.13\textheight]{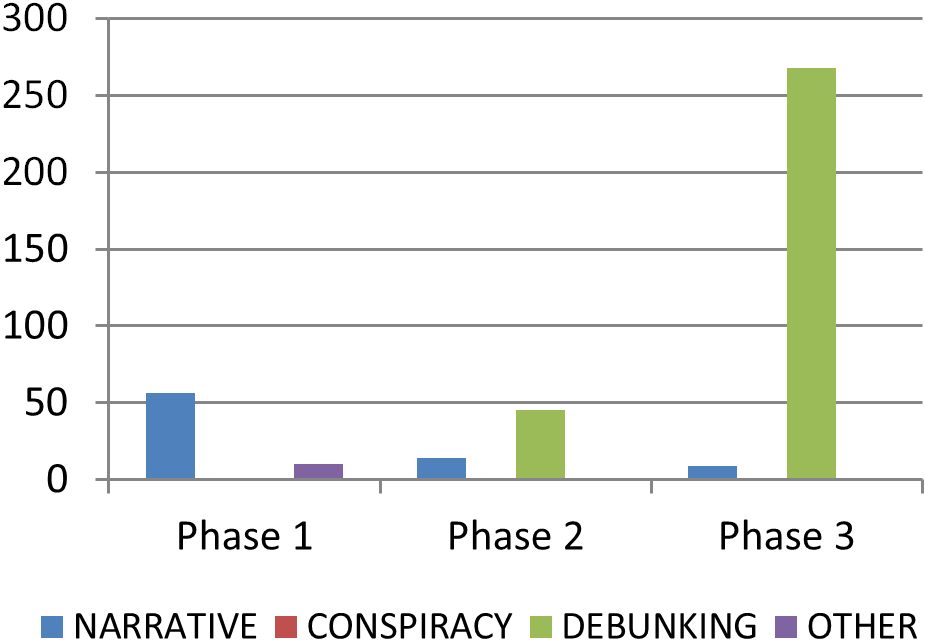}
    }
    \caption{URLs used by Opposers, Supporters and Unaffiliated accounts.}
    \label{fig:urls_by_phase}
\end{figure}

URLs posted by Opposers were concentrated in Phase~3 and were all in the DEBUNKING category, with nearly half attributed to Indiana University's Hoaxy service \citep{Shao_2016}, 
and nearly a quarter referring to the original ZDNet article \citep{Stilgherrian2020zdnet} 
(Figure~\ref{fig:cca_urls_by_phase}). In contrast, Supporters used many URLs in Phases~1 and~3, focusing mostly on articles 
emphasising the arson narrative, 
but with references to a number of climate change denial or right wing blogs and news sites (Figure~\ref{fig:ccd_urls_by_phase}).

Figure~\ref{fig:unaff_urls_by_phase} shows that the media coverage changed the content of the Unaffiliated discussion, 
from articles 
emphasising the arson
narratives in Phase~1 to 
Opposer-aligned articles 
in Phase~3. Although the activity of Supporters in Phase~3 increased significantly, the Unaffiliated members appeared to refer to Opposer-aligned external URLs much more often. 
This suggests that the new Unaffiliated accounts arriving in the final phase (discussed in Section~\ref{sec:community_timelines} above) held different opinions on the arson narrative from the Unaffiliated accounts active early in the discussion. In fact, it is possible they acted as bridges bringing in new Opposer accounts -- $411$ of the $585$, or approximately $70\%$ of Opposer accounts active in Phase~3 were were not active in earlier Phases.

\begin{figure}[th]
    \centering
    \includegraphics[width=.99\textwidth]{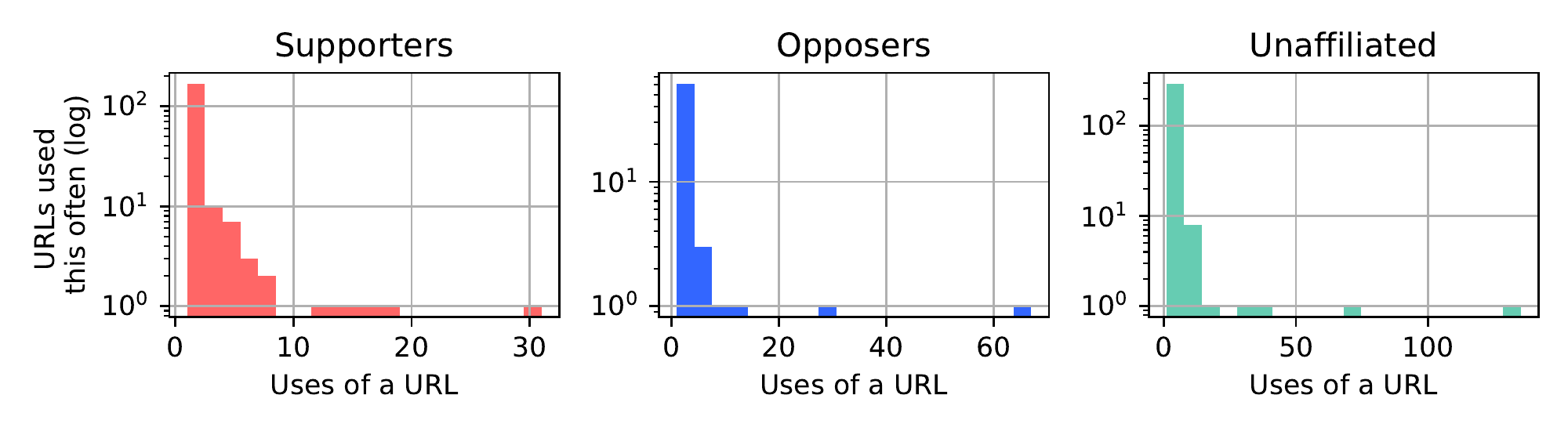}
    \caption{Distributions of URL use by Supporters, Opposers and Unaffiliated accounts.}
    \label{fig:url_use_distribution}
\end{figure}

Supporters used many more URLs than Opposers overall ($1{,}365$ to $399$) and nearly twice as many external URLs ($390$ to $212$). Supporters seemed to use many different URLs in Phase~3 and overall, but focused much more on particular URLs in Phase~1. 
Of the total number of unique URLs used in Phase~3 and overall, $263$ and $390$, respectively, only $77$ ($29.3\%$) and $132$ ($33.8\%$) appeared in the top ten, implying a wide variety of URLs were used. In contrast, in Phase~1, $72$ of $117$ appeared in the top ten ($61.5\%$), similar to Opposers' $141$ of $212$ ($66.5\%$), implying a greater focus on specific sources of information. In brief, it appears Opposers overall and Supporters in Phase~1 were 
focused in their choice of sources, but by Phase~3, Supporters had expanded their range 
considerably.
Ultimately, Supporters used $195$ URLs $390$ times (in total), Opposers used $68$ URLs $212$ times, and the Unaffiliated used $305$ URLs $817$ times, meaning a mean rate of URL use of $2.0$, $3.1$, and $2.7$, respectively, meaning Opposers were more focused in their URL use. This is evident in the distributions of URL uses in Figure~\ref{fig:url_use_distribution}, which Supporters use more URLs more often that Opposers, and Opposers focused many of their uses on a small number of URLs.

\subsection{Coordinated Dissemination}

To investigate whether coordinated dissemination of content was occurring, we performed co-retweet, co-hashtag and co-URL analyses \citep{WeberN2020coord}, searching for sub-communities of accounts that retweeted the same tweets, and shared the same hashtags, URLs, and URL domains within the same timeframes (denoted by $\gamma$). 
Regarding the URLs, Figure~\ref{fig:urls_by_phase} indicates the nature of article external links referred to, but not the distributions of the URLs or their domains, which is the aim of using these co-activity analyses. The analyses result in weighted networks consisting of the sub-communities as disconnected components of accounts, the edge weights of which indicate the frequency of co-linking or co-mentioning of a hashtag. Further, to examine how the sub-communities relate to one another, we can then re-introduce the URLs and domains as explicit `reason' nodes into these networks, making them bigraphs in which communities are joined according to these `reason' nodes \citep{Weber2021coord}.

\begin{figure}[t!]
    \centering
    \includegraphics[width=0.99\textwidth]{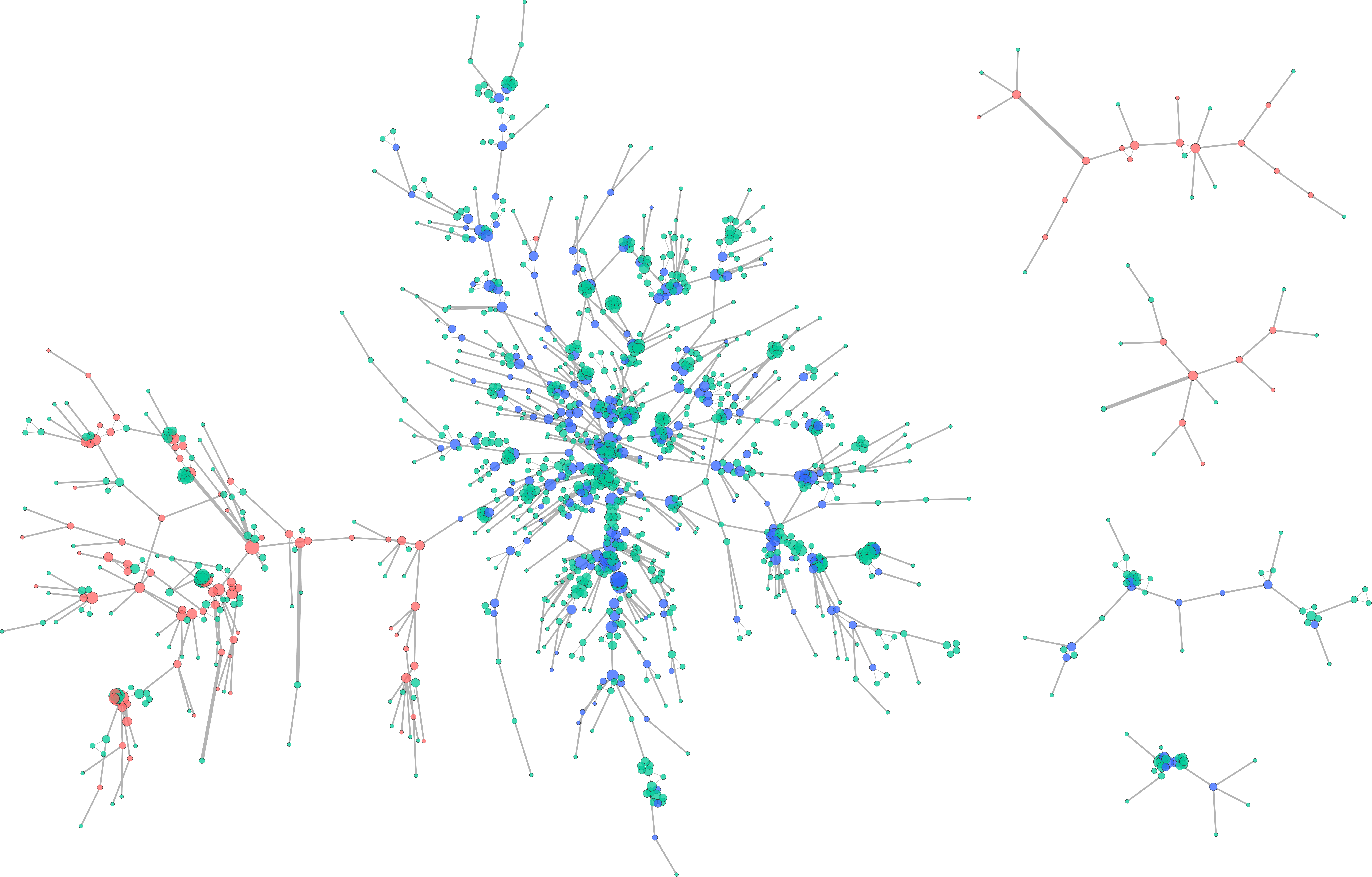}
    \caption{The five largest connected components of the co-retweet coordination network ($\gamma$=$1$ minute), limited to only Supporter and Opposer accounts, which are sized by indegree. Red nodes are Supporters, blue are Opposers, and edges are sized by frequency of co-retweeting.}
    \label{fig:co-rt-1m-graph}
\end{figure}

\subsubsection{Co-retweet analysis} The largest components of the co-retweet network ($\gamma$=$1$ minute) shown in Figure~\ref{fig:co-rt-1m-graph} show that the polarisation observed in the retweet network (in Figure~\ref{fig:retweets-polarised}) is still evident, as expected, but what is particularly notable is the absence of tight cliques amongst the Supporter nodes, which, as promoters of the arson narrative, were originally thought to include a large proportion of bots \citep{Stilgherrian2020zdnet,GrahamKeller2020conv}. Cliques would indicate accounts all retweeting the same tweets within the same timeframe, a signal associated with automation, but also with high popularity (i.e., increasing the number of interested accounts increases the chance that they co-retweet accidentally). Cliques are visible amongst the $103$ Opposers and many of the $966$ Unaffiliated accounts (and could also be due to simple popularity and coincidence), but rare amongst the $233$ Supporters. Instead their connection patterns imply real people seeing and retweeting each others retweets. For example, account A sees a tweet and retweets it, which is then seen by account B (within $1$ minute), and then account C sees that and retweets it as well, but longer than $1$ minute after A. A $1$ minute window is quite large for the purposes of identifying botnets, so this would indicate a lack of evidence of retweeting bots amongst the Supporters.

A further item to note is the degree of support offered by the Unaffiliated accounts, which co-retweet with Opposer accounts far more frequently than Supporter accounts in the coordination networks presented in Figure~\ref{fig:co-rt-1m-graph}. This observation raises the question of whether some of the Unaffiliated accounts may, in fact, be Opposers, but were simply not captured in the application of conductance cutting community detection to the retweet network, and they may have been captured with modification of the detection parameters.

\begin{figure}[t!]
    \centering
    \includegraphics[width=0.99\textwidth]{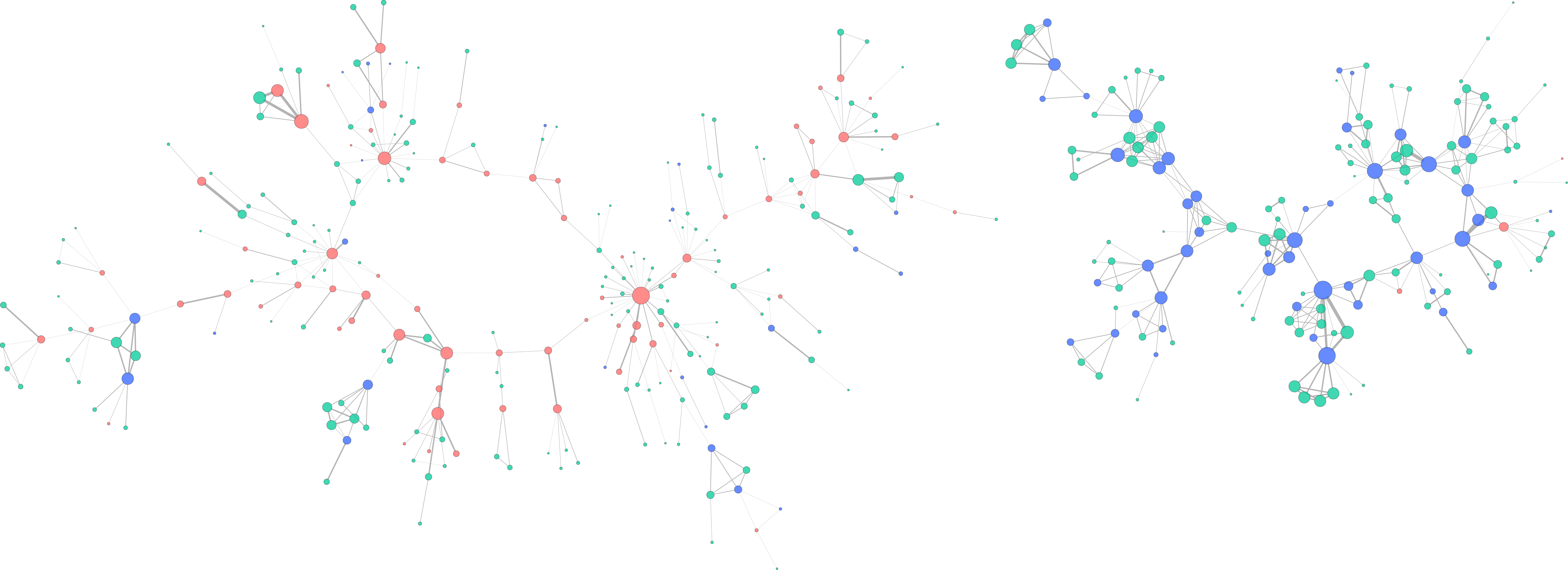}
    \caption{The two largest connected components of the co-hashtag coordination network ($\gamma$=$1$ minute, excluding \hashtag{ArsonEmergency}), with nodes sized by the number of tweets they posted in the discussion. Red nodes are Supporters, blue are Opposers, green are Unaffiliated, and edge widths are sized by the frequency of co-hashtag activity.}
    \label{fig:co-hashtag-1m-all}
\end{figure}

\subsubsection{Co-hashtag analysis} As using a hashtag in a tweet can increase its reach to observers of the hashtag as well as one's followers, coordinated promotion of a hashtag is a mechanism to disseminate one's message \citep{varol2017campaigndetection}, as well as pollute a discussion space \citep{woolley2016autopower,nasim2018real}. Given how frequently hashtags are used, we chose a tight timeframe of $1$ minute and excluded \hashtag{ArsonEmergency} from our co-hashtag analysis. The two largest components discovered highlight the polarisation between the Supporter and Opposer communities (Figure~\ref{fig:co-hashtag-1m-all}). The ring formation amongst the Supporters and small node sizes indicate less activity including a wider variety of hashtags. Opposers are more active and focused in the hashtags they used. These findings emphasise the findings in Section~\ref{sec:hashtag_analysis} but also highlight the support of Unaffiliated accounts, the most active of which appear to support the Opposers.

\begin{figure}[t!]
    \centering
    \subfloat[Co-URL coordination network including only Supporters (in red) and Opposers (in blue).\label{fig:co-url-10m-supopp}]{
        \includegraphics[width=0.7\textwidth]{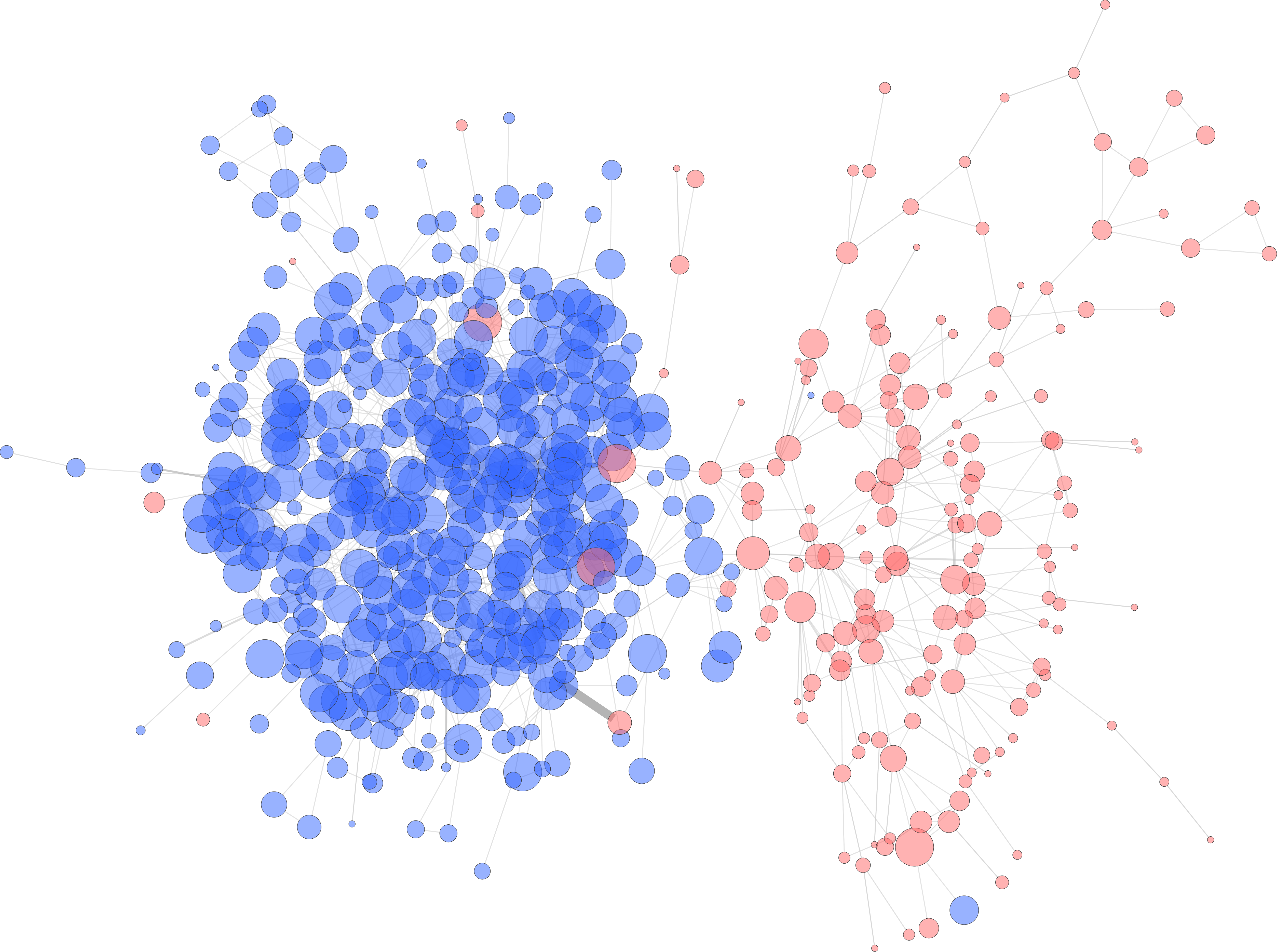}
    }
    \\
    \subfloat[Co-URL coordination network laid according to the network's quadrilateral Simmelian backbone~\citep{Serrano2009backbone,nocaj2014untangling}.\label{fig:co-url-10m-all}]{
        \includegraphics[width=0.93\textwidth]{resources/arson-co_urls-10m-agg_lcn-v2.pdf}
    }
    \caption{The coordination networks resulting from co-URL analysis ($\gamma$=$10$ minutes), with nodes sized by indegree. Red circular nodes are Supporters, blue are Opposers, and the green remainder are Unaffiliated accounts. Edge width and darkness indicates frequency of co-linking.}
    \label{fig:co-url-10m}
\end{figure}

\subsubsection{Co-URL and co-domain analysis} For human users, grassroots-style coordinated co-linking should be visible in `human' timeframes, such as within $10$ minutes, allowing time for users to see each others' tweets. The polarisation evident in the retweet network is also evident in the co-linking networks ($\gamma$=$10$ minutes) shown in Figure~\ref{fig:co-url-10m}, especially considering only the Supporter and Opposer networks (Figure~\ref{fig:co-url-10m-supopp}). When we examine the co-linking in context in Figure~\ref{fig:co-url-10m-all}, along with the contributions of Unaffiliated accounts, we can see that, again, Unaffiliated accounts co-acted with Opposer accounts far more often than Supporters, which appear relatively isolated, compared with the concentrated co-linking in the Opposer/Unaffiliated clusters on the right. Here, cliques represent groups of accounts sharing the same URLs, but it is unclear whether each clique represents a different URL or simply a different time window. To consider that, we need to introduce `reason' nodes, representing the shared URLs, to create account/URL bigraphs.

\begin{figure}[t!]
    \centering
    \includegraphics[width=0.99\textwidth]{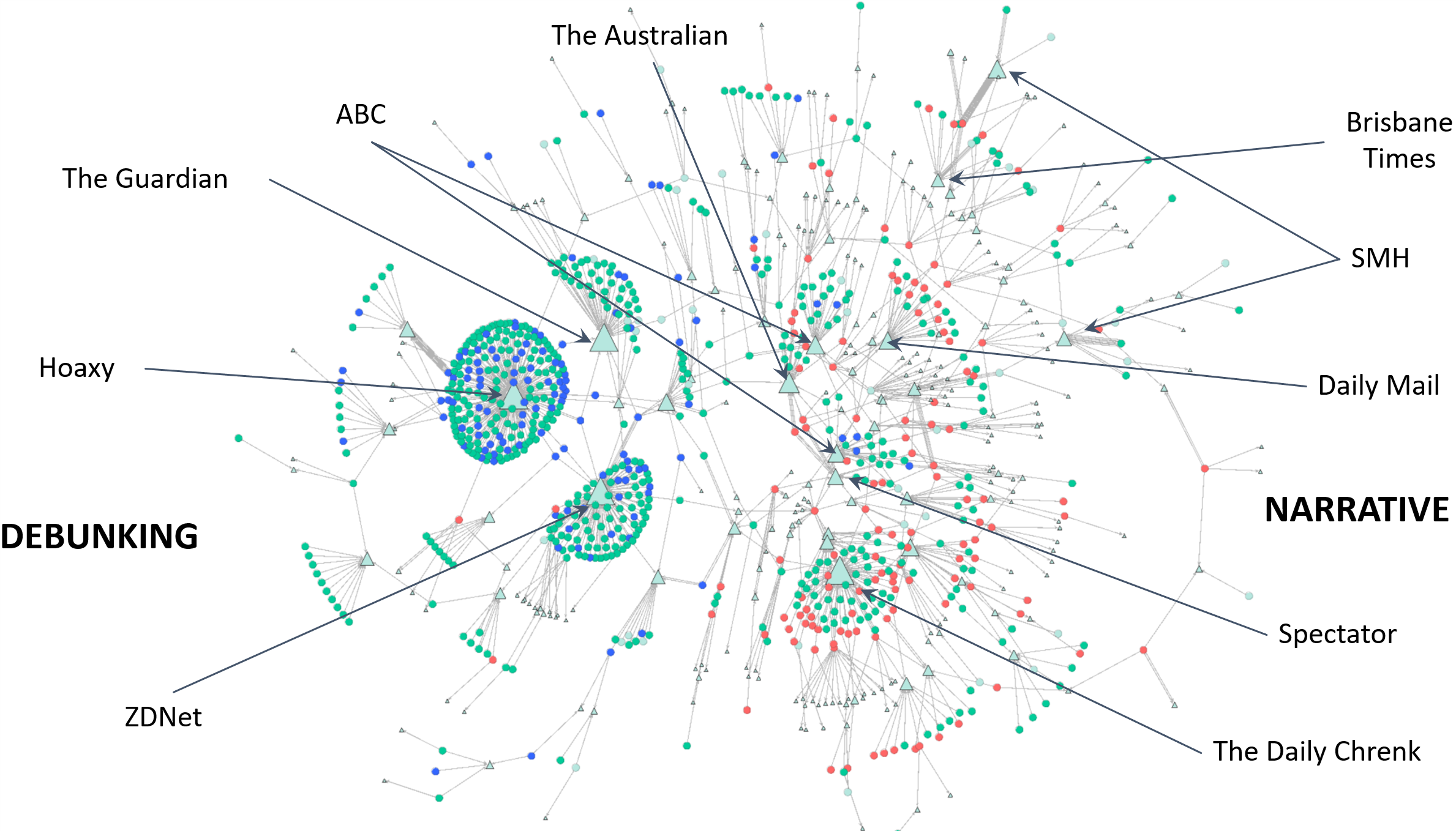}
    \caption{The account/URL bigraph resulting from co-URL analysis ($\gamma$=$10$ seconds), annotated with the websites hosting highly shared articles. Pale green triangular nodes are the URLs, sized by indegree. Red circular nodes are Supporters, blue are Opposers, and the green remainder are Unaffiliated accounts.}
    \label{fig:co-url-bigraph}
\end{figure}

\begin{figure}[ht!]
    \centering
    \includegraphics[width=0.99\textwidth]{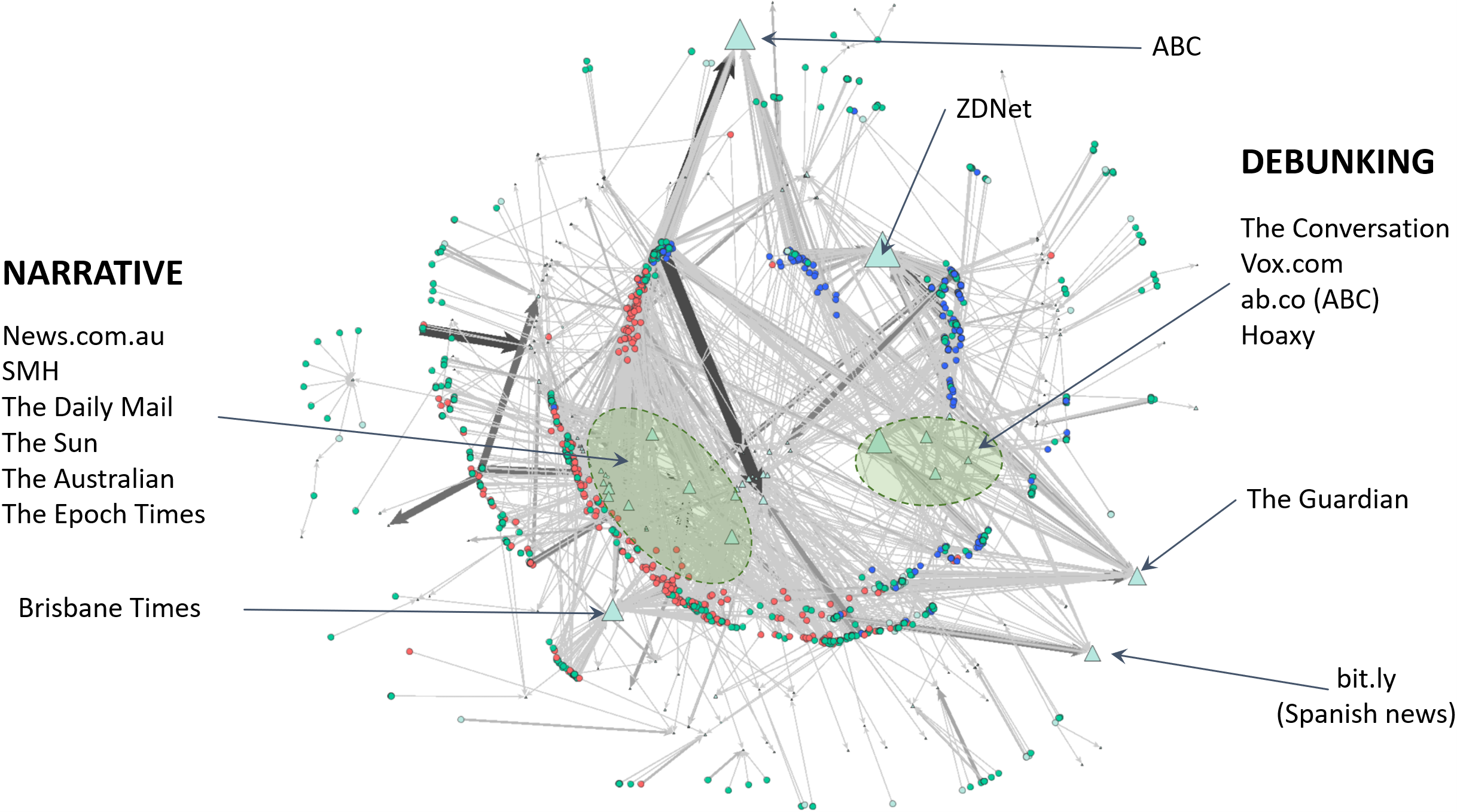}
    \caption{The account/domain bigraph resulting from co-domain analysis ($\gamma$=$10$ seconds), annotated with the websites hosting highly shared articles. Pale green triangular nodes are the URL domains, sized by indegree. Red circular nodes are Supporters, blue are Opposers, and the green remainder are Unaffiliated accounts. Two zones of contrasting highly linked to domains are highlighted, one primarily used to support the arson narrative, and one used primarily to debunk it.}
    \label{fig:co-domain-bigraph}
\end{figure}

Figure~\ref{fig:co-url-bigraph} shows the resulting account/URL bigraph, which includes annotations indicating the websites hosting the most shared articles (referred to by the URLs). As expected, there is clear polarisation around the URLs, but it is immediately also clear how focused the Opposer accounts were on a small number of URLs, similar to their use of hashtags. The blue Opposer nodes link mostly to three URLs: the original ZDNet article \citep{Stilgherrian2020zdnet}, the Hoaxy website \citep{Shao_2018}, and an article on The Guardian relating to online misinformation during the bushfires.\footnote{\url{https://www.theguardian.com/australia-news/2020/jan/08/twitter-bots-trolls-australian-bushfires-social-media-disinformation-campaign-false-claims}} The Supporter community's use of URLs is more dispersed, and includes MSM sites with the addition of a large cluster of Supporters and Unaffiliated accounts around an article on The Daily Chrenk, the website of an Australian blogger promoting the arson narrative. It is notable that two Australian Broadcasting Corporation (ABC) articles are so centrally located amongst the Supporters, as these were classified as DEBUNKING articles. When we consider the co-domain bigraph (Figure~\ref{fig:co-domain-bigraph}), however, it is clear that the ABC domain binds the polarised Supporter and Opposer communities together, along with, interestingly, The Guardian and the URL shortener \texttt{bit.ly}. One \texttt{bit.ly} link appeared much more frequently than others, and it resolved to a Spanish news article on online bushfire misinformation.\footnote{\url{https://www.muyinteresante.es/naturaleza/articulo/actualidad-las-fake-news-de-los-incendios-de-australia}} Highlighted in the co-domain bigraph are two zones of domains that appear mostly linked to one or the other of the Supporter and Opposer nodes, which are, again, appear polarised in the network. The domains in these zones appear aligned again with Opposers referring to domains hosting DEBUNKING URLs and Supporters referring to domains hosting NARRATIVE URLs. A few domains are referred to very frequently by individual nodes (visible as dark, large edges), and these are often social media sites, such as YouTube, Instagram, and Facebook.

The analyses of a variety of co-activities here emphasises the polarisation observed in the retweet network permeates the groups' collaborative efforts. Evidence indicates that Opposers, much less so than Supporters, engaged in coordinated action, however, given the significant contribution of Unaffiliated accounts, it is unclear whether this is deliberate or merely a reflection of high popularity (especially given the considerably greater number of Unaffiliated accounts active in the discussion).


\subsection{Locations}

\begin{figure}[th!]
    \centering
    \includegraphics[width=.99\textwidth]{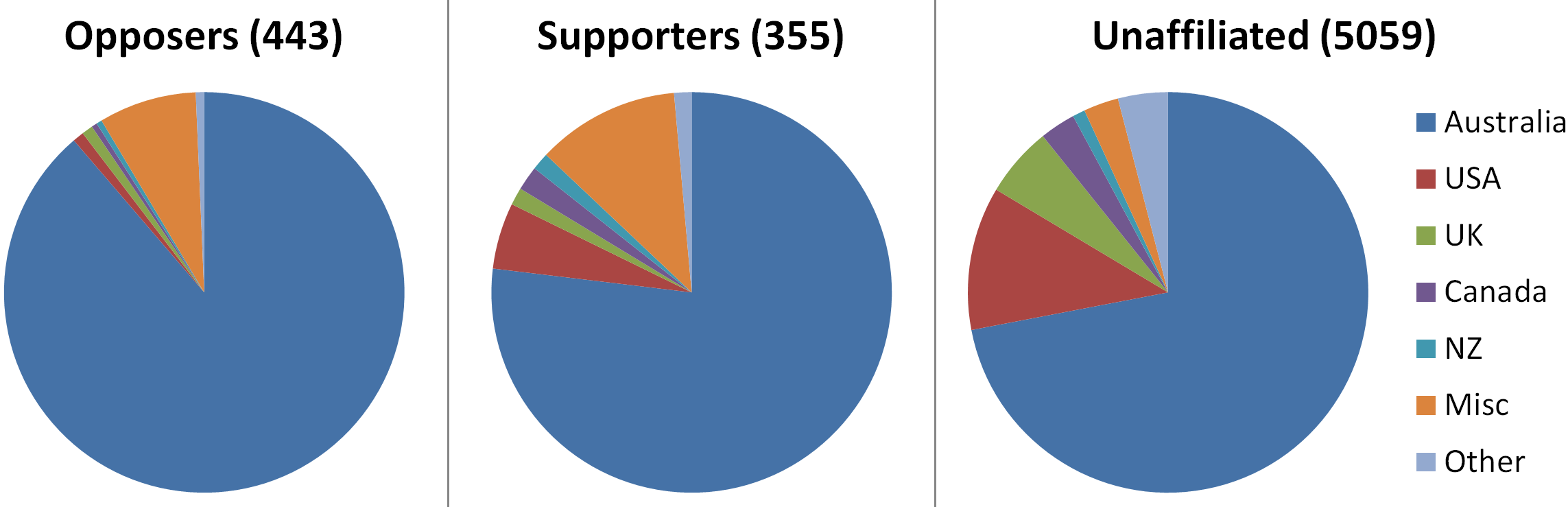}
    \caption{The self-reported locations of Supporter, Opposer and Unaffiliated accounts. The number in brackets indicates how many accounts were evaluated. The Miscellaneous category was used for locations which described a physical location but were vague, e.g., Earth, whereas Other was used for whimsical entries, e.g., ``Wherever your smartphone is.'' or ``Spot X''.}
    \label{fig:location_piecharts}
\end{figure}

Given the global effect of climate change, any prominent contentious discussion of it is likely to draw in participants from other timezones. Although the activity patterns in Figure~\ref{fig:arson-timeline} indicate the majority of activity aligns with Australian timezones, a deeper analysis of the self-reported account `location' fields in tweets revealed that only $88\%$ of active\footnote{We considered all Supporters, Opposers, plus all Unaffiliated accounts that tweeted at least three times, and who populated the field.} participants were Australian (Figure~\ref{fig:location_piecharts}). (Tweets can contain geolocation information but rarely do: only $127$ tweets in the `ArsonEmergency' dataset had any geolocation information, and $114$ were posted in Australia.) Based on the self-reported location, more Supporters declared locations outside Australia ($23\%$) than Opposers ($11\%$), but the biggest proportion of non-Australian participants were Unaffiliated, perhaps drawn in by the international news. It is unclear whether the international accounts were drawn in to aid the Supporters or Opposers in Phase~3, but we know the articles the Unaffiliated shared changed to DEBUNKING in that Phase, and Unaffiliated accounts appeared to coordinate with Opposers. 

More detail can be found in Appendix \ref{app:locations}.

\section{Botness Analysis}

The analysis reported in ZDNet \citep{Stilgherrian2020zdnet} indicated 
widespread bot-like behaviour 
by using the \texttt{tweetbotornot}\footnote{\url{https://github.com/mkearney/tweetbotornot}} R library. 
Our analysis had two goals: 1)~attempt to replicate Graham and Keller's findings in Phase 1 of our dataset; and 2)~examine the contribution of bot-like accounts detected in Phase~1 in the other phases. 
Specifically, we considered the questions:
\begin{itemize}
    \item Does another bot detection system find similar levels of bot-like behaviour?
    \item Does the behaviour of any bots from Phase~1 change in Phases~2 and~3?
\end{itemize}

We evaluated $2{,}512$ or $19.5\%$ of the accounts in the dataset using Botometer \citep{botornot2016}, including all Supporter and Opposer accounts, plus all Unaffiliated accounts that posted at least three tweets either side of Graham and Keller's analysis 
reaching 
the MSM (i.e., the start of Phase~3).





Botometer \citep{botornot2016} is an ensemble bot classifier for Twitter accounts, relying on over a thousand features drawn from six categories, which provides a structured analysis report of an account, rating various of its features for `botness'. 
The report includes a ``Complete Automation Probability'' (CAP), a Bayesian-informed probability that the account in question is ``fully automated'', as well as a rating that assumes an account is English-speaking which is different from the language-agnostic rating. 
This does not accommodate hybrid accounts \citep{GrimmeAA2018changingperspectives} and only uses English training data \citep{nasim2018real}, leading some researchers to use conservative ranges of CAP scores for high confidence that an account is human (\textless $0.2$) or bot (\textgreater $0.6$) \cite[e.g.,][]{rizoiu2018debatenight}. We adopt that categorisation.

\begin{table}[th]
    \centering
    \caption{Botness scores and contribution to the discussion across the phases by a subset of the accounts.}
    \label{tab:botness-scores-and-contribution-by-phase}
    \resizebox{0.99\textwidth}{!}{%
        \bgroup
        \setlength{\tabcolsep}{5pt}
        \begin{tabular}{@{}lcrrrrrrr@{}}
            \toprule
                      &           &       & \multicolumn{3}{c}{\textbf{Accounts active}} & \multicolumn{3}{c}{\textbf{Tweets contributed}} \\ 
            Category  & CAP       & Total & Phase~1 & Phase~2 & Phase~3         & Phase~1 & Phase~2 & Phase~3 \\
            \cmidrule(r){1-1} \cmidrule(lr){2-2} \cmidrule(lr){3-3} \cmidrule(lr){4-6} \cmidrule(l){7-9}
            Human	  & 0.0--0.2  & 2,426 &	    898 &     438 &           1,931 &   2,213 &     674 &  11,700 \\
            Undecided &	0.2--0.6  &    66 &      20 &       6 &              56 &      28 &      11 &     304 \\
            Bot	      & 0.6--1.0  &    20 &       9 &       4 &              11 &      23 &       6 &      84 \\
            \bottomrule
        \end{tabular}
        \egroup
    }
\end{table}

\begin{figure}[t!]
    \centering
    \subfloat[Tested accounts overall and in only Phase~1 (inset).\label{fig:botness_hist-all}]{%
        \includegraphics[width=0.99\textwidth]{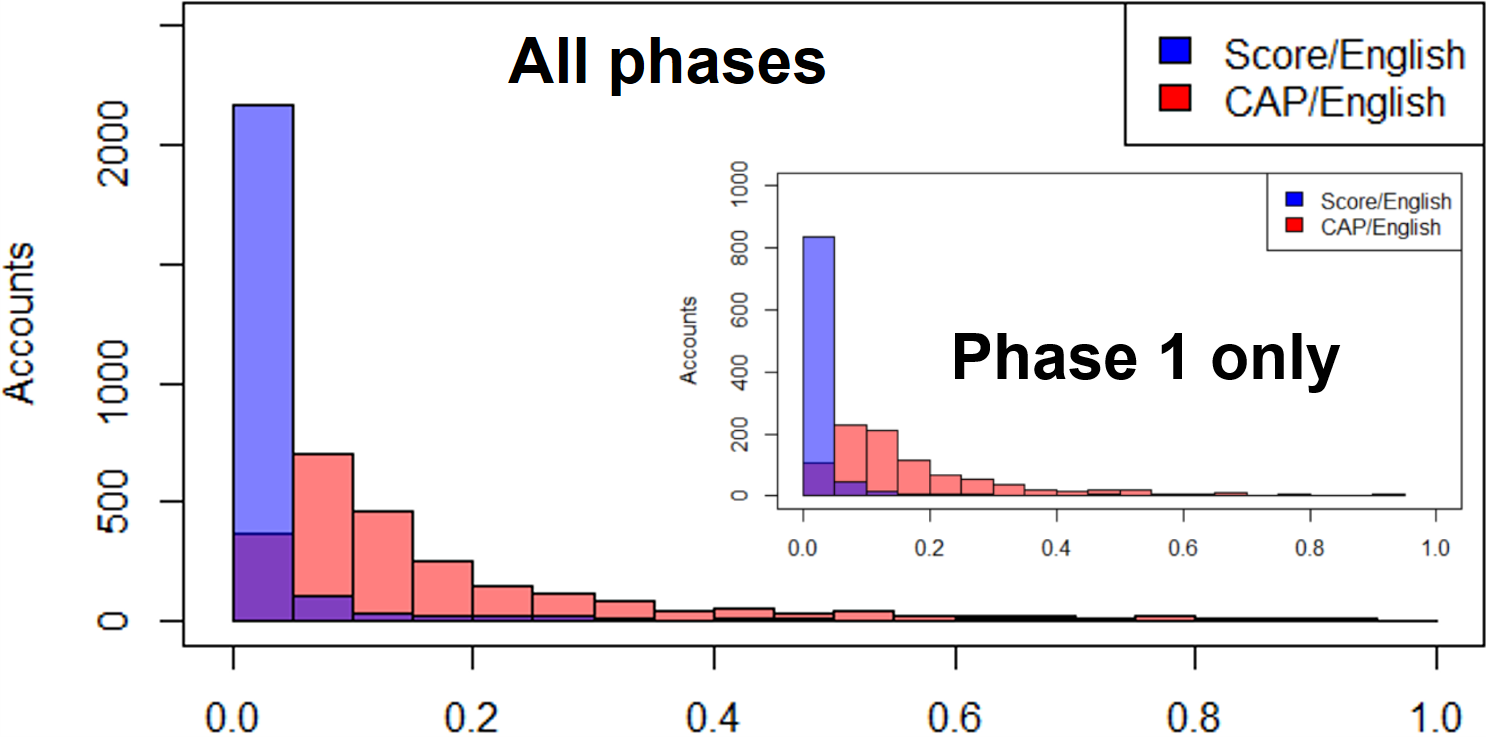}
    }
    \\
    \subfloat[Supporter accounts.\label{fig:botness_hist-sup}]{%
        \includegraphics[width=0.49\textwidth]{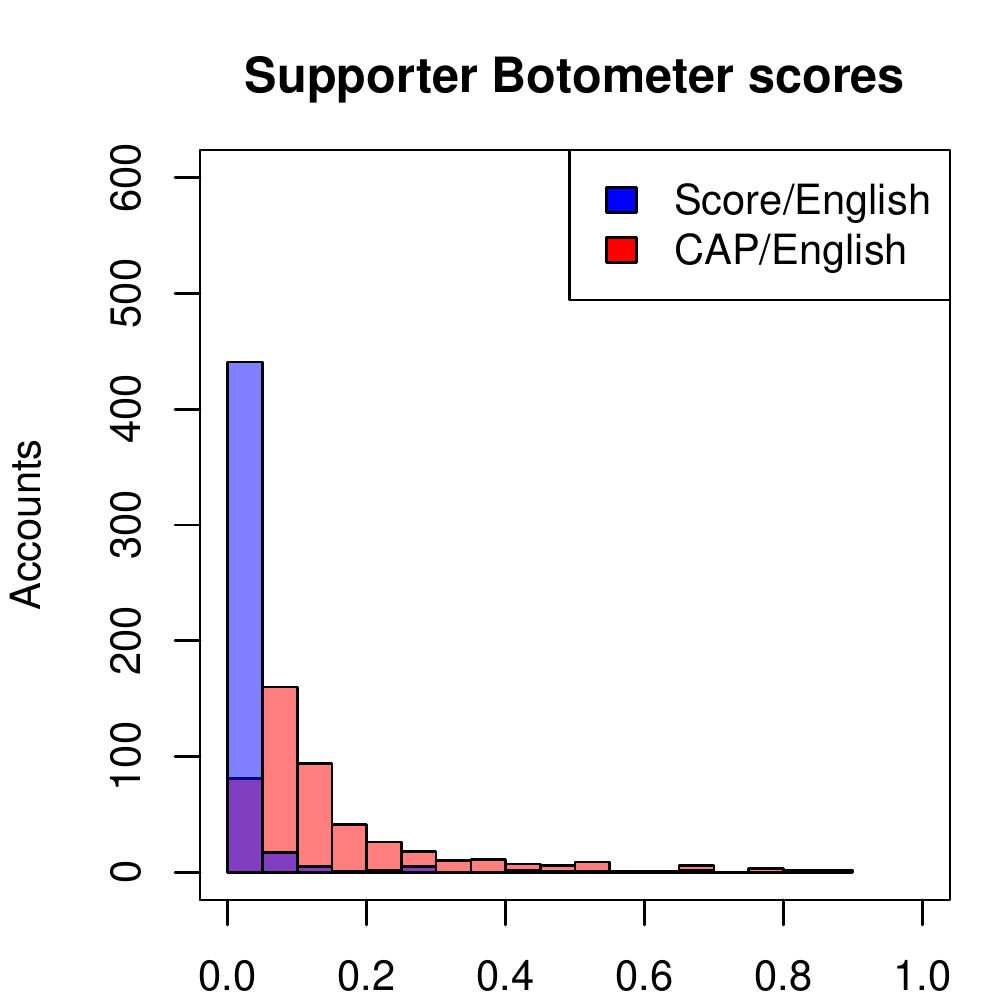}
    }
    \hfill
    \subfloat[Opposer accounts.\label{fig:botness_hist-opp}]{%
        \includegraphics[width=0.49\textwidth]{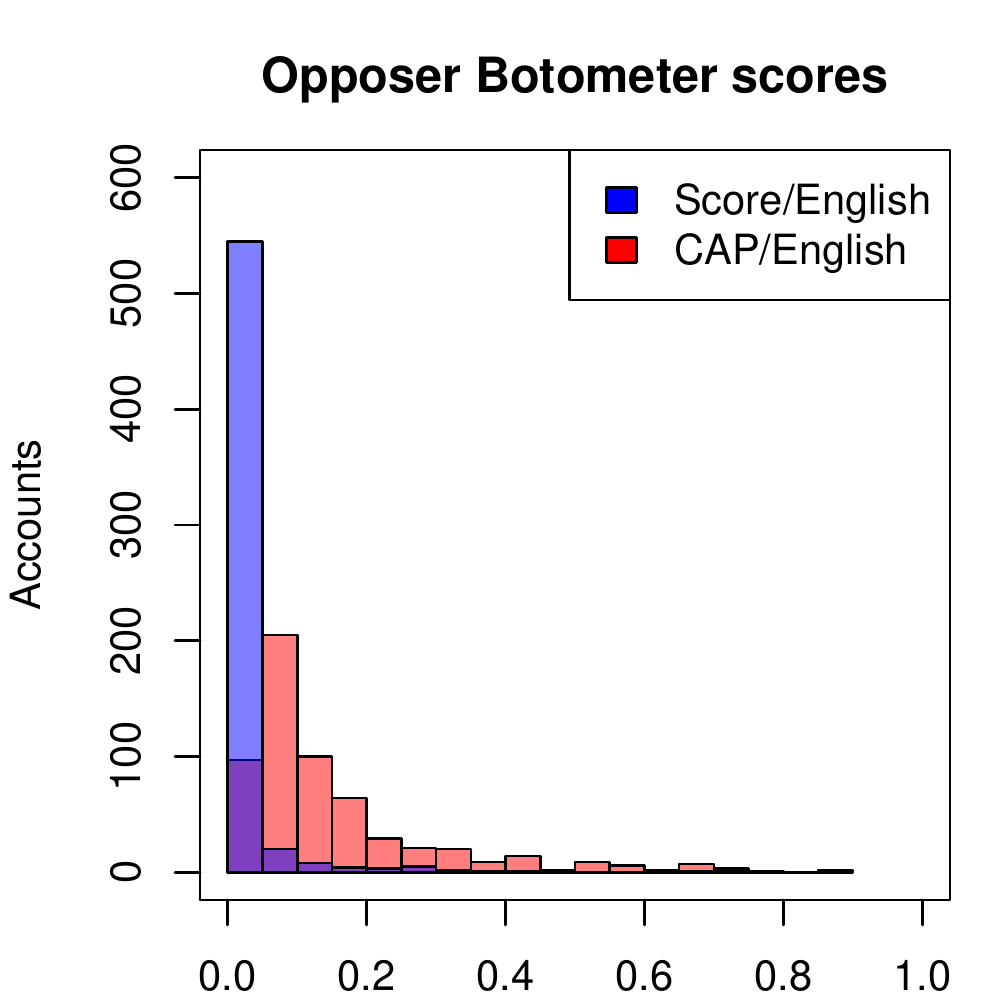}
    }
    \caption{The distribution of Botometer scores. The scores presented are the English score and the Complete Automation Probability, and all are heavily skewed towards low values (i.e., non-automated). The bars are semi-transparent, affecting their colour, to account for their overlap.}
    \label{fig:botness_histogram}
\end{figure}

Table~\ref{tab:botness-scores-and-contribution-by-phase} shows that the majority of accounts were human and contributed more than any automated or potentially automated accounts. The distributions of English and CAP scores for all tested accounts overall and only in Phase~1, when few Opposers were active, and separately for Supporters and Opposers are shown in Figure~\ref{fig:botness_histogram}. There are no significant differences between the ratings of the tested accounts overall and in Phase~1, nor between Supporters and Opposers. 
A t-test confirmed that the overall and Phase~1 score distributions do not have the same mean ($p < 0.05$ for both the CAP and English scores), and a Mann Whitney test confirmed that there is not enough information to be confident that the Supporter and Opposer score distributions are the same ($p > 0.05$ for both scores).

These results contrast with the reported findings \citep{Stilgherrian2020zdnet} is likely to be due to a number of reasons, 
but the primary one is differences in our datasets. 
Graham and Keller used the collection tool Twint (which avoids using the Twitter API and instead uses the Twitter web user interface (UI) directly) to focus on 
results from Twitter's web UI when searching for \hashtag{ArsonEmergency}. 
Only $812$ tweets appeared in both datasets, and even those were restricted to Phase~1. Of the $315$ accounts in common, $100$ were Supporters and $5$ Opposers, implying that those Supporter accounts had already been flagged by misinformation researchers as having previously engaged in questionable behaviour. The size of our dataset and the greater number of accounts we tested is likely to have skewed our Botometer results towards typical users. There are also differences between the bot analysis tools. 
Botometer's CAP score is focused on non-hybrid, English accounts, whereas \texttt{tweetbotornot} may provide a more general score, taking into account troll-like behaviour. The content and behaviour analysis discussed above certainly indicates Supporters engaged more with replies and quotes, consistent with other observed trolling behaviour \citep{kumar2018conflict,MaricontiSBCKLS2019cscw} or ``sincere activists'' \citep{StarbirdWilson2020}. 
Follow-up work by Graham and Keller's research group has focused on such ``activists'', finding that they appeared to coordinate their activities with prominent public figures and media outlets as part of a broader and longer-running disinformation campaign spanning the months surrounding the period we have focused on \citep{keller2020arson}.

Finally, 
it should be noted that at the time of writing the \texttt{tweetbotornot} library has been replaced with a new version in a completely separate library \texttt{tweetbotornot2}\footnote{\url{https://github.com/mkearney/tweetbotornot2}} in which the bot rating system has been changed and is now more conservative. In this way, the original findings in January 2020 may be been an artifact of the original implementation, however the polarised communities discovered since are certainly real and worthy of study.




\subsection{The Most Bot-like Accounts}

\begin{table}[th]
    \centering\small
    \caption{Supporter and Opposer accounts with a Botometer rating above 0.8. Counts of tweets, friends, and followers, and ages are as of the last tweet captured during the collection period in January, 2020. }
    \label{tab:affiliated_bots}
        \begin{tabular}{@{}lrrrrr@{}}
            \toprule
                             & \multicolumn{3}{c}{\textbf{Supporters}} & \multicolumn{2}{c}{\textbf{Opposers}} \\
                             &   Bot 1 &  Bot 2\footnotemark[1] &    Bot 3 &   Bot 1 &  Bot 2\footnotemark[2]  \\
            \cmidrule(r){1-1} \cmidrule(lr){2-4} \cmidrule(l){5-6}
            Contribution     &       5 &       9 &       59 &       4 &      4  \\
            Retweets         &       5 &       9 &       56 &       4 &      4  \\
            \cmidrule(r){1-1} \cmidrule(lr){2-4} \cmidrule(l){5-6}
            Age (in days)    &   1,081 &     680 &    1,087 &   1,424 &    925  \\
            Lifetime tweets  &  47,402 &  10,351 &  349,989 &  62,201 &     74  \\
            Tweets per day   &   43.85 &   15.22 &   321.98 &   43.68 &   0.08  \\
            Friends          &  17,590 &  13,226 &   25,457 &     633 &    392  \\
            Followers        &  16,507 &  13,072 &   24,873 &     497 &     55  \\
            Reputation       &   0.484 &   0.497 &    0.494 &   0.440 &  0.123  \\
            \bottomrule
        \end{tabular}
    \footnotetext[1]{This account was found to have been deleted when checked in October, 2020.} 
    \footnotetext[2]{This account was found to have been deleted when checked in December, 2020.}
\end{table}

\begin{figure}[!ht]
    \centering
    \subfloat[Supporter bot 1.\label{fig:supporter_bot1}]{%
        \includegraphics[height=0.27\textheight]{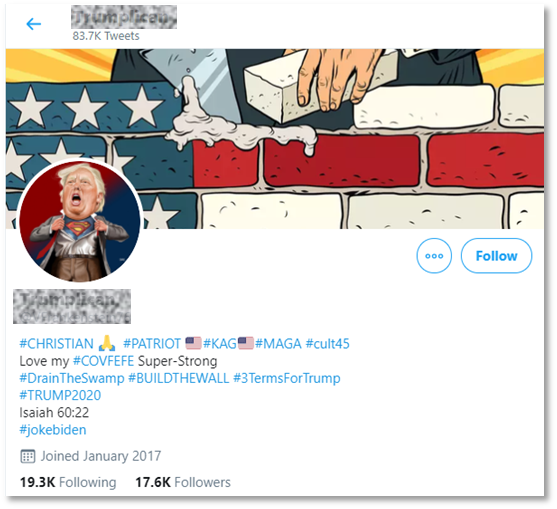}
    }
    \hfill
    \subfloat[Supporter bot 2, which was suspended---this mockup is based on data from the collection.\label{fig:supporter_bot2}]{%
        \includegraphics[height=0.27\textheight]{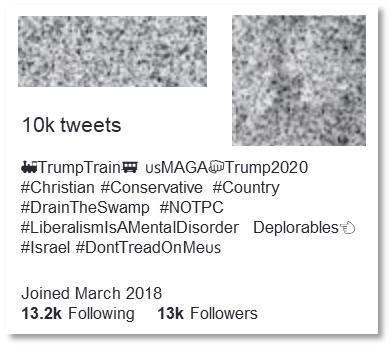}
    }
    \\ 
    \subfloat[Supporter bot 3.\label{fig:supporter_bot3}]{%
        \includegraphics[height=0.3\textheight]{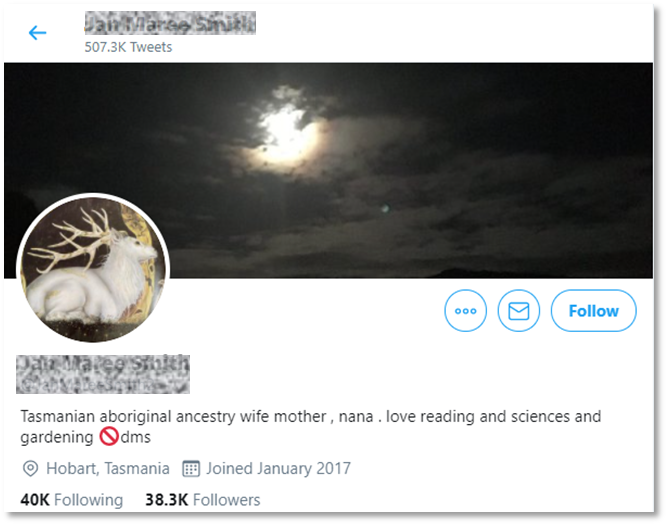}
    }
    \caption{Supporter accounts with a Botometer rating higher than 0.8, implying a high degree of bot-like traits. Personal details have been obscured. Screenshots of accounts were obtained in mid October, 2020.}
    \label{fig:supporter_bots}
\end{figure}

\begin{figure}[!ht]
    \centering
    \subfloat[Opposer bot 1.\label{fig:opposer_bot1}]{%
        \includegraphics[height=0.245\textheight]{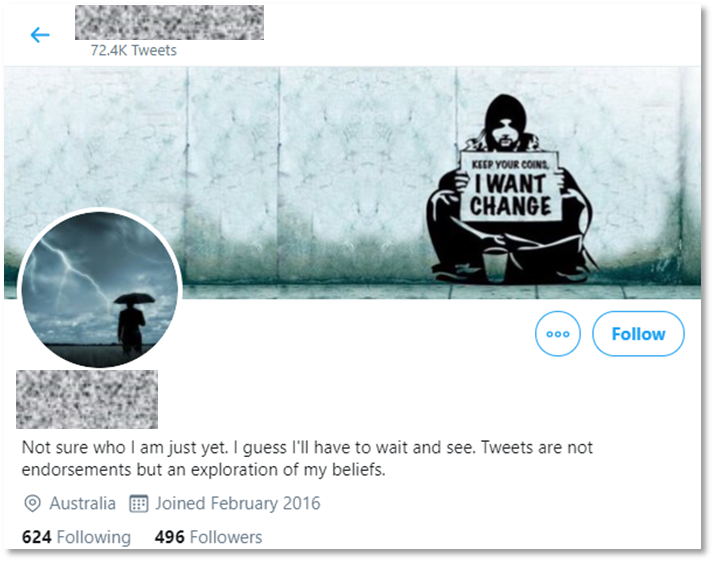}
    }
    \subfloat[Opposer bot 2.\label{fig:opposer_bot2}]{%
        \includegraphics[height=0.245\textheight]{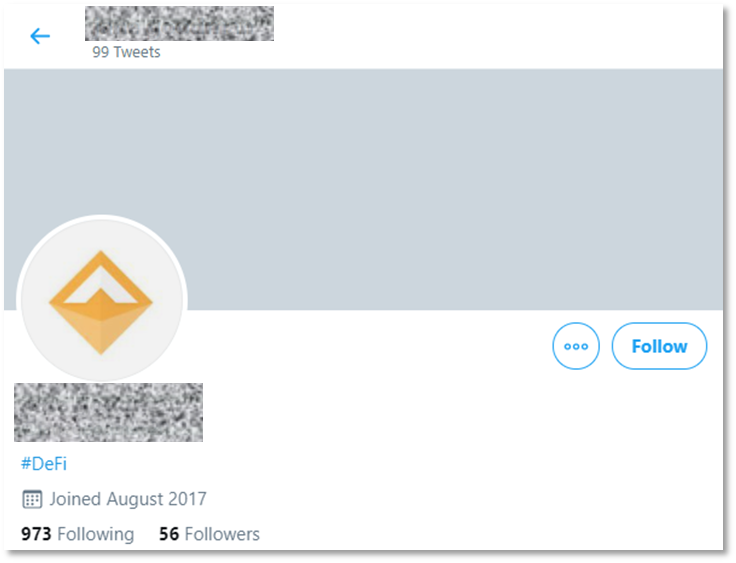}
    }
    \caption{Opposer accounts with a Botometer rating higher than 0.8, implying a high degree of bot-like traits. Personal details have been obscured. Screenshots of accounts were obtained in mid October, 2020. Bot~2 was suspended in December, 2020.}
    \label{fig:opposer_bots}
\end{figure}

Deeper analysis of the most bot-like accounts (those with a Botometer \citep{botornot2016} CAP rating of $0.8$ or more) revealed that the kinds of bot-like accounts present in each community differed significantly in a few primary respects (see Table~\ref{tab:affiliated_bots}). For convenience, we will refer to these accounts as ``bots'', but given they all present as genuine human users, they all also qualify as ``social bots'' \citep{Cresci2020} and therefore are likely to be tools for influence. 
The accounts were re-examined in October, 2020, and screenshots taken of their Twitter profiles (see Figures~\ref{fig:supporter_bots}\footnote{Supporter bot 2's account had been deleted, and so a mock-up based on the last known tweet in the ArsonEmergency corpus is presented in Figure~\ref{fig:supporter_bot2}.} and~\ref{fig:opposer_bots}). Two of the Supporter accounts appear to be American supporters of US President Donald Trump, while the third presents as an Australian indigenous woman from Tasmania who is also an active Trump supporter. The Opposer accounts include one with very little personal detail, mentioning only a hashtag for decentralised finance,\footnote{\textit{Decentralised finance}: a field of cryptocurrency in which blockchain technology is used to avoid financial institutions in transactions. \url{https://theconversation.com/decentralised-finance-calls-into-question-whether-the-crypto-industry-can-ever-be-regulated-151222}} in its description, and one that presents as a left-wing individual.

Together, the five accounts contributed $81$ tweets over the $18$ day collection period, $73$ by the Supporters (including $59$ from Bot 3) and $4$ each from the Opposer bots. This suggests they had very limited opportunity to have an impact on the discussion. All accounts had been active for at least eighteen months, up to a maximum (at the time of the collection) of nearly four years. The variations in posting rates highlight the fact that Botometer's ensemble classifier will catch accounts that do not have high posting rates (e.g., Opposer bot 2 only posted approximately $25$ tweets per year, but had been suspended by December, 2020). 
The \emph{reputation} score is defined by 
\begin{equation}
    reputation = \frac{\lvert followers \rvert}{\lvert friends \rvert + \lvert followers \rvert},
\end{equation}
and is a measure considered desirable enough worth manipulating through follower fishing \citep{Dawson2019}, yet even the bots' reputation scores 
are not very different (other than Opposer bot 2, which seems to be a rarely used account). In fact, the primary distinction between the Supporter and Opposer bots is the magnitude of their friend and follower counts. 

Supporter bots had an average of $18.8$k 
friends and $18.5$k 
followers compared with Opposer bots' averages of $512.5$ friends and $276$ followers. 
By October, 2020, over nine months, the two remaining Supporter bots, bots~1 and~3, had increased their friend and follower counts significantly: bot~1 had $1.7$k\footnote{Count changes are in thousands, as the figures are obtained from the profile screenshots.} more friends and $1.1$k more followers, while bot~3 had $14.5$k more friends and $13.4$k more followers. Over the same 
period, bot~1 had posted another $36.3$k tweets (a $77\%$ increase at more than $130$ tweets per day) and bot 3 had posted another $157.3$k tweets (a $45\%$ increase at nearly $600$ tweets per day). Bots~1 and~3 had been created $6$ days apart and, in January, 2020, both had been running for just over three years. In contrast, Opposer bot~1 had lost one follower and reduced the number of accounts it followed by $9$, but added just over $10$k tweets (approximately $37$ tweets per day), while Opposer bot~2 had increased the accounts it followed by $148\%$, added one follower and posted only $25$ tweets.

It is not clear why these accounts are so different. It is possible these accounts are, in fact, merely highly motivated people, who spend a significant amount of time curating their Twitter feeds to include material they prefer and then retweet almost everything they see to simply promote their preferred narrative. This accords with recent observations that Twitter increasingly consists of retweets of official sources and celebrities and tweets with URLs, and rather than being a town square of public discussion, it should be treated as an ``attention signal'', which highlights the ``stories, users and websites resonating'' at a given time \citep{Leetaru2019twitter}. These accounts appear driven to amplify that ``attention signal'' for ideological reasons, for the most part (Opposer bot~2's tweeting motivations are unclear). What also stands out is that the Supporter bots differ distinctly from the rest of the Supporter community who relied much less on retweets than the Opposer community.

\begin{figure}[!ht]
    \centering
    \includegraphics[width=0.99\textwidth]{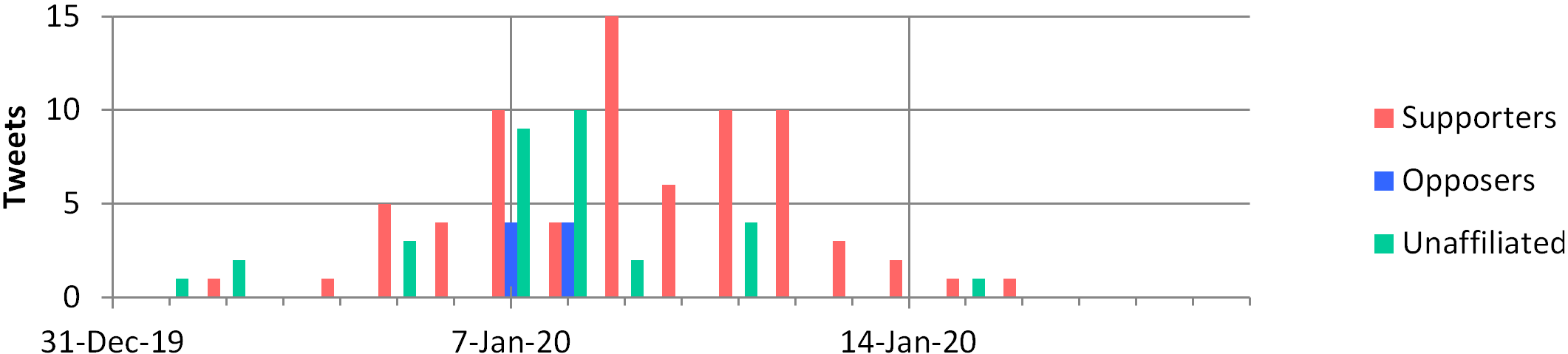}
    \caption{Tweets per day by the three Supporter, two Opposer and fifteen bot accounts.}
    \label{fig:bot_timeline}
\end{figure}

Figure~\ref{fig:bot_timeline} shows the activity patterns for the Supporter and Opposer bot accounts, and also for the $15$ Unaffiliated accounts that had been suspended when the bot analysis was conducted (at the end of January 2020). The Opposer contribution is small and occurs in Phase~2 and the first day of Phase~3, clearly responding to the MSM news, while the Supporter bots are active in the lead up to Phase~2 and well into Phase~3, engaging in the ongoing discussion, though their activity patterns indicate that if they are bots tweeting frequently, then their tweets mostly avoided using \hashtag{ArsonEmergency} (and thus were not captured in our collection). The Unaffiliated accounts are also mostly active only on the day the story reached the MSM and the following day, and their contribution was limited to only $32$ tweets.

\subsection{Inauthentic Behaviour}\label{sec:inauthentic_behaviour}

Aggressive language was observed in both Supporter and Observer content, but the hashtag and mention use provide the most insight into potential inauthentic behaviour \citep{weedon2017facebook}. Supporters used more hashtags and more mentions in tweets than Opposers in general (Table~\ref{tab:group-activity-by-interaction-type-and-phase}), and posted individual tweets with many more of each (the number of tweets with at least $14$ hashtags or $5$ mentions was $50$), though a small proportion of Unaffiliated accounts used even more hashtags in their tweets (a maximum of $27$). Supporters posted tweets consisting of only hashtags, mentions and a URL in various combinations (i.e., eschewing actual content) far more frequently than Supporters or Unaffiliated, on a per-account basis, particularly in Phase~3 (see Table~\ref{tab:inauthentic_tweet_text_patterns}). Using hashtags and mentions in these numbers is a way to increase the reach of your message (though, ironically, it often leaves little space for the message itself), but can also be used to attack others or pollute hashtag-based discussion communities \citep{woolley2016autopower,nasim2018real}.




\begin{table}[ht]
    \centering
    \caption{Frequency of inauthentic text patterns in the ArsonEmergency tweets (includes retweeted text).}
    \label{tab:inauthentic_tweet_text_patterns}
    \resizebox{\textwidth}{!}{%
        \begin{tabular}{@{}llrrrrrrrr@{}}
        \toprule
                     &                           & \multicolumn{2}{c}{\textbf{Overall}} & \multicolumn{2}{c}{\textbf{Phase 1}} & \multicolumn{2}{c}{\textbf{Phase 2}} & \multicolumn{2}{c}{\textbf{Phase 3}} \\ 
                       &                         & Count & \% of All     & Count & \% of All    & Count & \% of All    & Count & \% of All     \\
        \cmidrule(r){1-2} \cmidrule(lr){3-4} \cmidrule(lr){5-6} \cmidrule(lr){7-8} \cmidrule(l){9-10}
        Supporters   & \textit{All tweets}                & 6,972 & 100.0\% & 1,573 & 100.0\% & 121 & 100.0\% & 5,278 & 100.0\% \\
        \cmidrule(r){2-2} \cmidrule(lr){3-4} \cmidrule(lr){5-6} \cmidrule(lr){7-8} \cmidrule(l){9-10}
                     & Hashtag(s)                  & 20         & 0.3\%          & 1          & 0.1\%          & 0          & 0.0\%          & 19         & 0.4\%          \\
                     & Hashtag(s) + URL            & 669        & 9.6\%          & 160        & 10.2\%         & 7          & 5.8\%          & 502        & 9.5\%          \\
                     & Mention(s) + Hashtag(s)       & 340        & 4.9\%          & 60         & 3.8\%          & 3          & 2.5\%          & 277        & 5.2\%          \\
                     & Mention(s) + Hashtag(s) + URL & 73         & 1.0\%          & 12         & 0.8\%          & 2          & 1.7\%          & 59         & 1.1\%          \\
        \cmidrule(r){1-2} \cmidrule(lr){3-4} \cmidrule(lr){5-6} \cmidrule(lr){7-8} \cmidrule(l){9-10}
        Opposers     & \textit{All tweets}                & 3,587      & 100.0\%               & 33         & 100.0\%               & 327        & 100.0\%               & 3,227      & 100.0\%               \\
        \cmidrule(r){2-2} \cmidrule(lr){3-4} \cmidrule(lr){5-6} \cmidrule(lr){7-8} \cmidrule(l){9-10}
                     & Hashtag(s)                  & 0          & 0.0\%          & 0          & 0.0\%          & 0          & 0.0\%          & 0          & 0.0\%          \\
                     & Hashtag(s) + URL            & 47         & 1.3\%          & 1          & 3.0\%          & 3          & 0.9\%          & 43         & 1.3\%          \\
                     & Mention(s) + Hashtag(s)       & 0          & 0.0\%          & 0          & 0.0\%          & 0          & 0.0\%          & 0          & 0.0\%          \\
                     & Mention(s) + Hashtag(s) + URL & 5          & 0.1\%          & 0          & 0.0\%          & 0          & 0.0\%          & 5          & 0.2\%          \\
        \cmidrule(r){1-2} \cmidrule(lr){3-4} \cmidrule(lr){5-6} \cmidrule(lr){7-8} \cmidrule(l){9-10}
        Unaffiliated & \textit{All tweets}                & 16,987     &  100.0\%              & 1,961      & 100.0\%                & 759        & 100.0\%                & 14,267     & 100.0\%                \\
        \cmidrule(r){2-2} \cmidrule(lr){3-4} \cmidrule(lr){5-6} \cmidrule(lr){7-8} \cmidrule(l){9-10}
                     & Hashtag(s)                  & 34         & 0.2\%          & 2          & 0.1\%          & 0          & 0.0\%          & 32         & 0.2\%          \\
                     & Hashtag(s) + URL            & 629        & 3.7\%          & 181        & 9.2\%          & 14         & 1.8\%          & 434        & 3.0\%          \\
                     & Mention(s) + Hashtag(s)       & 180        & 1.1\%          & 35         & 1.8\%          & 8          & 1.1\%          & 137        & 1.0\%          \\
                     & Mention(s) + Hashtag(s) + URL & 102        & 0.6\%          & 18         & 0.9\%          & 1          & 0.1\%          & 83         & 0.6\%          \\ \bottomrule
        \end{tabular}%
    }
\end{table}

\begin{figure}[t!]
    \centering
    \includegraphics[width=0.99\textwidth]{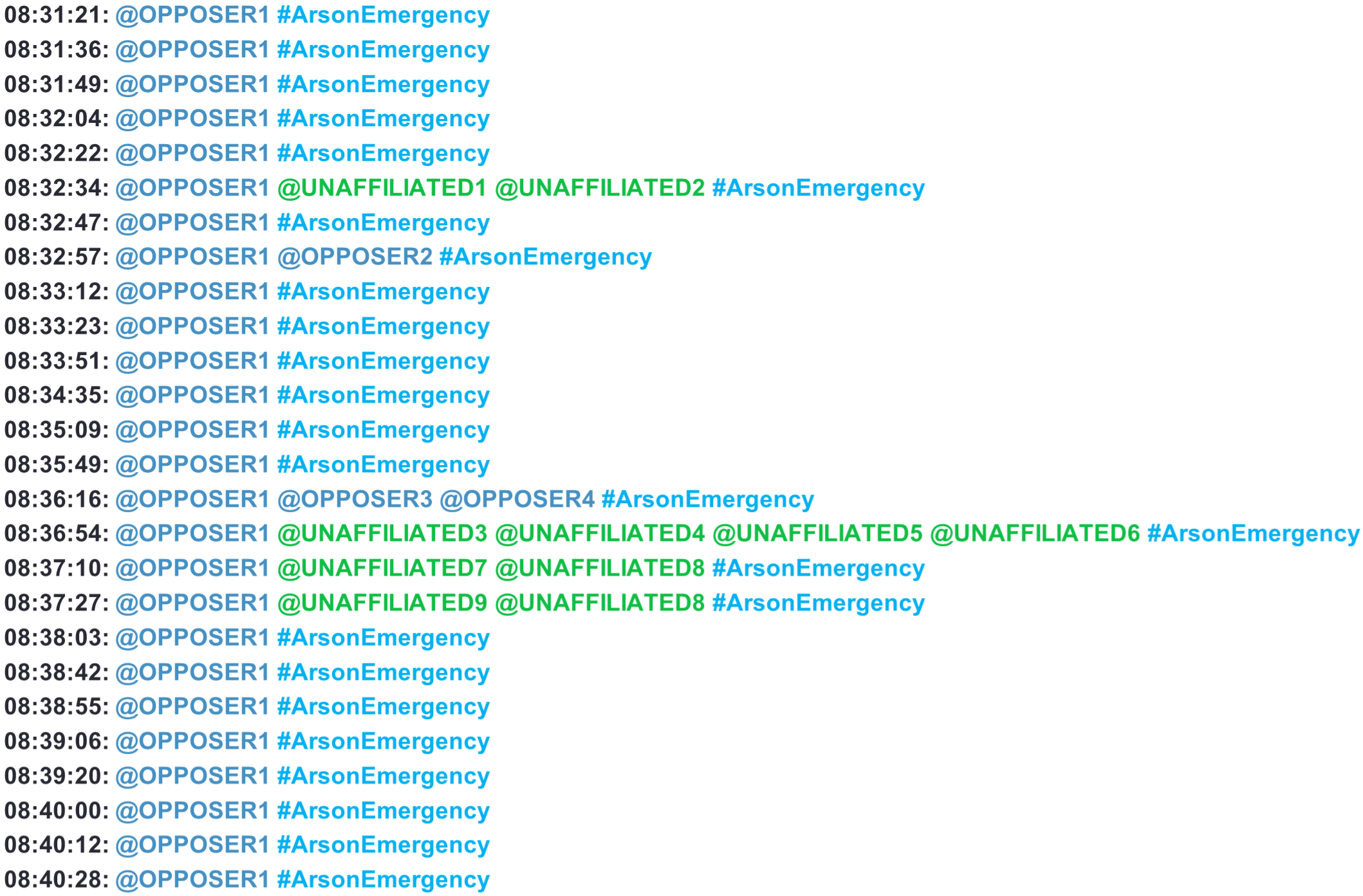}
    \caption{The timestamps and text of tweets posted by a single Supporter account targeting an Opposer account early in Phase~3. Account names are anonymised but named and coloured according to their affiliation.}
    \label{fig:troll_example}
\end{figure}

In one notable instance, a Supporter account posted $26$ highly repetitive tweets to an Opposer account within $9$ minutes, including only the \hashtag{ArsonEmergency} hashtag in the majority of them (Figure~\ref{fig:troll_example}). In six tweets, other accounts were mentioned, including prominent Opposer and Unaffiliated accounts, perhaps in the hope that they would engage by retweeting and thus draw in their own followers.

\section{Discussion}

Our discussion addresses the research questions we posed in Section~\ref{sec:research_questions}.

\begin{description}
    \item [\textbf{RQ1}] \emph{Discerning misinformation-sharing campaigns.} 
    
    Analysis 
    revealed two distinct polarised communities, each of which amplified particular narratives. The content posted by the most influential accounts in each of these communities shows Supporters were responsible for the majority of arson-related content, 
    while Opposers 
    countered the arson narrative, debunking the errors and false statements with official information from community authorities and fact-check articles. 
    Prior to the release of the ZDNet article, the discussion on the \hashtag{ArsonEmergency} hashtag was dominated by arson-related content. In that sense, the misinformation campaign was most effective in Phase~1, but only because its audience was small. Once the audience grew, as the hashtag received broader attention, the conversation became dominated by the Opposers' narrative and related official information.

    \item [\textbf{RQ2}] \emph{Differences in the spread of information across phases and other discussions.} 
    
    We regarded URL and hashtags as proxies for narrative and studied their dissemination, finding distinct differences between the groups and the their activity in different phases.
    In Phase~1, only Supporters and Unaffiliated shared URLs, the most popular of which were in the NARRATIVE category, but by the third Phase, the most popular URLs shared were DEBUNKING in nature by a ratio of $9$ to $1$, and NARRATIVE URLs were share only by Supporter accounts. Although it is unclear whether this change in sharing behaviour was due to changes in opinions or the influx of new accounts, there was certainly a changing of the guard. Of the $2{,}061$ accounts active in Phase~1, less than $40\%$ ($787$) remained active in Phase~3. While most Phase~1 Supporters ($339$ of $360$) posted in Phase~3, many fewer Unaffiliated accounts did ($427$ of $1{,}680$) indicating that the Supporters lost the support of most of the Phase~1 Unaffiliated accounts.
    
    The diversity of URL and hashtag use also changed from Phase~1 to Phase~3: while the number of active Supporters grew modestly from $360$ to $474$, the number of unique external URLs they used more than doubled, from 
    $193$ to $321$. 
    Opposers and Unaffiliated used more unique URLs in Phase~3 ($492$ and $4{,}368$, respectively), but Figures~\ref{fig:url_use_distribution} and~\ref{fig:co-url-bigraph} shows they focused on a small set of URLs more than Supporters did.
    
    The number of hashtags Supporters used increased from $191$ hashtags used $5{,}382$ times to $543$ hashtags used $14{,}472$ times. This implies Supporters attempted to connect \hashtag{ArsonEmergency} with other hashtag-based communities, which could have been to in order to promote their message widely, to co-opt existing discussion spaces, or due to non-Australian contributors being unfamiliar with which hashtags would be relevant to the mostly Australian audience. From Phase~1 to Phase~3, Opposer activity increased from $34$ hashtags used $150$ times to $200$ hashtags used $9{,}549$ times, and Figure~\ref{fig:cca_hashtag_co-mentions} shows Opposers focused the majority of their discussion on a comparatively small number of hashtags. 
    
    The \hashtag{ArsonEmergency} discussion's growth rate followed a similar pattern to a related hashtag that appeared around the same time (\hashtag{AustraliaFire}), but it was clearly different from that of a well-established discussion (\hashtag{brexit}).
    

    \item [\textbf{RQ3}] \emph{Behavioural differences over time and the impact of media coverage.} 
    
    Supporters were more active in Phase~1 and~3 and used 
    more types of interaction than Opposers, 
    especially 
    replies and quotes, implying a significant degree of engagement, whether as trolls 
    or as ``sincere activists'' \citep{StarbirdWilson2020}. 
    Opposers and Supporters made up the majority of retweeted accounts overall, and made up $22$ of the top $25$ accounts retweeted by Unaffiliated accounts in Phase~3. 
    Supporters' use of interaction types remained steady from Phase~1 to~3. 
    While behaviour remained relatively similar, activity grew for both groups after the story reached the MSM. The vast majority of accounts shared 
    articles debunking the false narratives. The publication of the ZDNet article \citep{Stilgherrian2020zdnet} also affected activity, spurring Opposers and others to share the analysis it reported.

    \item [\textbf{RQ4}] \emph{Position of communities in the discussion network.} 
    
    Supporter efforts to engage with others in the discussion resulted in them being deeply embedded in the discussion's reply, mention and quote networks and having correspondingly high centrality values. Our $k$-core analysis showed they were evenly distributed throughout the networks, from the periphery to the cores. Despite Opposers staying more on the periphery of the networks, they maintained high closeness and eigenvector centrality scores, meaning they stayed connected to more of the network than Supporters and certainly to more important nodes in the network. Correspondingly, this may imply that Supporters, though being highly connected, were not connecting as efficiently as Opposers, in order to spread their narrative. Both Opposer and Supporter groups were highly insular with respect to each other, across a variety of network analyses, but they connected strongly to the broader community according to E-I indices and assortativity.

    \item [\textbf{RQ5}] \emph{Content dissemination and coordinated activity.} 
    
    Analyses of hashtag and URL use revealed further evidence of the gap between Supporters and Opposers, not just in terms of connectivity, as discussed above, but also in terms of narrative. Supporters used a variety of hashtags to reach greater audiences, to disrupt existing communication channels, or to otherwise harass. In doing so, they exhibited less evidence of coordination than Opposers, who were focused in both the hashtags and URLs they used, supported by or in concert with the much greater number of Unaffiliated accounts. Analysis of co-activities (namely co-retweeting, and co-URL and co-hashtag instances) suggested a lack of botnets in the discussion and that some Unaffiliated and Opposers were coordinating their URL sharing, appearing together in cliques that are often attributed to automation \cite[e.g.,][]{Pacheco2020www}. The apparent coordination could, however, be attributed to high levels of popularity driven by increased activity in Phase~3 (i.e., coincidence due to high numbers of discussion participants), and the co-activities of Supporters indicated the presence of genuine human users more than any automated coordination. Further analysis using account/URL bigraphs showed that Opposers and Unaffiliated were focused on sharing a small set of URLs, compared with Supporters' greater variety. These findings imply the Supporter community members, for all they attempted to engage with others via replies, mentions and hashtags, becoming deeply embedded in the interaction networks, remained relatively isolated from a narrative perspective.

    \item [\textbf{RQ6}] \emph{Support from non-Australian accounts.} 
    
    Based on manual inspection of accounts' free text `location' fields, the Supporter group included more non-Australian than Opposers, with the greatest number of non-Australian accounts Unaffiliated with either, but the vast majority of all groups indicated they were located in Australia ($> 70\%$). Despite the large number of Unaffiliated accounts present in Phase~1 ($1{,}680$), the majority joined the discussion in Phase~3, likely bringing in the majority of non-Australian accounts. Investigations of content dissemination also revealed that Opposers received the majority of Unaffiliated support, resulting in a majority of debunking article shares in Phase~3 from a majority of narrative-aligned article shares in Phase~1, so it is possible that this also included non-Australian support. Given most accounts do not report their location, and locations have not been verified, this conclusion remains speculative.

    \item [\textbf{RQ7}] \emph{Support from bots and trolls.} 
    
    We found very few bots and their impact was limited: 
    only $0.8\%$ ($20$ of $2{,}512$) had a Botometer \citep{botornot2016} CAP score above $0.6$ while $96.6\%$ ($2{,}426$) were highly likely to be human (CAP $< 0.2$). In contrast, Graham and Keller had found many more bots ($46\%$) and fewer humans ($< 20\%$) in their smaller sample \citep{Stilgherrian2020zdnet,GrahamKeller2020conv}. The affiliated `bot' accounts, on closer examination, may not all have been automated, but the ones with bot-like posting rates could certainly be classed as `social bots' \citep{Cresci2020} given their appearance as genuine human users. 
    Aggressive language was observed in both affiliated groups, but troll-like tweet text patterns including only hashtags, mentions and URLs (i.e., without content terms) were employed far more often by Supporters. 
    Distinguishing deliberate baiting from honest enthusiasm (even with swearing) is non-trivial \citep{StarbirdAW2019cscw,StarbirdWilson2020}, but identifying targeted tweets lacking content is a more tractable approach to detect inauthentic and potentially malicious behaviour.
\end{description}

Further research is required to examine the dynamic aspects of the social and interaction structures formed by groups involved in spreading misinformation to learn more about how to better address the challenge they pose to society.
Future work will draw more on social network analysis based on interaction patterns and content \citep{bagrow2019information} as well as developing a richer, more nuanced understanding of the Supporter community itself, including revisiting the polarised accounts over a longer time period and consideration of linguistic differences. A particularly challenge is determining a social media user's intent when they post or repost content, which could help distinguish between disinformation intended to deceive, and merely biased presentation of data or misinformation that aligns with the user's worldview.


\section{Conclusion}

The study of polarised groups, their structure and their behaviour, during times of crisis can provide insight into how misinformation can enter and be maintained in online discussions, as well as provide clues as to how it can be removed.
The \hashtag{ArsonEmergency} activity on Twitter in early 2020 provides a unique microcosm to study the growth of a misinformation campaign before and after it was widely known. 
Here we have shown that polarised groups can communicate over social media in very different ways while discussing the same issue. In effect, these behaviours can be considered communication strategies, given they are used to promote a narrative and represent attempts to convince others to accept their ideas. 
Supporters of the arson narrative used direct engagement with mentions and replies to reach individuals and hashtags to reach groups with a wide range of URLs to promote their message, while Opposers focused on using retweets and a select set of URLs to counter their message. Supporter activities resulted in them being deeply embedded and distributed in the interaction networks, yet Opposers maintained high centrality and were supported by and appeared to coordinate with active Unaffiliated accounts. 
The counteraction appears to have been successful, with the predominant class of articles shared shift from narrative-aligned in Phase~1 to debunking articles in Phase~3. 
Graham and Keller's efforts to draw attention to the \hashtag{ArsonEmergency} discussion~\citep{Stilgherrian2020zdnet}, and the subsequent associated MSM attention, is likely to have contributed to this effect, given the significant increase in discussion participants in Phase~3. This highlights the value in publicising research into misinformation promotion activities. 

We speculate that the communication patterns documented in this study could be communication strategies discoverable in other misinformation-related discussions, such as those relating to vaccine conspiracies \citep{Broniatowski2018}, COVID-19 anti-lockdown regulations \citep{loucaides2021}, challenging election results \citep{Scott2021capitolriots,Ng2021} or QAnon \citep{soufan2021qanon}, and could help inform the design and development of counter-strategies. 


\backmatter

\bmhead{Supplementary information}

This paper includes appendices with further detail of analyses conducted.

\bmhead{Acknowledgments}

The authors acknowledge support from the Australian Research Council's Discovery Projects funding scheme (project DP210103700) and thank Graham and Keller for access to their datasets for comparison.

\section*{Declarations}

\subsection*{Funding}
The authors acknowledge support from the Australian Research Council's Discovery Projects funding scheme (project DP210103700).

\subsection*{Competing interests}
The authors have no relevant funding, financial or non-financial interests to disclose.

\subsection*{Ethics approval}
All data was collected, stored and analysed in accordance with Protocol H\nobreakdash-2018\nobreakdash-045 as approved by the University of Adelaide's human research ethics committee.



\subsection*{Availability of data and materials}

The datasets collected and analysed during the current study (the identifiers of the tweets, as per Twitter's terms and conditions) are available at \url{https://github.com/weberdc/socmed_sna}.

\subsection*{Code availability}

The code used in this work are available at \url{https://github.com/weberdc/socmed_sna}.

\subsection*{Authors' contributions}

\begin{itemize}
    \item Derek Weber: Conceptualisation, Formal Analysis, Methodology, Software, Investigation, Data Curation, Writing – Original Draft, Writing – Review \& Editing, Visualisation
    \item Lucia Falzon: Supervision, Writing – Original Draft, Writing – Original Draft, Writing – Review \& Editing, Validation
    \item Lewis Mitchell: Funding, Supervision, Writing – Review \& Editing, Validation
    \item Mehwish Nasim: Supervision, Conceptualisation, Formal Analysis, Methodology, Investigation, Writing – Original Draft, Writing – Review \& Editing, Visualisation
\end{itemize}

\begin{appendices}

\section{Location Analysis}\label{app:locations}

In exploring the discussion of any contentious regional topic on social media, it is sensible to consider from where contributors come. People from different countries may bring different opinions to the table, and when such discussions may help shape public policy, there is the potential for malign foreign interference. The simplest approach is to consider the `lang' field in the tweet metadata,\footnote{The `language', `utc\_offset' and `timezone' fields within the `user' field of tweets have been deprecated: \url{https://developer.twitter.com/en/docs/tweets/data-dictionary/overview/user-object}.} which is assigned by Twitter. Across every group and phase, roughly 99\% of the tweets had a language code of `en' (English) or `und' (undefined). Manual inspection of the largest `und' proportion ($1{,}007$ tweets by Supporters in Phase~3, $19.1$\% of those tweets) revealed the tweets' content comprised almost entirely of @mentions and hashtags.

\begin{table}[ht]
    \centering
    \caption{The self-reported locations of accounts, categorised by country by hand. Only non-empty locations were used, and only those used multiple times by Unaffiliated accounts were considered (i.e., unique Unaffiliated locations were ignored).}
    \label{tab:location-by-group}
    \resizebox{\columnwidth}{!}{%
        \begin{tabular}{@{}lrrrrrr@{}}
            \toprule
                          & \multicolumn{2}{c}{\textbf{Opposer}} & \multicolumn{2}{c}{\textbf{Supporter}} & \multicolumn{2}{c}{\textbf{Unaffilated}}  \\
            Country       & Counts & Proportion     & Counts & Proportion     & Counts & Proportion   \\
            \cmidrule(r){1-1} \cmidrule(l){2-3}       \cmidrule(l){4-5}         \cmidrule(l){6-7}
            Australia     &    393 &         88.7\% &    273 &         76.9\% &  3,642 &       72.0\% \\
            USA           &      4 &          0.9\% &     19 &          5.4\% &    586 &       11.6\% \\
            UK            &      4 &          0.9\% &      5 &          1.4\% &    287 &        5.7\% \\
            Canada        &      2 &          0.5\% &      7 &          2.0\% &    146 &        2.9\% \\
            NZ            &      2 &          0.5\% &      5 &          1.4\% &     51 &        1.0\% \\
            Miscellaneous &     35 &          7.9\% &     41 &         11.5\% &    143 &        2.8\% \\
            Other         &      3 &          0.7\% &      5 &          1.4\% &    204 &        4.0\% \\
            \cmidrule(r){1-1} \cmidrule(l){2-3}        \cmidrule(l){4-5}            \cmidrule(l){6-7}
            Total         &    443 &        100.0\% &    355 &        100.0\% &  5,059 &      100.0\% \\
            \bottomrule
        \end{tabular}
    }
\end{table}

To learn more, we examined the `location' field in the `user' objects in the tweets. This is a free text field users can populate as they wish and contains a great variety of information, not all of which is accurate, but the majority of populated fields are at least meaningful locations ($88\%$). We manually coded the `location' for each Supporter and Opposer account and then the `location' values that appeared more than once for the Unaffiliated accounts (Table~\ref{tab:location-by-group}). The majority of contributors in each group is from Australia, but the Supporters and Unaffiliated accounts included more non-Australian but English-speaking contributions than Opposers. The larger proportion of American and UK contributions in the Unaffiliated accounts may be due to an influx of highly-motivated users who joined the discussion after Graham's analysis~\citep{Stilgherrian2020zdnet} reached the MSM. It is thought that climate change is less settled in those countries.\footnote{\url{https://www.theguardian.com/environment/2019/may/07/us-hotbed-climate-change-denial-international-poll}} This is borne out by the increased number of unique Unaffiliated accounts in Phase~3.

\section{Exploration of Inauthentic Behaviour}

Aggressive and profane language was observed in content posted by both Supporters and Opposers, but our observations includes behaviour that could be regarded as inauthentic~\citep{weedon2017facebook}, including trolling. We examined the frequency of hashtags and mentions appearing in tweets by Supporters, Opposers and the remainder of accounts, as well as identifying inflammatory behaviour through manual inspection. 

The $288$ Supporters and $149$ Opposers in the mention network connected to Opposers and Supporters, respectively, slightly more than they mentioned themselves, with $710$ edges (E-I Index of $-0.14$). When Unaffiliated accounts are considered (resulting in a mention network of $3{,}206$ nodes and $5{,}825$ edges, a subset of the one shown in 
Figure~\ref{fig:arson-mentions-network} (main paper) 
which omits Unaffiliated---Unaffiliated edges), the combined E-I Index for Supporters and Opposers rises to $0.7$, suggesting a clear preference to mention Unaffiliated accounts. 

\begin{figure}[ht]
    \centering
    \includegraphics[width=0.8\textwidth]{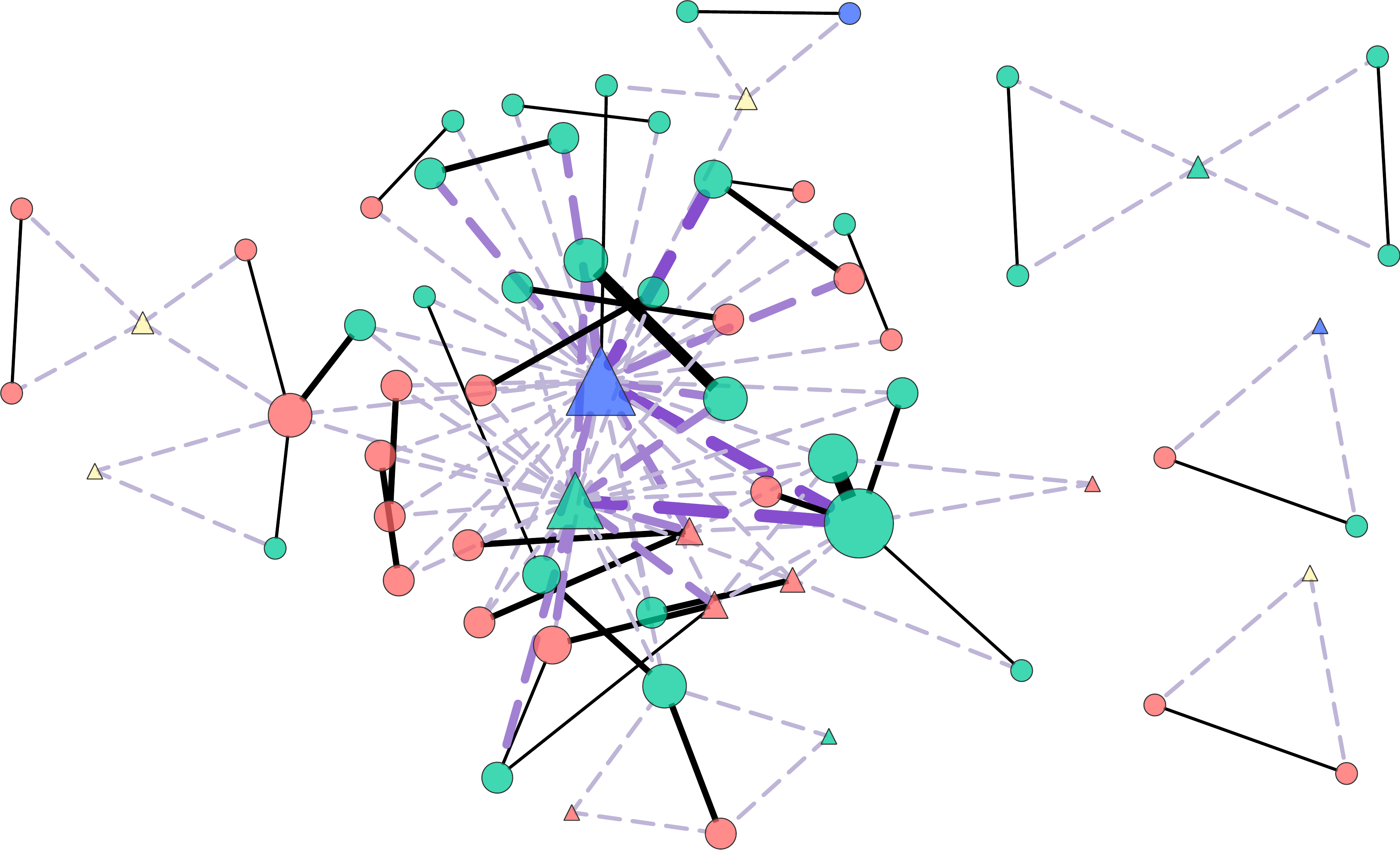}
    \caption{The account/mention bigraph resulting from a co-mention analysis, connecting accounts with black edges when they mentioned the same account within 60 seconds. Purple edges connect accounts with the accounts they mention, which are shown as triangles. Node colour indicates affiliation: red nodes are Supporters; blue nodes are Opposers; green nodes are Unaffiliated accounts; and yellow nodes are accounts that were mentioned but did not post a tweet in the dataset. Node size indicates the number of tweets they contributed to the corpus or, for mentioned accounts, their degree (reflecting the number of times they were mentioned). }
    \label{fig:arson_co-mention_acct-reason_network}
\end{figure}

An analysis of contemporaneous co-mentions also reveals that Supporter accounts mentioned the same accounts in quick succession much more frequently than Opposers, but that one prominent Opposer account was mentioned by many other accounts (Figure~\ref{fig:arson_co-mention_acct-reason_network}). It is clear the highly mentioned Opposer is a target for accounts, with many pairs of co-mentioners mentioning only the Opposer. A second (Unaffiliated) account is also highly mentioned, lying just below the Opposer account, though it appears mentioned more often by Supporter accounts, while the Opposer is more often mentioned by Unaffiliated accounts. The Opposer account is a prominent left-wing online personality mentioned more than $2400$ times in the dataset, while the Unaffiliated account had been suspended by the end of January 2020, just after the collection period, and was mentioned over $350$ times in the dataset. The largest Unaffiliated mentioning account (circular green node, on the right of the large connected component) appears to support the arson narrative and also promotes a number of QAnon-related hashtags~\citep{soufan2021qanon}.

\begin{figure}[t!]
    \centering
    \includegraphics[width=0.8\textwidth]{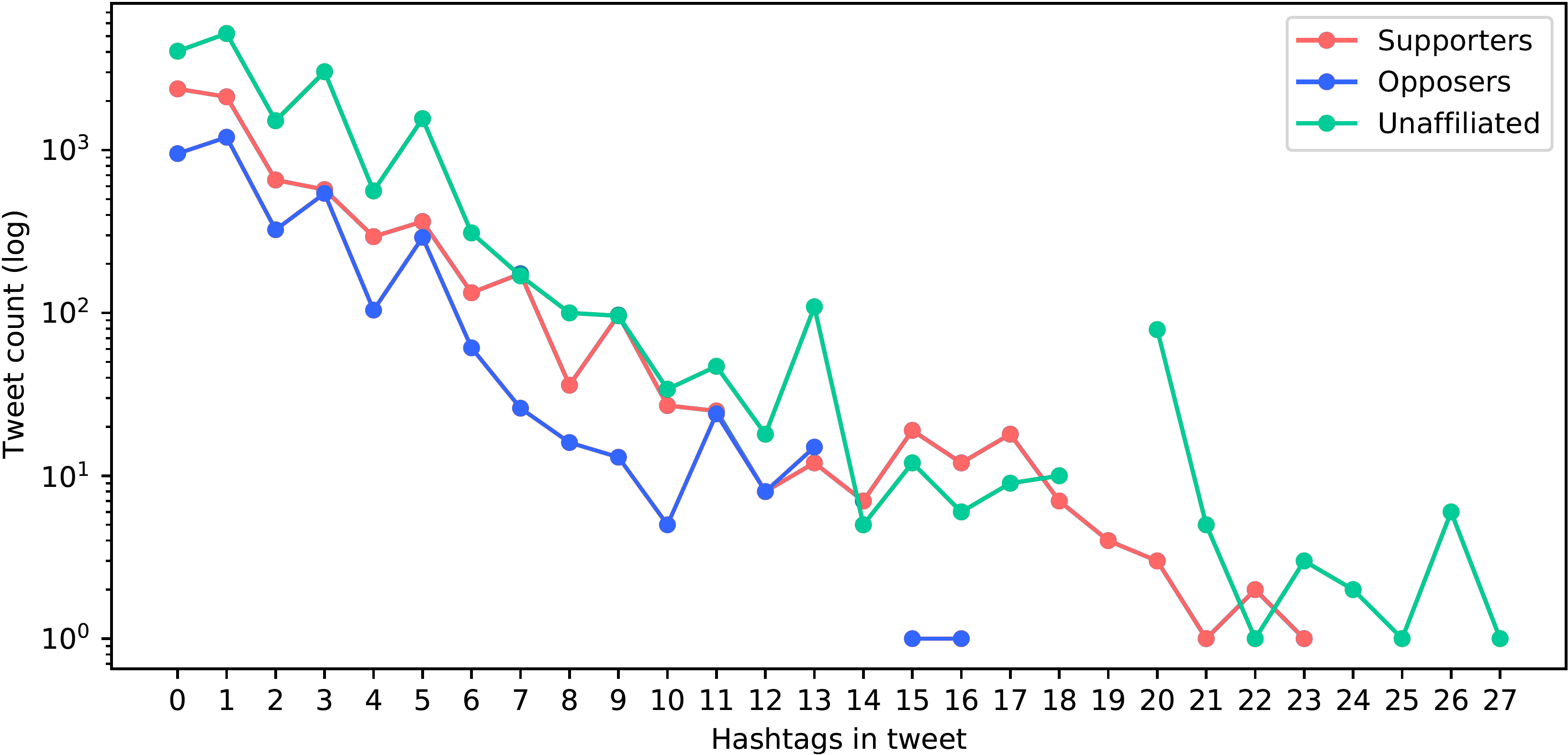}
    \caption{The distribution of hashtag uses amongst all ArsonEmergency tweets.}
    \label{fig:hashtag_use_distribution}
\end{figure}

Tweets that include many hashtags or mentions can stand out in a timeline, because the vast majority of tweets include very few, if any. By including many hashtags, a tweet may be seen by anyone searching by those hashtags, thereby increasing its potential audience. Including many mentions may be a way to draw other participants into an ongoing conversation or at least inform them of an opinion or other information. Figure~\ref{fig:hashtag_use_distribution} shows that all groups trended similarly, and that Supporters posted more tweets with many hashtags than Opposers did (although they tweeted nearly twice as often). Unaffiliated accounts used the most hashtags in tweets, with more than $100$ Unaffiliated tweets including $19$ or more hashtags. Given the great numbers of Unaffiliated accounts and tweets, these can be regarded as outliers (making up less than $1\%$ of their contribution).

\begin{figure}[t!]
    \centering
    \includegraphics[width=0.8\textwidth]{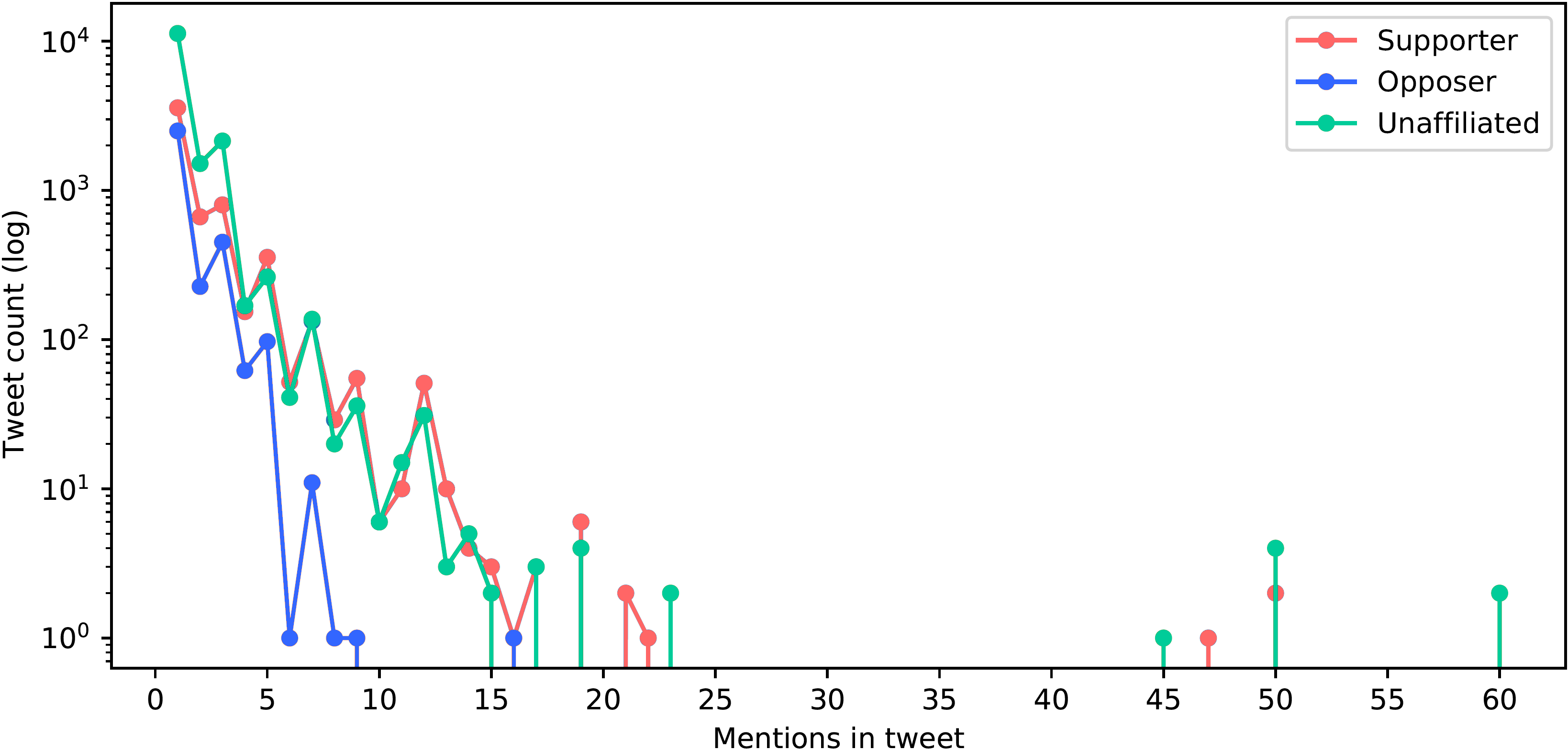}
    \caption{The distribution of mention uses amongst all ArsonEmergency tweets.}
    \label{fig:mention_use_distribution}
\end{figure}

Supporters used many more mentions than Opposers more often (Figure~\ref{fig:mention_use_distribution}). Opposers only used a maximum of $5$ mentions on fewer than $10$ occasions, while Supporters did the same more than $50$ times. In fact, Supporters used more than $5$ mentions in $369$ tweets. 
In a few tweets, $45$ or more mentions appear, however analysis of this phenomenon has revealed that Twitter accumulates mentions from tweets that have been replied to. One reply tweet including $50$ mentions was a simple reply into a reply chain that stretched back to 2018. Many replies in the chain had mentioned one or two other accounts, and they were then incorporated as implicit mentions in any replies to them. Unfortunately, from the point of view of the data provided by the Twitter API, it is unclear whether mentions in a reply are manually added by the respondent or included implicitly, as they simply appear at the start of the tweet text.

Although using many hashtags and mentions may expose inauthentic behaviour, trolling involves broad or direct attacks or simple provocation, and is exposed through use of platform features as well as the content of posts. Patterns of activity that appeared provocative included repetitions of tweets consisting of only:

\begin{itemize}
    \item one or more hashtags;
    \item one or more hashtags and a trailing URL;
    \item one or more mentions with one or more hashtags; and
    \item one or more mentions with one or more hashtags and a trailing URL.
\end{itemize}

\begin{figure}[t]
    \centering
    \includegraphics[width=0.99\textwidth]{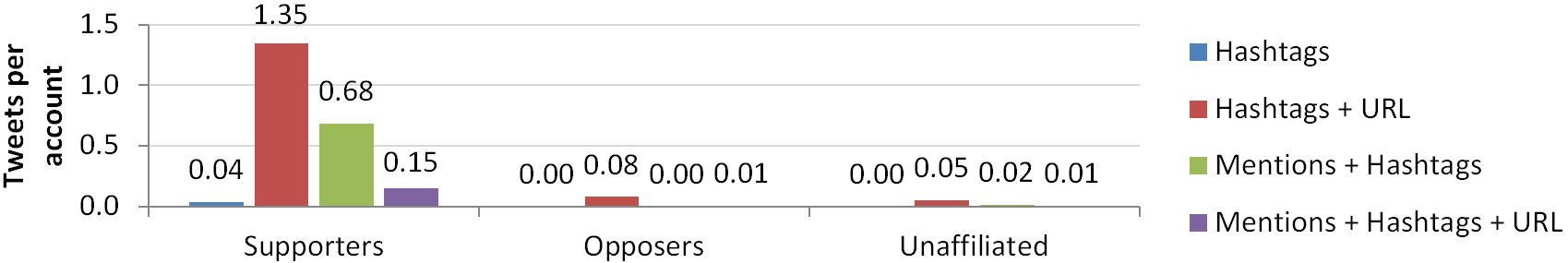}
    \caption{Rates of use of inauthentic tweet text patterns per account for the $497$ Supporters, $593$ Opposers and $11{,}782$ Unaffiliated accounts over the entire ArsonEmergency dataset.}
    \label{fig:inauth_text_pattern_tweet_rates}
\end{figure}

The frequencies of the occurrence of these text patterns in tweets by each group, in each phase and overall, is shown above in Table~\ref{tab:inauthentic_tweet_text_patterns}. The majority of these behaviours were present in Phase~3. 
Although Unaffiliated accounts certainly used some of these patterns, Supporters made much more use of them, particularly more than Opposers (Figure~\ref{fig:inauth_text_pattern_tweet_rates}). Many of the instances of hashtags followed by a URL are instances of quote tweets, where the URL is the link to the quoted tweet. These are attempts to disseminate the quoted tweet to a broader audience (engaged through the hashtags).

Finally, inspection of the ten most retweeted tweet contributors revealed that three were Supporters, one was Unaffiliated, and the remainder were Opposers (including five of the top six).

\section{Hashtag Use}

As expected, the most prominently used hashtag for all communities was \hashtag{ArsonEmergency}, however it is clear that there are other commonly occurring hashtags. Table~\ref{tab:top_hashtags} shows the top ten hashtags used by the Supporters, Opposers and Unaffiliated in each phase, as well as the number of tweets in which they appeared. 

In Phase~1, it is clear that the Supporters are trying to engage with existing climate change emergency discussion communities, as well as the media (\hashtag{7news}) and broader political discussion (\hashtag{auspol}). The few Opposer tweets seem to be poking fun at the discussion (e.g., \hashtag{RelevanceDepravationEmergency}, \hashtag{PoliticalBSEmergency}), while the Unaffiliated tweets are very broadly about the bushfires, but \hashtag{ClimateChangeHoax} is the third most used hashtag.

In the brief Phase~2, Supporters appear to be more concentrated in their promotion of the arson narrative (using \hashtag{ClimateCriminals} and \hashtag{ecoterrorism}) into the \hashtag{auspol} political discussion. Opposers seem to focus almost exclusively on using \hashtag{ArsonEmergency} rather than any other hashtags, while the Unaffiliated still follow, to some extent, the Supporters' lead with hashtags related to the arson narrative.

Finally, in Phase~3, Supporters focus mostly on just \hashtag{ArsonEmergency}, briefly linking to blaming an environmental political party and references to hoaxes, and even reversing the attack and accusing others of being \hashtag{ArsonDeniers}. Opposers are firmly focused on \hashtag{ArsonEmergency} but start referring to an individual prominent in the media industry commonly seen as advocating against dealing with climate change. By this stage, the Unaffiliated accounts are starting to follow the Opposers' lead discussing emergency- and fire-related hashtags.

\begin{table}[ht]
    \centering
    \caption{The top ten hashtags used by the Supporters, Opposers, and Unaffiliated communities in each phase. Hashtags have been compared without considering case in the same way Twitter does. The tag $anon_1$ in Phase~3 refers to the same redacted identity in 
    Figure~\ref{fig:cca_hashtag_co-mentions} (main paper).
    }
    \label{tab:top_hashtags}
    \resizebox{\textwidth}{!}{%
        \begin{tabular}{@{}llrlrlr@{}}
            \toprule
                    \multirow{15}{*}{\rotatebox[origin=c]{90}{\textbf{Phase 1}}} 
                    & \multicolumn{2}{c}{\textbf{Supporters}} & \multicolumn{2}{c}{\textbf{Opposers}}       & \multicolumn{2}{c}{\textbf{Unaffiliated}}     \\
                    & \multicolumn{2}{c}{\emph{1,573 Tweets}} & \multicolumn{2}{c}{\emph{33 Tweets}} & \multicolumn{2}{c}{\emph{1,961 Tweets}} \\
                    \cmidrule(r){2-3}                \cmidrule(lr){4-5}                   \cmidrule(l){6-7}
                    & Hashtag              & Count   & Hashtag                       & Count & Hashtag                    & Count  \\
                    \cmidrule(r){2-3}                \cmidrule(lr){4-5}                   \cmidrule(l){6-7}
                    & arsonemergency       & 2,086   & arsonemergency                & 43    & arsonemergency             & 2,534  \\
                    & auspol               & 574     & auspol                        & 9     & auspol                     & 1,012  \\
                    & climatechangehoax    & 232     & bushfires                     & 7     & climatechangehoax          & 682    \\
                    & climateemergency     & 230     & tresspassemergency            & 6     & climatechange              & 611    \\
                    & climatechange        & 191     & lootingemergency              & 6     & australiaburns             & 307    \\
                    & 7news                & 126     & bandeemergency                & 6     & australiaburning           & 227    \\
                    & vicfires             & 111     & theftemergency                & 5     & climateemergency           & 186    \\
                    & victoria             & 107     & relevancedepravationemergency & 4     & australiabushfires         & 142    \\
                    & nswfires             & 90      & politicalbsemergency          & 4     & bushfireemergency          & 133    \\
                    & globalwarming        & 84      & denialmachine                 & 4     & australianfires            & 78     \\
            \midrule
            \multirow{15}{*}{\rotatebox[origin=c]{90}{\textbf{Phase 2}}} & \multicolumn{2}{c}{\emph{121 Tweets}} & \multicolumn{2}{c}{\emph{327 Tweets}} & \multicolumn{2}{c}{\emph{759 Tweets}} \\
                    \cmidrule(r){2-3}                \cmidrule(lr){4-5}                   \cmidrule(l){6-7}
                    & Hashtag              & Count   & Hashtag                       & Count & Hashtag                    & Count  \\
                    \cmidrule(r){2-3}                \cmidrule(lr){4-5}                   \cmidrule(l){6-7}
                    & arsonemergency       & 142     & arsonemergency                & 487   & arsonemergency             & 1,135  \\
                    & auspol               & 79      & auspol                        & 36    & auspol                     & 194    \\
                    & bushfiresaustralia   & 51      & climateemergency              & 11    & bushfiresaustralia         & 110    \\
                    & climateemergency     & 26      & scottyfrommarketing           & 9     & climateemergency           & 53     \\
                    & climatecriminals     & 23      & australianbushfires           & 9     & climatecriminals           & 34     \\
                    & climatechange        & 8       & australiaisburning            & 9     & climatechange              & 23     \\
                    & victoria             & 7       & dontgetderailed               & 7     & climatechangehoax          & 18     \\
                    & ecoterrorism         & 6       & arsonmyarse                   & 7     & scottyfrommarketing        & 16     \\
                    & australiaisburning   & 6       & stupidemergency               & 6     & australianbushfires        & 15     \\
                    & australiaburning     & 6       & australiabushfire             & 6     & astroturfing               & 15     \\
            \midrule
            \multirow{15}{*}{\rotatebox[origin=c]{90}{\textbf{Phase 3}}} & \multicolumn{2}{c}{\emph{5,278 Tweets}} & \multicolumn{2}{c}{\emph{3,227 Tweets}} & \multicolumn{2}{c}{\emph{14,267 Tweets}} \\
                    \cmidrule(r){2-3}                \cmidrule(lr){4-5}                   \cmidrule(l){6-7}
                    & Hashtag              & Count   & Hashtag                       & Count & Hashtag                    & Count  \\
                    \cmidrule(r){2-3}                \cmidrule(lr){4-5}                   \cmidrule(l){6-7}
                    & arsonemergency       & 7,731   & arsonemergency                & 5,070 & arsonemergency             & 21,194 \\
                    & auspol               & 534     & australiafires                & 649   & australiafires             & 2,747  \\
                    & climateemergency     & 477     & climateemergency              & 601   & climateemergency           & 2,566  \\
                    & itsthegreensfault    & 270     & $anon_1$                      & 427   & $anon_1$                   & 1,778  \\
                    & climatechangehoax    & 270     & bushfires                     & 251   & australianbushfiredisaster & 1,101  \\
                    & climatechange        & 226     & auspol                        & 210   & auspol                     & 1,011  \\
                    & climatehoax          & 220     & australianbushfiredisaster    & 152   & climatechangehoax          & 758    \\
                    & climatecriminals     & 177     & climatechange                 & 140   & australianbushfires        & 739    \\
                    & bushfires            & 176     & fakenews                      & 137   & climatechange              & 721    \\
                    & arsondeniers         & 169     & australianbushfires           & 101   & bushfires                  & 664    \\

            \bottomrule
        \end{tabular}%
    }
\end{table}

\end{appendices}

\bibliography{ref}



\end{document}